\documentclass[12pt, journal, twocolumn]{IEEEtran}
\usepackage{amssymb,amsmath}
\usepackage{cite}
\usepackage{graphicx,subfigure}
\usepackage{psfrag}
\usepackage{url}
\usepackage{booktabs}
\usepackage[latin1]{inputenc}
\usepackage{multirow}
\usepackage{caption}
\usepackage[table]{xcolor}
\usepackage[absolute,overlay]{textpos}
\usepackage[latin1]{inputenc}
\usepackage{color}
\usepackage{hyperref}

\usepackage{array}
\newcolumntype{L}[1]{>{\raggedright\let\newline\\\arraybackslash\hspace{0pt}}m{#1}}
\newcolumntype{R}[1]{>{\raggedleft\let\newline\\\arraybackslash\hspace{0pt}}m{#1}}
\newcolumntype{C}[1]{>{\centering\let\newline\\\arraybackslash\hspace{0pt}}m{#1}}
\usepackage{arydshln}
\usepackage{hhline}
\DeclareTextCommandDefault{\textcopyright}{\textcircled{c}}

\begin{document}	
 \title{Energy-Sustainable IoT Connectivity:\\ 	Vision, Technological Enablers, Challenges, and Future Directions}
	\author{\IEEEauthorblockN{Onel L. A. L\'opez,~\IEEEmembership{Member,~IEEE,}
			Osmel M. Rosabal,~\IEEEmembership{Graduate Student Member,~IEEE,}
			David Ruiz-Guirola,~\IEEEmembership{Graduate Student Member,~IEEE,}
			Prasoon Raghuwanshi,~\IEEEmembership{Student Member,~IEEE,}
                Konstantin Mikhaylov,~\IEEEmembership{Senior~Member,~IEEE,}
                Lauri Lov\'en,~\IEEEmembership{Senior~Member,~IEEE,}
			Sridhar Iyer,~\IEEEmembership{Senior~Member,~IEEE}
		}
		\thanks{O. L\'opez, O. Rosabal, D. Ruiz-Guirola, P. Raghuwanshi, and K. Mikhaylov
  are with the Centre for Wireless Communications (CWC), while L. Lov\'en is with the Center for Ubiquitous Computing (UBICOMP), all at the University of Oulu, Finland. S. Iyer is with the Department of CSE(AI), KLE Technological University Dr. MSSCET, India. Emails: \{onel.alcarazlopez, osmel.martinezrosabal, david.ruizguirola, prasoon.raghuwanshi, konstantin.mikhaylov, lauri.loven\}@oulu.fi, sridhariyer1983@klescet.ac.in.}
		\thanks{This work is partially supported by the Finnish Foundation for Technology Promotion, Academy of Finland (Grants 348515, 341111 and 346208 (6G Flagship)), by the European Commission through the Horizon Europe/JU SNS project Hexa-X-II (Grant Agreement no. 101095759) and the ECSEL JU FRACTAL project (Grant 877056), receiving support from the EU Horizon 2020 programme and Spain, Italy, Austria, Germany, France, Finland, Switzerland, and the Neural pub/sub research project, funded by Business Finland with diary number 8754/31/2022.}
        \thanks{This work has been accepted for publication in IEEE Open Journal of the Communications Society. $\textcopyright$ 2023 IEEE. Personal use of this material is permitted. Permission from IEEE must be obtained for all other uses, in any current or future media, including reprinting/republishing this material for advertising or promotional purposes, creating new collective works, for resale or redistribution to servers or lists, or reuse of any copyrighted component of this work in other works.}}				
	\maketitle
	\begin{abstract}
 Technology solutions must effectively balance economic growth, social equity, and environmental integrity to achieve a sustainable society. Notably, although the Internet of Things (IoT) paradigm constitutes a key sustainability enabler, critical issues such as the increasing maintenance operations, energy consumption, and manufacturing/disposal of IoT devices have long-term negative economic, societal, and environmental impacts and must be efficiently addressed.  This calls for self-sustainable IoT ecosystems requiring minimal external resources and intervention, effectively utilizing renewable energy sources, and recycling materials whenever possible, thus encompassing energy sustainability. In this work, we focus on energy-sustainable IoT during the operation phase, although our discussions sometimes extend to other sustainability aspects and IoT lifecycle phases. Specifically, we provide a fresh look at energy-sustainable IoT and identify energy provision, transfer, and energy efficiency as the three main energy-related processes whose harmonious coexistence pushes toward realizing self-sustainable IoT systems. Their main related technologies, recent advances, challenges, and research directions are also discussed. Moreover, we overview relevant performance metrics to assess the energy-sustainability potential of a certain technique, technology, device, or network, together with target values for the next generation of wireless systems, and discuss protocol, integration, and implementation issues.  Overall, this paper offers insights that are valuable for advancing sustainability goals for present and future generations.
	\end{abstract}
 
\vspace{-2mm} 
\section{INTRODUCTION}\label{Intro}
\begin{figure}[t!]
        \centering
        \includegraphics[width=0.8\columnwidth]{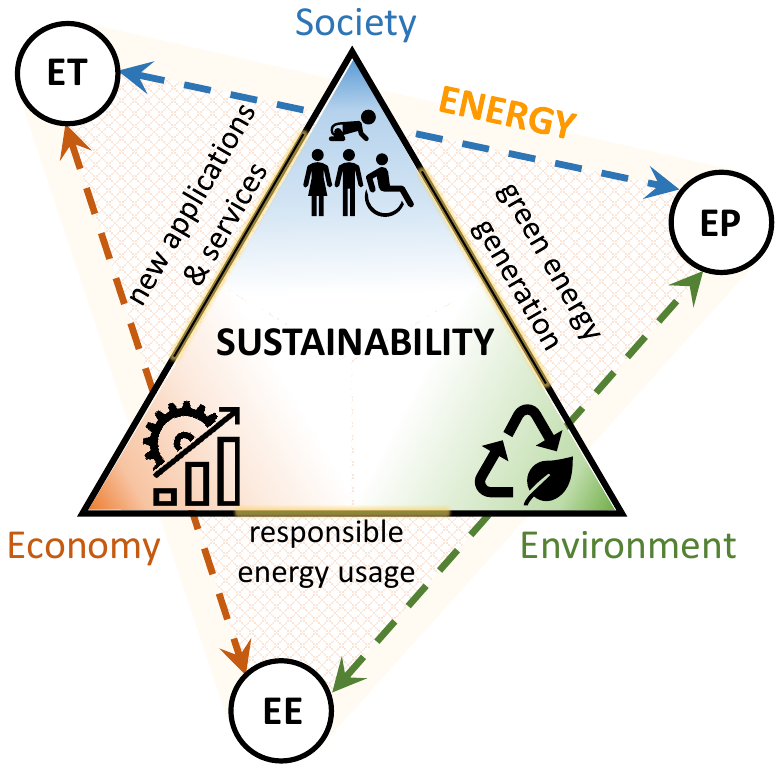}
        \caption{Sustainability triangle (with economy, environment, and society as corners) and energy sustainability supported by EP, ET, and EE processes and related technologies.}
        \label{Sustainability}      
        \vspace{-7mm}    
    \end{figure}

\IEEEPARstart{T}{he} vision of a sustainable society is relentlessly pursued by industry, academic research,  and regulatory bodies following the famous United Nations' sustainable development goals introduced in 2015\footnote{Refer to \url{https://sdgs.un.org/goals}.}. In general, sustainability is stimulated by \emph{simultaneously developing the economy, promoting social equity, and protecting the integrity of the environment for current and future generations} \cite{Munasinghe.1993}, as
captured by the so-called sustainability triangle illustrated in Figure~\ref{Sustainability}. Therefore, associating sustainability solely with environmentally conscious (also referred to as ``green'') practices is a widespread misconception since sustainability comprises Economic, Societal (Equity), and Environmental factors (also known as the 3 E's of sustainability). 

Quantifying/measuring and developing standardized metrics and benchmarks for sustainability is quite a challenging endeavor because i) sustainability is a subjective concept that can be assessed differently by various stakeholders; ii) the interdependencies, trade-offs, and feedback loops between the multiple interconnected systems and performance corners (economic, societal, and environmental) must be captured; and iii) pursuing specific long-term perspectives while measuring and tracking progress in the short term may not be easy and viable in general. This is why, despite numerous attempts over the years, e.g., \cite{Martins.2007,Boulanger.2008,Vivien.2008,Blackburn.2012,Ratiu.2014,Mura.2018,Corona.2019,Walzberg.2021}, none of them have gained significant traction.

 Regardless of the diverse assessment criteria and methods, sustainability enablers are generally well-established and include Internet of Things (IoT) technologies \cite{mao2017survey, Meng.2018, Alsamhi.2019, Sandro.2020, Hossein.2020, Hu.2020, Cetinkaya.2020, Marja.2020,Mahmood.2020,MahmoodAlves.2020,Mahmood.2021,Alves.2021,Lopez.2021, Albreem.2021,Bradu.2022,Sabovic.2023, Rahmani.2023}. Specifically, IoT solutions promote i) economic development (by enabling automation, facilitating accurate decision-making via massive data collection and real-time processing, optimizing resource usage and efficiency, improving maintenance processes, and reducing systems' downtime); ii) social equity and well-being (by providing enhanced safety and security measures against potential safety hazards or security breaches and by facilitating flexibility and barrier-free experience to the end users via the ubiquitous provision of services and the exploitation of remote-controlled or completely autonomous systems); and iii) environment protection (by optimizing energy usage and reducing waste/pollution); thus advancing sustainability goals. 

Notably, IoT systems are turning wireless connectivity into a basic utility, like water and electricity, while becoming increasingly massive \cite{Mahmood.2020,Mahmood.2021}. Specifically, the number of IoT devices is forecast to almost triple from 9.7 billion in 2020 to more than 29 billion IoT devices in 2030\footnote{Refer to \url{https://www.statista.com/statistics/1183457/iot-connected-devices-worldwide}.} to support 
sustainable agriculture \cite{Venkatesan.2017,Sandro.2020,Maroli.2021,Khan.2021,Albreem.2021}, transportation \cite{mao2017survey,Sandro.2020,Hossein.2020}, manufacturing \cite{Meng.2018,Rahmani.2023}, wireless charging \cite{Hu.2020,Cetinkaya.2020,Sandro.2020,Lopez.2021,Alves.2021,Rahmani.2023},  e-health \cite{Alsamhi.2019,Sandro.2020,Grua.2021,Albreem.2021}, cities \cite{Zhao.2013,Alsamhi.2019,Sandro.2020,Hossein.2020,Belli.2020,Avinash.2021},  water/air/energy management \cite{Narendran.2017,Sandro.2020,Hu.2020,Hossein.2020,Avinash.2021,Aivazidou.2021,Albreem.2021}, computing \cite{mao2017survey,Alsamhi.2019,Butt.2020,Hossein.2020,Hu.2020,Law.2021}, and many other use cases. Interestingly, this massive IoT growth poses sustainability challenges on its own mostly due to the increased maintenance operations, energy consumption (EC) to charge the IoT devices but also to store and process the massive amount of IoT data at the data and computing centers, and manufacturing/disposal of IoT devices, which may have long-term negative economical, societal, and environmental impacts.  This calls for self-sustainable IoT ecosystems, which require minimal external resources or intervention and can function independently by utilizing renewable energy sources and recycling materials whenever possible.
Notice that truly self-sustainable IoT approaches must consider sustainability aspects along the entire IoT product lifecycle, i.e., planning, manufacturing, deployment, operation (including the target use case), maintenance, and disposal.
\begin{table*}[t!]
            \caption{Representative Survey/Overview/Tutorial Papers Related to Energy-Sustainable IoT Connectivity (2017-2023)}
            \vspace{-1mm}
            \centering            
            \begin{tabular}{C{0.6cm}C{0.39cm}C{0.39cm}C{0.39cm}L{4.5cm}L{8.9cm}}
                \toprule
                 \textbf{Ref.} & \multicolumn{3}{c}{\textbf{Energy Process}} & \textbf{Technology Enablers}  & \textbf{Main Contributions}  \\  
                 & EP & ET & EE & & \\
                \midrule
                \cite{mao2017survey} & X & & X & green mobile edge computing (MEC) & $\bullet$ survey of the state-of-the-art on joint radio-and-computational resource management in MEC \newline $\bullet$ discussion on MEC challenges and future research directions, including MEC system deployment, cache-enabled MEC, mobility management for MEC, green MEC, and privacy-aware MEC \\ \hhline{~-----}
        \cite{Meng.2018} & X & & X & energy harvesting (EH), smart grids,  cloud computing, RFID, machine learning (ML)  & $\bullet$ review of  smart manufacturing and sustainable energy industry \newline $\bullet$ discussion of prominent technologies to support sustainable industrial IoT  \\ \hhline{~-----}
        \cite{Alsamhi.2019} & X & & X & EH, RFID, wake-up radio (WuR), cloud computing, data/communication reduction  & $\bullet$  survey of potential technologies for greening the IoT and relevant research directions \\ \hhline{~-----}
             \cite{Sandro.2020} & X & & X & smart microgrids, IoT for waste \& energy management, ML,  low-power wide-area networks,  hardware (HW) and software (SW) duty cycling   & $\bullet$ review of the state of the art research and advances on IoT technologies in/for energy management, environment monitoring, transportation, e-health, and smart cities \\ \hhline{~-----}
            \cite{Hossein.2020} & X & & X & EH, low-power wide-area networks, Bluetooth low energy (BLE), MEC, smart grids  & $\bullet$ review of applications of IoT in energy systems, especially in the context of smart grids \newline $\bullet$ discussion of IoT-enabling technologies and challenges in the energy sector, including privacy and security \\  \hhline{~-----}
            \cite{Hu.2020} & X & X & X & EH, wireless ET (WET), smart grids, MEC, unmanned aerial vehicle (UAV)-assisted communications & $\bullet$ overview of advances/prospects on the utilization, redistribution, trading, and planning of harvested energy in future wireless networks \newline $\bullet$ review of wireless power and information transfer technologies \\ \hhline{~-----}
            \cite{Cetinkaya.2020} & X & X & & EH, WET  & $\bullet$ proposal of a self-sufficient IoT architecture that adopts only single- and double-hop energy and data transitions to enable efficient energy sharing and reduced data traffic  \newline $\bullet$ discussion of enablers and identification of future research directions  \\ \hhline{~-----}
              \cite{Lopez.2021}  & X & X & & green radio-frequency (RF)  WET, distributed ledger technology (DLT) for energy trading, ultra-low power receivers, enhanced energy transmitters, metasurface-assisted RF-WET  & $\bullet$ discussion of RF-WET for enabling self-sustainable IoT \newline $\bullet$ overview of the main architectures, challenges, and techniques for efficient and scalable RF-WET \newline $\bullet$ outline of key research directions for realizing RF-powered 6G IoT\\ \hhline{~-----}
              \cite{Albreem.2021} & X & & X & RFID, EH, TinyML HW, ML, MEC & $\bullet$ overview of enablers, architectures, technologies, energy models, and strategies for green IoT \newline $\bullet$ discussion of effective behavioral change models and strategies to create energy-awareness among users and IoT service providers  \\ \hhline{~-----}
              \cite{Sabovic.2023} & X & & X & EH, energy-aware TinyML, MEC & $\bullet$ discussions on the incorporation of TinyML algorithms and application tasks on battery-less IoT devices \newline $\bullet$  study of several inference strategies, including local/cloud computing \newline $\bullet$ testing a battery-free person detection prototype \\ \hhline{~-----}
              \cite{Rahmani.2023} & X & X & X & EH, RF harvesters, WET, backscattering &  $\bullet$ overview of HW-related research trends, application use cases, and enabling technologies for sustainable IoT systems \newline $\bullet$ high-level discussion of eco-friendly manufacturing, sustainable WET, and low-power wireless connectivity solutions   \\
                \bottomrule
            \end{tabular}
            \label{art}
            \end{table*}
            \vspace{-2mm}
 \subsection{\uppercase{Energy-Sustainable IoT Ecosystem}} 	
Undoubtedly, one of the critical aspects of self-sustainable IoT systems is energy. Self-sustainable solutions must be energy-sustainable, which entails green\footnote{Green energy refers to energy from renewable sources.} energy autonomy when considering solely the operation phase of an IoT product lifecycle. Energy-sustainable IoT operation is precisely the focus of this work, although our discussions sometimes extend to other sustainability aspects and IoT lifecycle phases. Readers interested in specific discussions on sustainability at planning, manufacturing, deployment, maintenance, and disposal phases are encouraged to refer to \cite{Meng.2018,Malgorzata.2019,Keivanpour.2019,Jasiulewicz.2020,Avinash.2021,Kumar.2021,Pascal.2022,Marym.2022,Rahmani.2023}. 

The main energy processes that are present in an energy-sustainable IoT ecosystem are illustrated in Figure~\ref{Sustainability} together with the sustainability triangle and discussed below.
\begin{figure*}[t!]
        \centering
        \includegraphics[width=.9\linewidth]{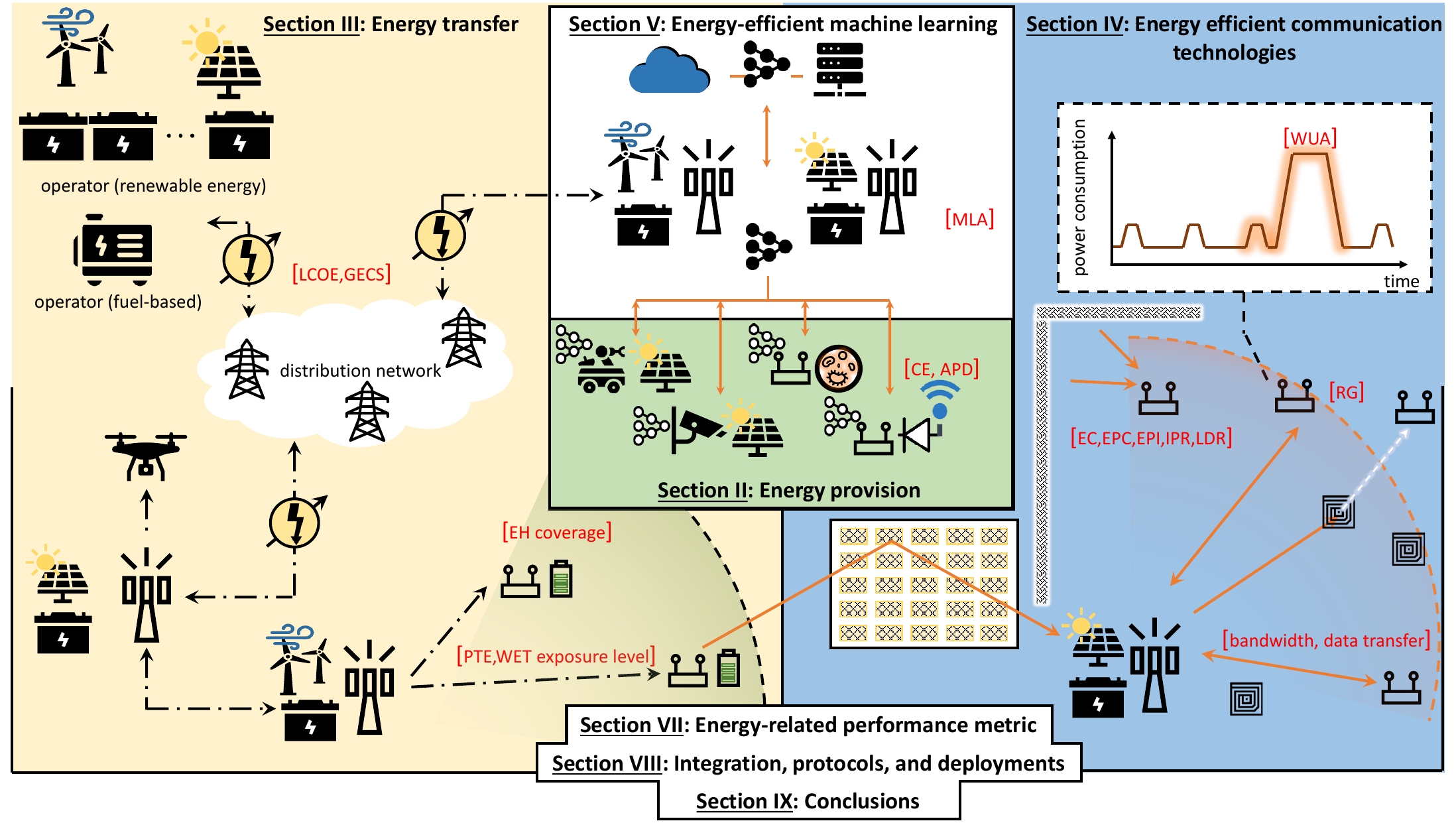}
        \vspace{-1mm}
        \caption{Integration of the EP, ET, and EE processes throughout the structure of the paper. The terms in brackets refer to KPIs, which are discussed in Section~\ref{KPI}.}
        \label{fig:Intro}
        \vspace{-7mm}
    \end{figure*}

    \textbf{Energy provision (EP)} refers to the charging process(es) exploiting green energy sources. Currently, the information and communication technology industry accounts for $1-1.5\%$ of the global electricity use \cite{iea2022data}, a figure that could rise to nearly $4\%$ considering the increasing traffic demands at the transmission networks and the computation requirements at the data centers \cite{freitag2021real}. Consequently, the implementation of sustainable energy in this sector will have a significant impact on the environment considering that the current average global carbon intensity factor\footnote{Refer to \url{https://ourworldindata.org/grapher/carbon-intensity-electricity}.} is approximately $0.441$ kg of CO$_2$ emissions per kWh. Furthermore, certain EP implementations enable the production of electricity at a local level, thereby making electricity (and also information and communication technology services) accessible to remote communities.
    
    \textbf{Energy transfer (ET)} refers to the intentional movement of energy from one device/system to another. This is required, for instance, when green energy sources are not available for direct exploitation (or provide insufficient energy). In such cases, another device/system powered by green sources can transfer such required energy. The global market projections indicate a significant increase in the value of ET technologies, reflecting the growing interest in this field. As per \cite{wetmarket2021}, the global market of ET is projected to surge from USD $\$5,705.1$ million in $2020$ to $\$35,226.4$ million by 2030, with a resulting average annual growth rate of $21.3\%$. Consequently, technological breakthroughs in ET will generate new business opportunities, enable previously infeasible IoT use cases, and facilitate the ubiquity of ET services across the globe. 
    
    \textbf{Energy efficiency (EE)} refers to the ability of a device/system/process to perform its intended function with minimal energy and thus is a measure of how effectively energy is used to achieve a desired outcome \cite{Sioshansi.2013}. 
    Obviously, EE directly impacts the environment and economy sustainability corners as it pushes to minimize EC and processing, i.e., toward more responsible energy usage. Notably, an ambitious goal for the sixth generation (6G) of wireless systems is to realize $10-100$-fold EE gains with respect to the current fifth-generation (5G) networks \cite{Jiang.2021}. Notice that EE designs are often low-complexity/cost, thus promoting sustainability in other IoT lifecycle phases as well.

The harmonious coexistence of the above processes ultimately pushes to zero dirty-EC,  thus, realizing truly energy-sustainable IoT systems.   
\begin{figure*}[t!]
\captionof{table}{List of Acronyms}
\vspace{-2mm}
\label{acronym}
\begin{minipage}{.05\linewidth}
\ \ 
\end{minipage}
\begin{minipage}{1.9\linewidth}
\footnotesize
    \begin{tabular}{L{1.7cm}L{5.8cm}L{0.0cm}L{1.7cm}L{5.9cm}}
         \toprule
                \textbf{   Acronym} & \  \textbf{Definition} &  & \textbf{Acronym} & \textbf{Definition} \\ 
                \midrule
    \end{tabular}
    \end{minipage}\\
 \begin{minipage}{.48\linewidth}
\footnotesize
 \begin{tabular}{L{1.6cm}L{6.2cm}|}   
3GPP & 3rd Generation Partnership Project \\
5G/6G & fifth/sixth generation \\ 
AHE & average harvested energy \\
APD & average power density \\
BC & backscatter communication \\
BCAP & battery capacity \\
BL & battery lifetime \\
BLE & Bluetooth low energy \\
BoI & bandwidth of influence \\
BS & base station \\
CA/CV & coverage area/volume \\
CE & conversion efficiency \\
CEE & component EE \\
CEP & complex event processing \\
CPV & concentrator photovoltaic \\
CSI & channel state information \\
CU/DU & centralized/distributed unit \\
DC & direct-current \\
DL & deep learning \\
DLT & distributed ledger technology \\
DRL/DSL & deep RL/SL \\
EC & energy consumption \\
ED & energy density \\
EE/EP/ET & energy efficiency/provision/transfer \\   
EER & EE ratio  \\
EH & energy harvesting \\
EM, EMF & electromagnetic, electromagnetic field \\
ENO & energy-neutral operation \\
EOP & energy outage probability \\
EPC/EPI & energy proportionality coefficient/index \\
FL & federated learning \\
FLOPS & floating-point operations per second \\
FN/FP & false negative/positive \\
GECS & green EC share \\
GLB & geographical load balancing \\
GRS & grid reliability/stability \\
HW/SW & hardware/software \\
INFT & inference time \\
IoT & Internet of Things\\
IPR & idle-to-peak power ratio \\            
IRS & intelligent reflective surface \\          
KPI & key performance indicator \\
LA & localization accuracy \\
LCOE & levelized cost of electricity \\
LIS & large intelligent surface 
\end{tabular}
\end{minipage}
\begin{minipage}{.48\linewidth}
  \footnotesize
    \begin{tabular}{L{1.7cm}L{5.9cm}}
            LOS & line-of-sight \\ 
            MAE & mean absolute error \\
            MCU & microcontroller \\
            MDHE & meta distribution of the harvested energy \\
            MDS & model size \\
            MEC & mobile edge computing \\
            MFC & microbial fuel cell \\
            MIMO & multiple-input multiple-output \\
            MIPS & million instructions per second \\
            ML & machine learning \\
            MLA & ML accuracy \\
            MPE & maximum level of exposure \\      
            NAS & neural architecture search \\
            NHE & net harvested energy \\
            NFV & network functions virtualization \\            
            NN, FANN & neural network, fast neural network\\
            O-RAN & Open-RAN \\
            PB/PC & power beacon/consumption \\
            PFA & probability of false-alarm \\
            PHY & physical layer \\
            PKD & peak demand \\
            PMD & probability of miss-detection \\
            PTE & power transfer efficiency \\
            PULP & parallel ultra-low power \\
            PV & photovoltaic cell \\           
            QoS & quality of service \\
            RAN/RAT & radio access network/technology \\
            RF & radio-frequency \\
            RG & range \\            
            RL & reinforcement learning \\
            RMSE & root mean square error \\
            RU & radio unit \\  
            SL & supervised learning \\
            STD & standard deviation \\
            TEG & thermoelectric generator \\
            THP & throughput \\
            TN/TP & true negative/positive \\
            TX/RX & transmit/receive \\
            UAV & unmanned aerial vehicle \\
            UE & user equipment \\
            UL & unsupervised learning \\
            VEH & vibration-based EH \\     
            WET & wireless ET \\
            WUA & wake-up accuracy \\
            WuR/WuS & wake-up radio/signal
            \end{tabular}
      \end{minipage}\\   
  \centering
    \rule{0.98\linewidth}{0.4pt}
\vspace{-5mm}
\end{figure*}
   \vspace{-4mm}
\subsection{\uppercase{Contributions \& Organization of this Work}}
The realization of energy-sustainable IoT connectivity depends critically on the technological advancements in the EP, ET, and EE processes, and their holistic integration. Notably, the related state-of-the-art discussions are sparse in the literature as most works focus on a specific enabling technology. Only a few other works, such as those listed in Table~\ref{art}, have a wider scope, although still limited in terms of technology enablers, recent advances, key performance indicators (KPIs), challenges, and/or research directions, and their link to sustainability. Moreover, they generally lack clarity regarding the role or classification of the discussed technologies/approaches toward supporting (self)-sustainable IoT connectivity, e.g., by associating them with the corresponding EP, ET, and EE processes. 

Our work aims to address the aforementioned research gap by making the following contributions:
\begin{itemize}
\vspace{-1mm}
    \item We provide a fresh look at energy-sustainable IoT. Moreover, we identify EP, ET, and EE as the three main energy-related processes whose harmonious coexistence pushes toward realizing self-sustainable IoT systems. This is mostly covered in Section~\ref{Intro}.
    \item We discuss the main technologies, recent advances, and challenges and associated research directions for EP (Section~\ref{EP}), ET (Section~\ref{ETr}), and EE (including communication (Section~\ref{EE}) and learning/computation (Section~\ref{ML})-related aspects processes to support sustainable IoT connectivity. 
    \item We present a set of energy-related performance metrics to assess the absolute/relative performance potential of a certain technique, technology, device, or network. They are classified as i) energy conversion and transfer, ii) energy storage and consumption, iii) EE ratio (EER), and iv) other metrics. Also, we overview related KPIs for the next generation of wireless systems. This is comprehensively covered in Section~\ref{KPI}.
    \item We discuss the integration of the surveyed technologies, suitable protocols and cross-layer designs, scalability issues, tools and testing/validation frameworks, and real-world IoT implementations targeting energy-sustainable IoT ecosystems. This, and corresponding challenges and opportunities, are covered in Section~\ref{integration}. 
\end{itemize}
In addition to these, we summarize the key relevant challenges and research directions both per technology and at the system level in Section~\ref{conclusion}. Figure~\ref{fig:Intro} illustrates how the structure of the paper covers the integration of the EP, ET, and EE processes. Finally, Table~\ref{acronym} lists the acronyms used throughout this article in alphabetical order.
\vspace{-2mm}
\section{\uppercase{Energy Provision (EP)}}\label{EP} 
The need for a reliable energy supply to power the IoT has made EH technologies appealing to reduce/eliminate the need for batteries. There are readily available energy sources in our environment such as sunlight, wind, heat, electromagnetic (EM) radiation, and many others. Table~\ref{tab:EPComparison} summarizes the advantages, limitations, and attractive use cases for the EH technologies covered in this section.

  \begin{table}[t!]
        \caption{Comparison among EP technologies}
        \vspace{-1mm}
        \label{tab:EPComparison}
        \centering
        \begin{tabular}{p{0.9cm} p{0.75cm}  p{2.9cm} p{2.5cm}}
            \toprule
             \textbf{Source} & \textbf{CoC$^a$} & \textbf{Limitations} &\textbf{Use cases} \\
             \midrule
             Light & M & Sensitive to blockage & Remote applications, indoor sensors  \\
             Heat & M-H & Sensitive to thermal stress & Wearables, infrastructure monitoring \\
             Microbes & M-VH & Low stability/lifespan & Smart farming, wastewater treatment \\
             Vibration & VH-H & Sensitive to mechanical fatigue & Wearables, infrastructure monitoring \\
             Flow & H & Sensitive to weather  & Medium to large EH systems \\
             RF & L & Very low output power & Ultra-low power devices \\
             \bottomrule
        \end{tabular}
        \begin{flushleft}{\footnotesize{$^a$ CoC refers to complexity and cost, which is classified as very high (VH), high (H), moderate (M), low (L), and very low (VL) }}\end{flushleft}
        \vspace{-4mm}
    \end{table}
  
\begin{figure}[t!]
        \centering
        \includegraphics[width=\columnwidth]{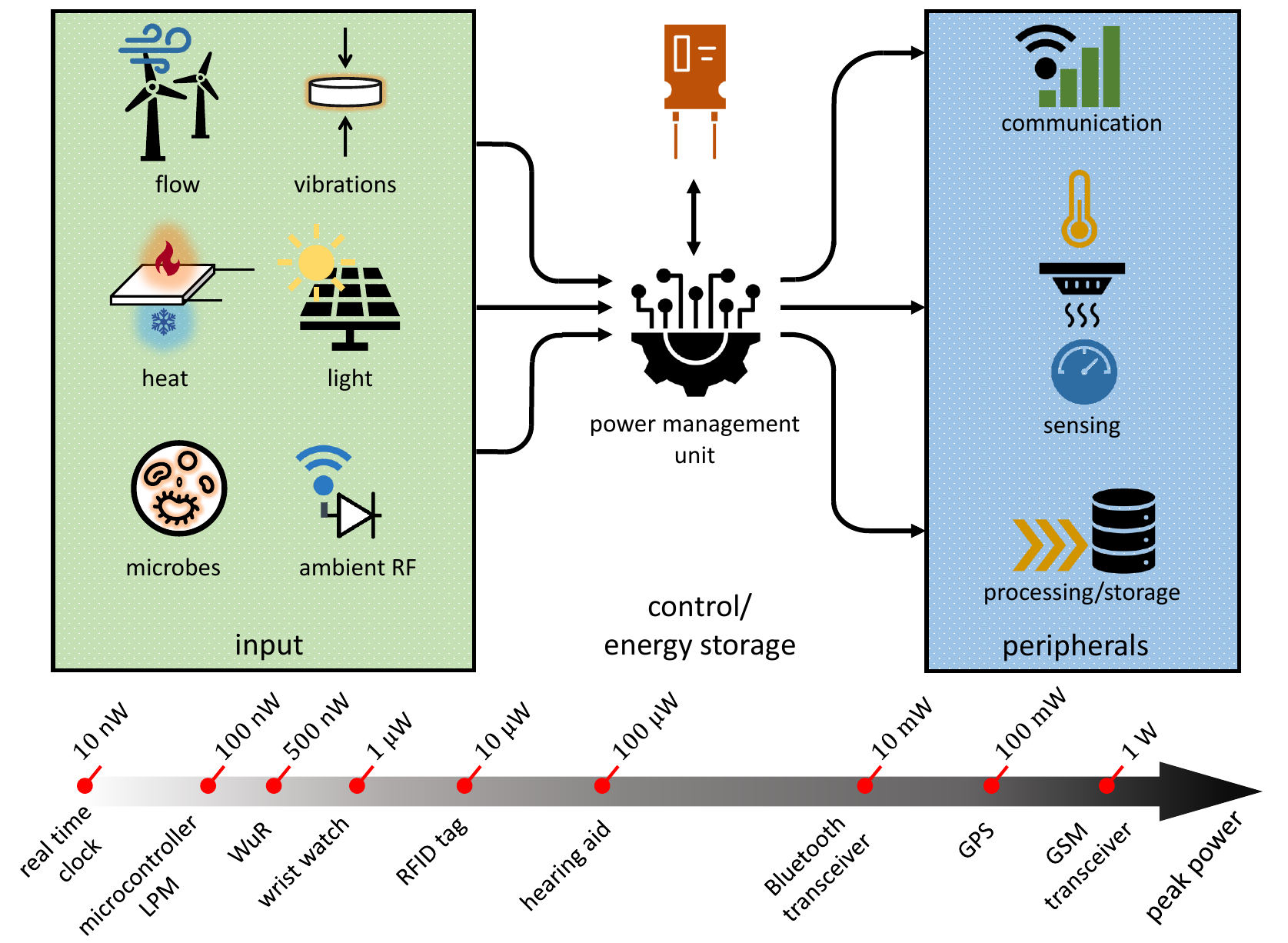}
        \vspace{-5mm}
        \caption{Anatomy of an EH device.}
        \label{fig:anatomyEHdevice}
        \vspace{-6mm}
    \end{figure}

 To power devices' electronics, EH circuits must transform the various energy forms into electric energy. For this, the architecture of a generic EH device, illustrated in Figure~\ref{fig:anatomyEHdevice}, consists of three main components:
    \begin{itemize}
        \item transducer, which turns a physical variable (or its variations) into electricity;
        \item power management unit, which accommodates the output signal of the transducer to power the devices' electronics and/or charge the storage element. One important function of this block is to dynamically match the output impedance of the transducer to a variable input impedance of a rectifying circuit or other device's peripherals;
        \item energy storage component, e.g., a battery or a supercapacitor, which buffers the variations of the ambient energy. We refer the reader to Section~\ref{KPI}-\ref{subsec:energystorageKPIs} for more details on this component.
    \end{itemize}

\vspace{-4.5mm}
    \subsection{\uppercase{Light-based EH}}
\begin{figure}
        \centering        
        \includegraphics[width=\columnwidth]{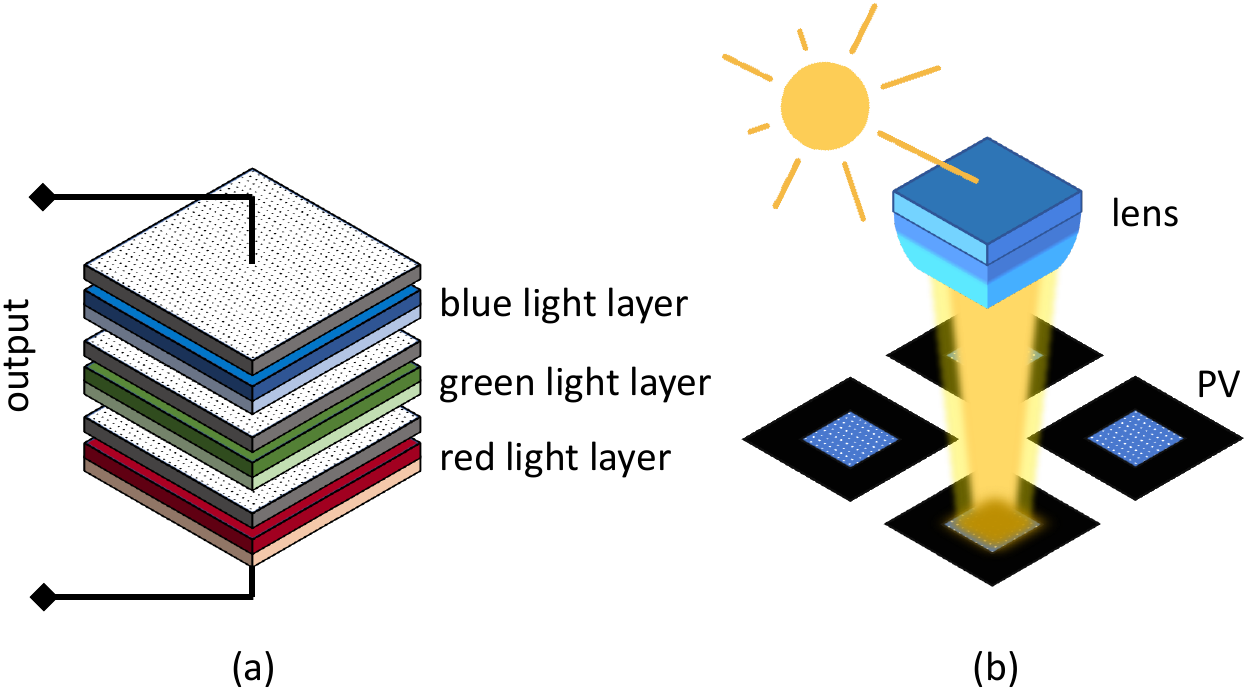}
        \caption{Efficient light-based EH: (a) multi-junction PV and (b) CPVs.}
        \vspace{-8mm}
        \label{fig:lightEH}
    \end{figure}

    Light-based EH systems exploit the photovoltaic effect to generate electricity from light using photovoltaic cells (PVs). Notice that sunlight can provide $100~\text{mW/cm}^2$ of average power density (APD) outdoors (only during the daytime), whereas combined artificial light and indirect sunlight can illuminate indoors with $100~\mu \text{W/cm}^2$ APD \cite{Sanislav.2021}. This opens the opportunity for indoor sensors illuminated by artificial light sources \cite{Shore.2021}, a mixture of artificial and sunlight, e.g., in asset tracking \cite{Phan.2022}, and for those in remote places exposed to sunlight with high probability, e.g., sensor and base stations (BSs) in smart agriculture applications \cite{Liu.2022}. However, these reference values can change depending on the geographic location, weather conditions, and indoor working hours. Increasing the conversion efficiency (CE) of PVs under different ambient light conditions is key for reducing the footprint, increasing the cost-effectiveness, and accelerating the adoption of light-based EH systems. In this front, multi-junction PVs, which, as shown in Figure~\ref{fig:lightEH}a, are constructed by stacking multiple p-n junctions, become appealing as each layer is optimized for harvesting at different wavelength regions, thus, increasing total harvested energy \cite{yamaguchi2021multi}. However, multi-junction PVs are challenging and expensive to manufacture, which motivated researchers to propose the replacement of some silicon layers with perovskite PVs. Perovskite-based multi-junction cells facilitate tuning the optical properties of the PVs for improved response in the solar spectrum and lower the manufacturing costs considerably \cite{torabi2019progress}. Unfortunately, perovskite PVs' instability in real-life conditions significantly degrades their CE over the long-term (compared to silicon PVs), which makes them (currently) a less favorable option for widespread use \cite{azadinia2021maximizing}.

    Concentrator photovoltaics (CPVs) are an alternate solution to boost the harvested energy. This system, as shown in Figure~\ref{fig:lightEH}b, relies on a built-in light concentrator to focus the incoming light from multiple directions on a very small semiconductor area \cite{vaidya2022immersion}. This eliminates the need for complex electro-mechanical tracking mechanisms, which increase the operational expenditure, aka OpEx, and limit the possible use cases, such as in tiny Bluetooth transmitters, due to the resulting heavier design and moving parts \cite{Soonsawad.2021}. Note that a high concentration of sunlight increases the operating temperature of CPVs, thus decreasing their efficiency. 
    
    In general, the seamless integration of light-based EH with the corresponding application is required. This is favored by CPVs, which allow the miniaturization of light-based EH systems to the millimeter and sub-millimeter scale with reduced manufacturing costs \cite{alves2019thin}. Recent developments even allow the implementation of flexible CPVs accommodating multiple PV modules in a small surface, leading to improved performance and seamless integration with a multitude of device designs \cite{Sato.2021}. Moreover, thin-film PVs improve the integration capabilities by allowing the deployment of light-based EH systems on complex-shaped surfaces such as cars' roofs and umbrellas \cite{li2021review}. In this regard, the recent introduction of organic 
    and perovskite PVs permits printing PVs on practically any surface (e.g., check the IoT device in \cite{Perera.2021}) without impacting their mechanical properties or transparency \cite{wang2019large}. Despite their current low efficiency, the abundance of required raw materials, low-cost and scalable manufacturing process, and the possibility of using biodegradable materials, make these cells a green choice for future EH IoT implementations.

\vspace{-3mm}
    \subsection{\uppercase{Heat-based EH}}
    
\begin{figure}
        \centering
        \includegraphics[width=\columnwidth]{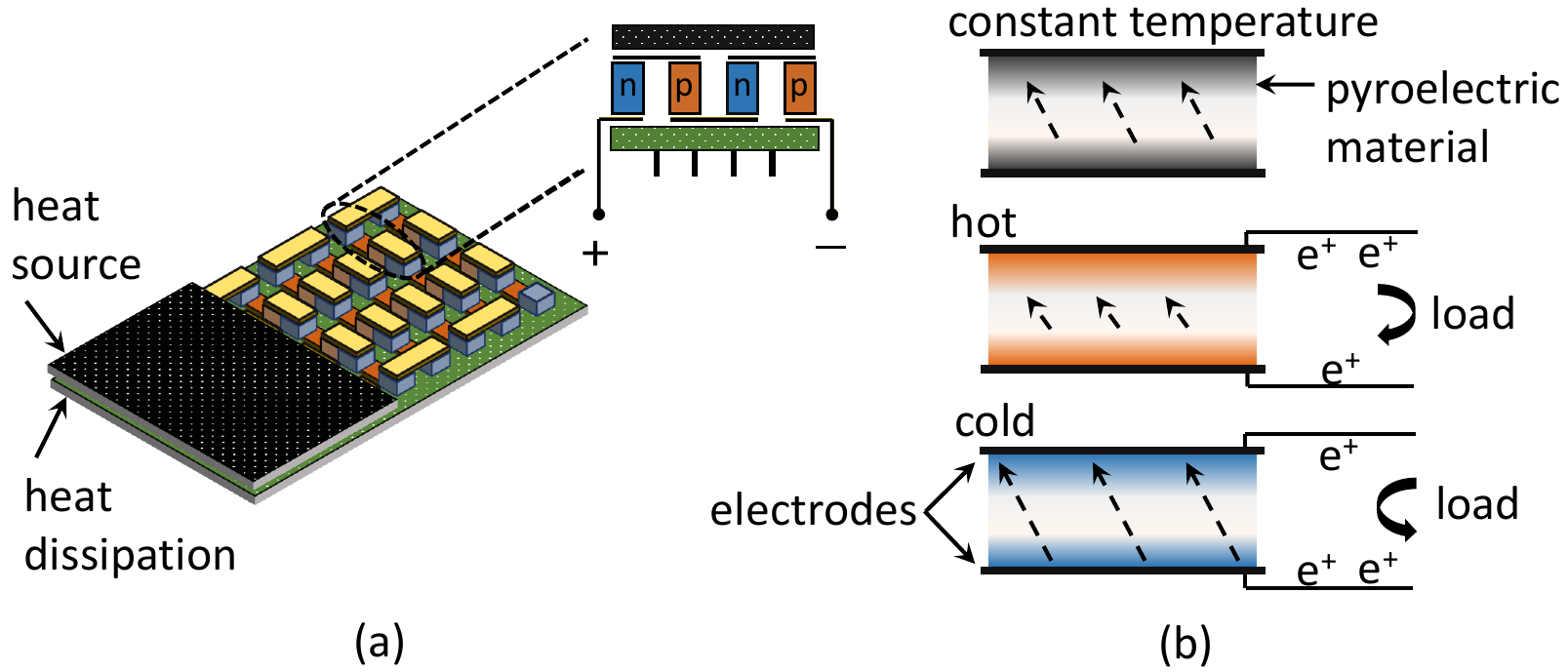}
        \caption{Basic architecture of (a) heat-based EH and (b) pyroelectric generator.}
        \vspace{-5mm}
        \label{fig:thermalEH}
    \end{figure}

    Heat-based EH exploits temperature changes in the surrounding environment \cite{9440395}. Specifically, thermoelectric generators (TEGs) turn spatial gradients of temperature on the device's surface into electric energy. TEGs work on the principle of the Seebeck effect, due to which a pair of dissimilar materials exposed to different temperatures generate electricity \cite{zhu2022review}. The basic TEG's architecture is illustrated in Figure~\ref{fig:thermalEH}a and consists of a set of n- and p-type semiconductors connected electrically in series, and thermally in parallel. To maximize the output power, one side is connected to a heat sink which increases the temperature difference with respect to the side exposed to the heat source. In practice, an array of thin-film TEGs becomes more appealing as one can accommodate more devices per unit volume, thus increasing the harvested energy \cite{Wang2.2022}. Moreover, the miniaturization of the TEGs makes heat-based EH difficult since a very small surface area may not suffice to capture spatial temperature gradients.
    
    Pyroelectric generators are an alternate implementation of heat-based EH that can transform temporal temperature variations into electricity. In principle, pyroelectric materials have an asymmetric crystallographic structure whose charges' disposition changes in response to a temperature variation \cite{kishore2021harvesting}. Figure~\ref{fig:thermalEH}b illustrates the basic architecture and operating principle of a pyroelectric generator. When the material is heated, its lattice structure expands creating more space for the charges to move, which weakens the polarization. On the contrary, the entire structure shrinks when the material is cooled, thus reinforcing the lattice asymmetry and increasing the spontaneous polarization of the material. In either case, the temperature variations will cause a flow of electrons between the electrodes to compensate for the new charges' disposition in the crystal.

    Heat-based EH technologies are especially useful for waste heat recovery, e.g., in automotive applications \cite{8676092}, for powering wearables (using the temperature changes in the human body) \cite{nozariasbmarz2020review}, and in industrial facilities \cite{albert2022waste}.
	\vspace{-2mm}
    \subsection{\uppercase{Microbial fuel cells (MFCs)}} 
    MFCs exploit bioelectrochemical conversion to harness the energy resulting from the metabolic processes of microorganisms to generate electricity. MFCs are useful in such scenarios where the devices interact with organic matter, such as compost, wastewater, and ponds \cite{palanisamy2019comprehensive}.

\begin{figure}
        \centering
        \includegraphics[width=.8\columnwidth]{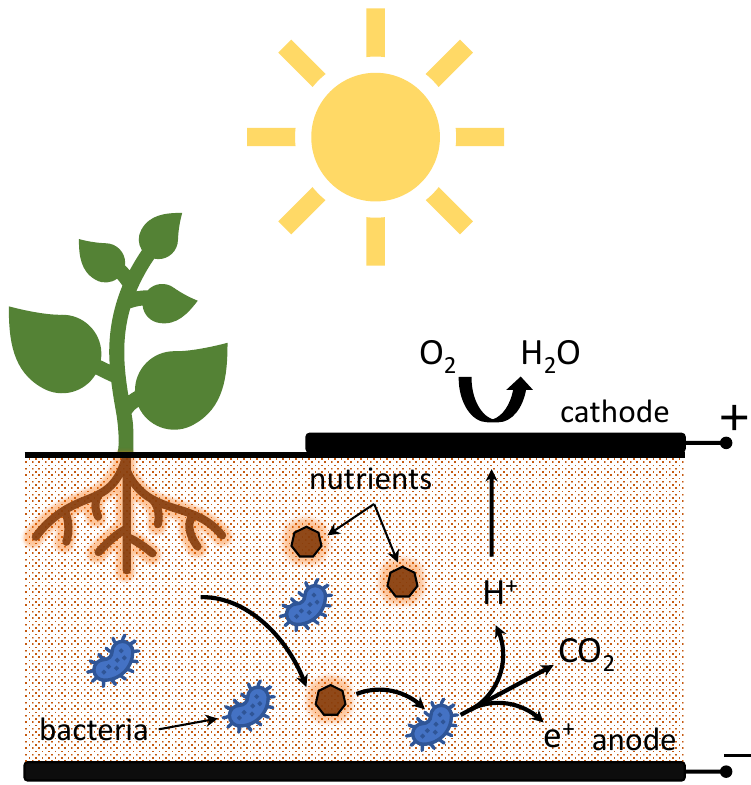}
        \caption{Plant MFC operating principle.}
        \label{fig:PMFC}
        \vspace{-7mm}
    \end{figure}

    Notice that MFCs require a continuous supply of fresh materials to prevent the microorganism from depriving the available organic matter and hence stopping the electricity generation. To solve this problem, plant MFCs leverage the plants' root exudates and rhizodeposits, resulting from the photosynthesis process, to continuously feed the microorganisms \cite{maddalwar2021plant}, as illustrated in Figure~\ref{fig:PMFC}. 
    Plant MFCs can be regarded as a form of converting the sunlight stored as chemical energy in plants into electricity. Factors such as plant species, weather conditions, soil nutrients, and microbial diversity heavily determine the lifetime and output power of this technology \cite{maddalwar2021plant}. 

    Further development directions focus on boosting the output power and lowering the manufacturing costs of current MFCs. In this regard, the authors in \cite{ramya2022review} showed that stacking multiple MFCs boosts the CE compared to individual MFCs. Moreover, the authors in \cite{yaqoob2020development,yaqoob2021modern} discussed how to build cost-effective MFCs as well as key properties of the components such as high conductivity, large surface area, porosity, durability, and corrosion resistance.
    \vspace{-2mm}
    \subsection{\uppercase{Vibration-based EH (VEH)}}
\begin{figure}
        \centering
        \includegraphics[width=\columnwidth]{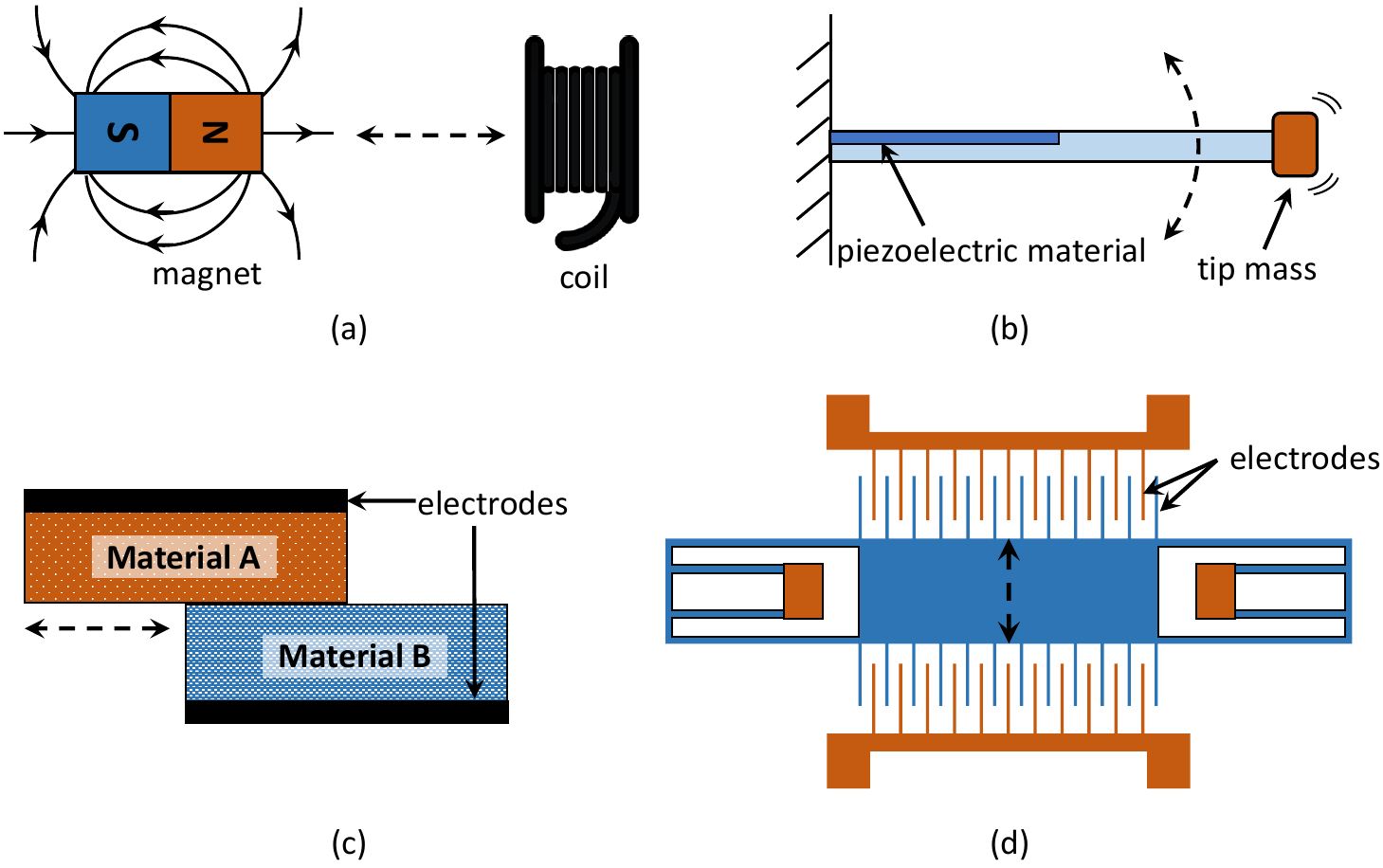}
        \caption{Basic working principle of VEH: (a) vibration EM, (b) piezoelectric, (c) triboelectric, and (d) electrostatic.}
        \label{fig:mechanicalEH}
        \vspace{-5mm}
    \end{figure}

    VEH is the process of converting mechanical energy in the form of vibration (including acoustic vibrations), impact, deformation, or friction into electric energy. As Figure~\ref{fig:mechanicalEH} shows, there are four main types of mechanical-based EH:
    \begin{itemize}
        \item Piezoelectric energy harvesters, which exploit the direct piezoelectric effect, i.e., certain materials, such as quartz, generate electrical charges under mechanical stress \cite{8762143}. The output power depends on the relative directions of the electric field and the stress/strain for the polarization direction of the piezoelectric material. 
        \item Triboelectric energy harvesters, which exploit the triboelectric effect, i.e., certain materials generate electrical charges when they contact and separate from another material. Due to their relatively small form factor and high APD, these harvesters are widely used in wearables to harness energy from human motion \cite{khalid2019review}.
        \item Electrostatic energy harvesters, which are variable capacitor structures. Specifically, their capacitance changes when the plates' overlap area or the gap distance varies in response to an external force. The basic operating principle consists of exciting the capacitor with an external power supply when the capacitance is maximum. Then, the energy is harvested when the potential energy stored in the capacitor increases in response to a smaller capacitance value \cite{Li.2022}. Due to the high polarization voltage required in this method, some electrostatic energy harvesters incorporate electrets, which are quasi-permanent electric dipoles that can hold electrical charges for years.  
        \item Vibration EM energy harvesters, which exploit Faraday's law of induction, i.e., the relative movement between a wire coil and a permanent magnet produces an electromotive force.
    \end{itemize}

    VEH presents a significant challenge: the resonant frequency of the EH device must match the input environmental frequency to maximize the output power \cite{9305188}. However, ambient energy frequency can vary considerably, demanding the adjustment of the resonant frequency to ensure a reliable energy supply. There are various approaches in the literature to achieve this, including manual mechanical tuning \cite{9777715}, mechanical self-tuning \cite{Wang3.2023}, and electronic self-tuning \cite{Morel.2021}. Another solution is to design the device to resonate at multiple frequencies, thereby expanding its frequency response spectrum. For this, hybrid energy harvesters \cite{Feng.2023} or arrays of multiple harvesters \cite{kumari2021design} can be used, with each device resonating at a different frequency.

    Another challenge of VEH is the relatively low frequency of ambient vibrations with respect to the resonant frequency of the devices. To address this issue,  frequency up-conversion mechanisms have been proposed. For instance, by applying a strong external force for a brief period, e.g., using a free-moving mass, the system may vibrate at its natural resonant frequency \cite{li2022frequency}. Alternatively, one can lower the harvester's natural resonant frequency by reducing the stiffness of its moving parts. This can be achieved, for instance, by replacing them with fluids \cite{Jackson.2020}. 

    Finally, ambient VEH is often the result of multiple external forces acting in different directions. Thus, allowing devices to harvest energy from multiple directions, e.g., using two-dimensional \cite{zhao2019direction} and three-dimensional \cite{shi2020piezo} EH devices, can boost the total output power. 
\vspace{-3mm}
    \subsection{\uppercase{Flow-based EH}} Flow-based EH leverages the kinetic energy of naturally or artificially originated fluid flows, such as water streams and wind currents, to generate electricity. Notice that in most use cases, the underlying physical principles of VEH are also applicable to flow-based EH. Therefore, this subsection focuses on wind-based EH and, in particular, on their form factor, blade design, and energy-efficient design for medium- to large-scale electricity generation.
    
\begin{figure}
        \centering
        \includegraphics[width=\columnwidth]{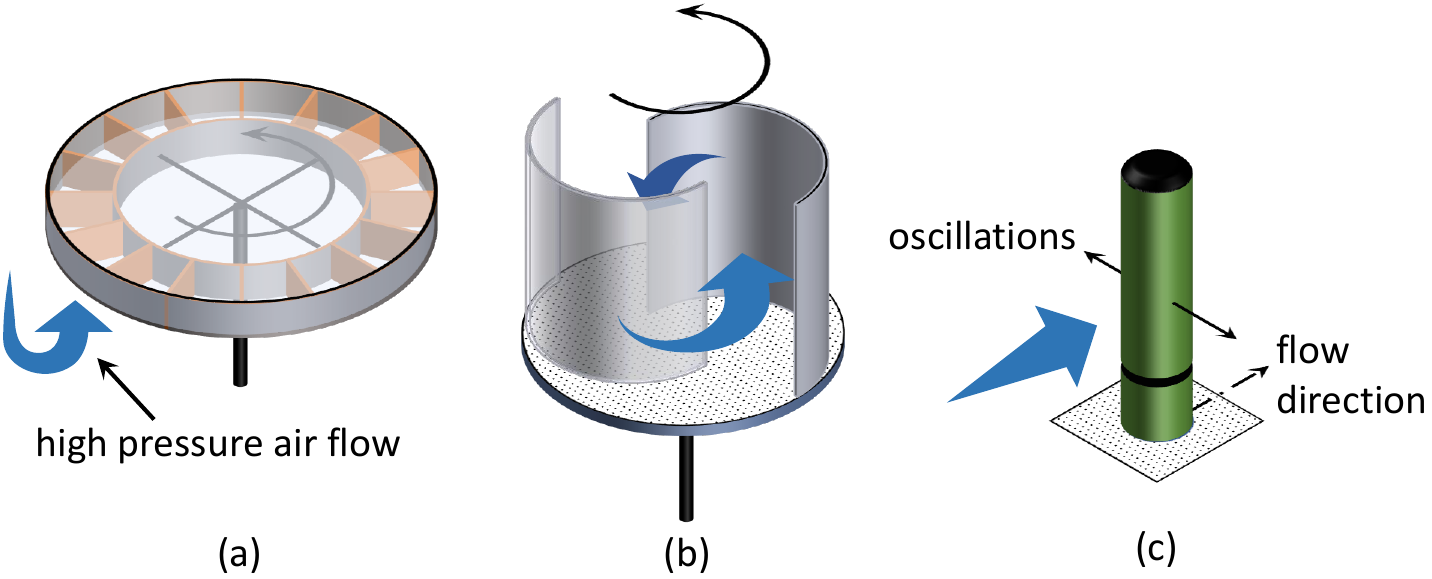}
        \caption{Wind turbine architectures: (a) Powerpods internal blades system, (b) Savonius turbines, and (c) Vortex Bladeless generator.}
        \label{fig:flowEH}
        \vspace{-6mm}
    \end{figure}

    Traditional wind turbines leverage the kinetic energy of the wind to turn the rotor of an electric generator. Wind turbines are usually deployed in farms for large-scale electricity generation. However, massive IoT deployments are more likely to appear in cities where installing a huge wind turbine is not always feasible. Thus, research and industry efforts have been focused on building compact and cost-effective designs of wind-based EH systems. 
    A significant issue with traditional wind turbine design is the potential threat to wildlife posed by the rotating blades. To address this, Halcium has created the PowerPods\footnote{For more information please check \url{https://www.halcium.com/}.}, which are portable wind turbines designed for residential areas. The blades of the PowerPods are contained within the pod, as shown in Figure~\ref{fig:flowEH}a, making them safe for people, pets, and wildlife in close proximity. The shell of the PowerPods captures wind from multiple directions and channels it through small exits, increasing the wind speed and therefore the harvested energy. Different from traditional wind turbines, PowerPods utilize vertical-axis blades whose low aerodynamic noise pollution and ability to operate under unstable wind flow (without needing a yawing mechanism) make them appealing for urban environments \cite{zhao2022review}. Vertical-axis turbines are also the basis of the so-called Savonius turbines (shown in Figure~\ref{fig:flowEH}b) which are reliable and cost-effective systems that can operate under turbulent wind flows and stormy weather \cite{ebrahimpour2019numerical}. Notice that traditional wind turbines require a braking mechanism to operate under such conditions as high wind speeds can cause the system to shatter becoming a hazard for the nearby areas.
    
    Another development area in wind-based EH involves blade-less turbines, e.g., Vortex Bladeless\footnote{For more information please check \url{https://vortexbladeless.com/}.} shown in Figure~\ref{fig:flowEH}c, which leverages the vortex shedding phenomenon \cite{villarreal2018viv}. Vortex shedding occurs when a fluid flow separates from a solid surface and creates vortices. In the case of the Vortex Bladeless, the vortex shedding effect causes a vertical mast to oscillate in resonance with the wind flow. One advantage of this design is the reduced maintenance costs, as there are no moving parts that can wear out through friction. However, the system can experience mechanical fatigue and stress at the base of the oscillating mast. Additionally, the smaller footprint of the Vortex Bladeless compared to traditional wind turbines allows for the installation of multiple units in the same area, potentially compensating for their individual lower performance.
\vspace{-3mm}
    \subsection{\uppercase{RF-EH}}
    RF-EH technology utilizes a rectenna to convert EM waves into direct-current (DC) power, which can be harnessed to power and/or charge small electronic devices and batteries. The rectenna is composed of a receiving antenna and a rectifier plus a matching network to maximize the power transfer efficiency (PTE). The main sources of ambient RF energy are classified as \cite{Alves.2021}: i) static, which corresponds to energy transmissions that remain relatively stable over time, e.g., from television and radio transmitters, allowing long-term predictability of the energy supply, thus favoring the network planning, and ii) dynamic, which supplies time-varying energy over a certain region, either because of fluctuating transmit power levels or mobility, e.g., from WiFi/mobile access points or devices, thus should be adaptive and allow operation over multiple frequencies. 

    \begin{figure}
        \centering
        \includegraphics[width=0.9\columnwidth]{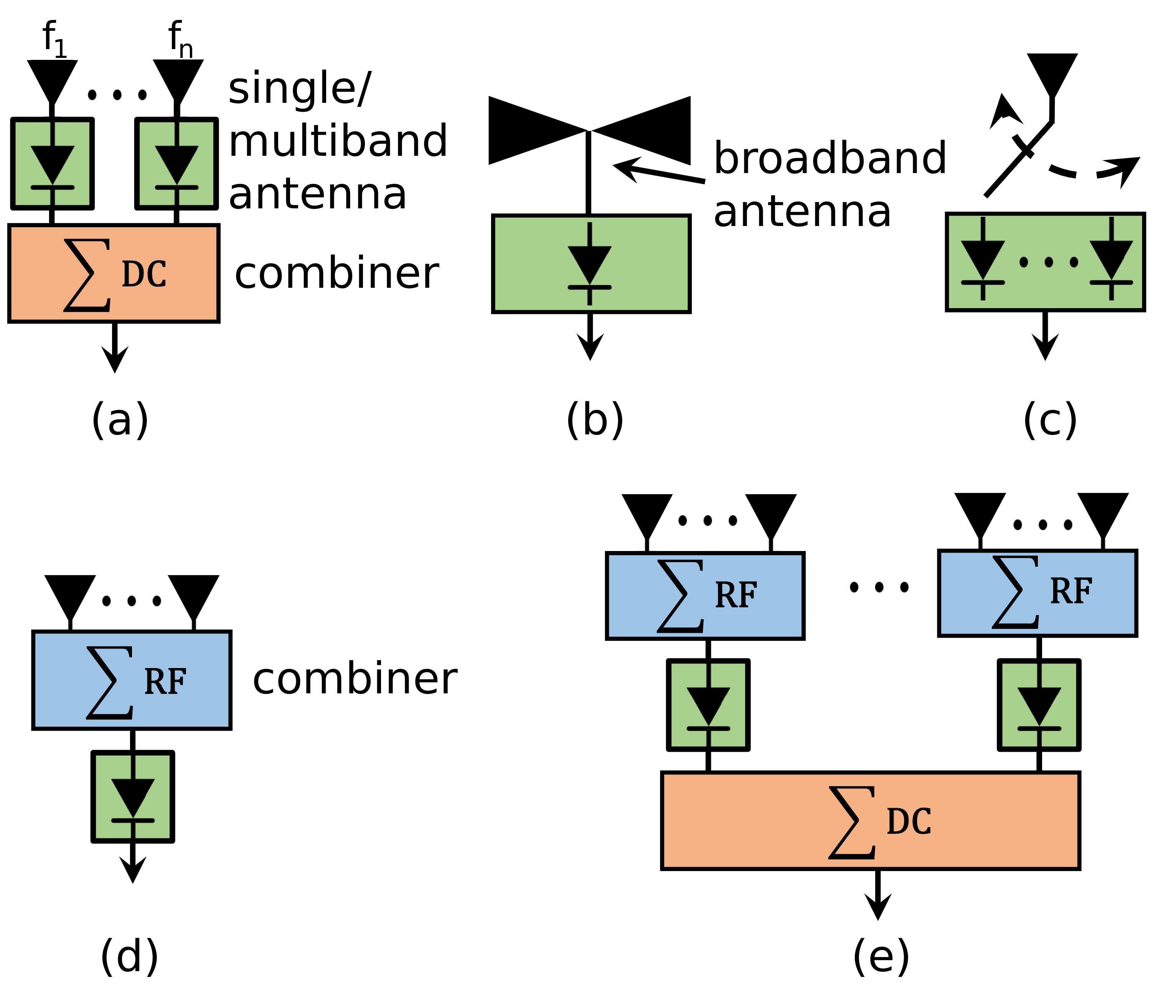}
        \caption{Architecture of (a) DC combining, (b) broadband, (c) extended dynamic range, (d) RF combiner, and (e) hybrid combiner.}
        \label{fig:RFEH}
        \vspace{-8mm}
    \end{figure}
    
    The main challenge of ambient RF-EH circuits lies in how to guarantee an appropriate performance given mostly unpredictable EM conditions. Notice that the receive signal parameters, e.g., carrier frequency, bandwidth, polarization, and antenna orientation determine the amount of harvested energy. Hence, key desirable features of RF-EH circuits include performing EH i) from multiple bands, which can be accomplished by using wideband, multi-band, or tunable receivers \cite{Alves.2021}, ii) from multiple spatial directions, for which omnidirectional antennas are preferred, iii) from randomly polarized signals, which is mostly implemented by some sort of circularly polarized antenna; and iv) from a variety of input power levels, in which case the RF-EH adjust the sensitivity and saturation levels to obtain the most from the ambient energy source. Figs.~\ref{fig:RFEH}a-c illustrate the basic architecture of some of the aforementioned receivers.

     \begin{table*}[t!]
    \caption{Comparison among WET technologies}
    \vspace{-1mm}
    \label{tab:WETComparison}
    \centering
    \begin{tabular}{l l l l l l}
        \toprule
         \textbf{Technology} & \textbf{Efficiency} & \textbf{Complexity} & \textbf{Range (RG)} & \textbf{Scalability} & \textbf{Standardization} \\
         \midrule
         RF-WET & Medium & Moderate & Long & Very good &  AirFuel Alliance [AirFuel RF], ARIB STD-T113, \\
         & & & & & NFC Forum [Wireless Charging Specification 2.0] \\
         Inductive & Very high & High & Medium & Moderate & Wireless Power Consortium [Qi,Ki standards], SAE [J2954,J2847/6], \\
         & & & & & IEC 61980-3:2022, ISO 19363:2020, ISO 15118-20:2022, \\
         & & & & & AirFuel Alliance [AirFuel resonant] \\ 
         Capacitive & High & Moderate & Medium & Limited & ARIB STD-T113 \\
         Laser & Low & High & Long & Moderate & - \\
         Acoustic & Very low & High & Very short & Very limited & - \\
         \bottomrule \vspace{-7mm}
    \end{tabular}
\end{table*}

    Multi-antenna RF-EH circuits, as shown in Figure~\ref{fig:RFEH}a, Figure~\ref{fig:RFEH}d, and Figure~\ref{fig:RFEH}e, are another alternative to boost the amount of harvested energy. In such case, the incoming signals can be combined i) in the RF domain; ii) in the DC domain, or iii) hybridly in both domains \cite{lopez2021dynamic}. Multi-antenna rectennas also improve the spatial selectivity of the antenna in the direction of the maximum incident signal, if properly adjusted. However, for ambient RF-EH the optimal receive beamforming is difficult to attain due to the non-dedicated nature of the transmissions. To cope with this problem, several low-complexity solutions, including codebook-based beamforming, were proposed in \cite{lopez2021dynamic} such that an ambient RF-EH device sweeps a phase shift table in an initial phase seeking the configuration that yields the best performance for exploitation in a second phase.
    
    Magnetostrictive antennas are an alternative to rectennas. In such a case, the transducer is built with multiferroic structures that are able to sense the magnetic component of EM waves and, in response, produce vibrations. Then, a mechanically connected piezoelectric EH converts the vibrations into electrical energy. This allows for realizing an ultra-compact receiver, which is up to two orders of magnitude smaller than state-of-the-art rectennas \cite{song2021up}. Metamaterials-aided designs are another solution for realizing ultra-compact RF-EH implementations, due to the sub-wavelength periodicity of the unit cells within the lattice structure. Metasurfaces-aided RF-EH boosts the CE as it allows wider beamwidths, polarization-independent operation, built-in and less lossy matching networks, and higher antenna gains \cite{9627147}. 
    \vspace{-3mm}
	\section{\uppercase{Energy Transfer (ET)}}\label{ETr}
ET refers to the intentional transmission of energy using dedicated transmitters to power EH devices, and thus it is mostly/inherently wireless, leading to WET. In contrast to ambient EH, WET provides a controllable and predictable energy supply and the tools for increasing the end-to-end power CE. In Table~\ref{tab:WETComparison}, we compare the current development state of the WET technologies discussed in this section.
\vspace{-3mm}
    \subsection{\uppercase{Laser-based WET}} 
\begin{figure}[t!]
        \centering
        \includegraphics[width=\columnwidth]{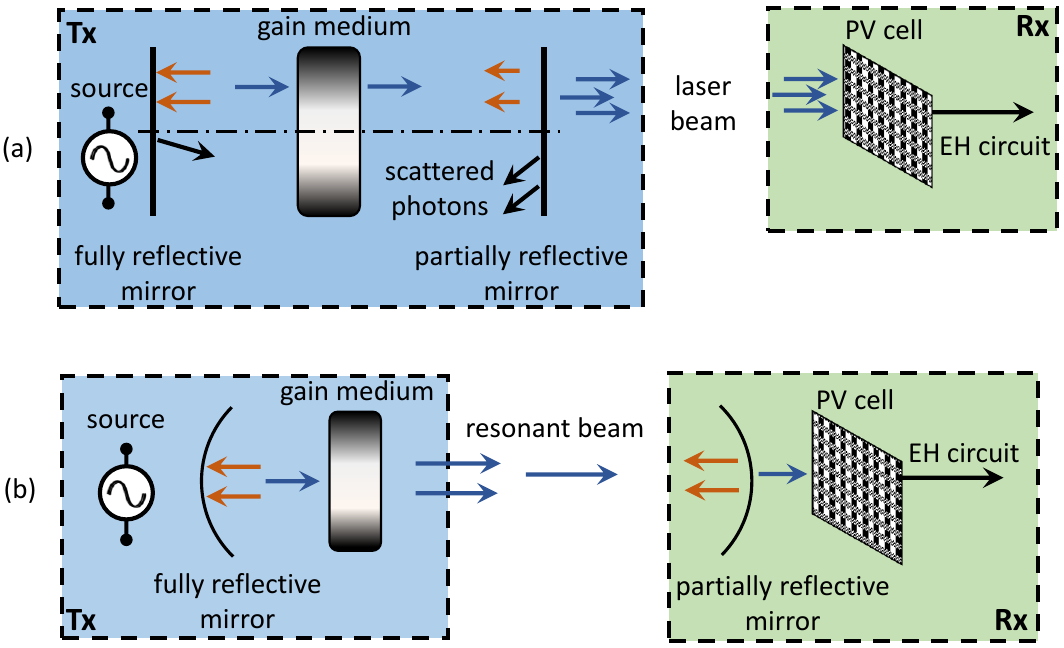}
        \caption{Laser-based WET architectures: (a) laser power beaming and (b) distributed resonant beam charging. }
        \label{fig:laserWET}
        \vspace{-8mm}
    \end{figure}

    The most common implementations of laser-based WET are laser power beaming and distributed resonant beam charging. In the former case, shown in Figure~\ref{fig:laserWET}a, a laser diode at the transmitter steers an optical beam towards a PV cell at the receiver (similar working principle as in conventional solar power systems). Meanwhile, in the latter, the receiver bounces back a portion of the incident energy towards a gain medium at the transmitter, which amplifies the light and initiates a resonant beam \cite{8618313}, as illustrated in Figure~\ref{fig:laserWET}b. 
    
    Although the above two implementations are regarded as long-RG WET technologies, laser power beaming provides superior coverage, in the order of several km \cite{8404085}, and may be key for intra-satellite WET applications \cite{9945839}, charging UAVs \cite{9777271}, and energy transmission from solar-powered satellites to terrestrial stations \cite{Jaffe.2022}. However, the performance of the laser power beaming heavily depends on the line-of-sight (LOS) and the atmospheric conditions. Besides, laser power beaming raises safety concerns due to the potential harm to the living species' tissue. Notably, lowering the transmission frequency is appealing to provide a hazardous-free system, although incurring high maintenance cost and lower efficiency (and therefore trading efficiency for safety) \cite{8404085}. Alternatively, it is advisable to monitor the periphery of the beam path for detecting intruders \cite{Jaffe.2022}. 
    
    Differently, resonant beam charging is inherently safe as any interruption of the link causes the resonance to stop immediately. In addition, the system is capable of supporting mobile users without the need for a tracking system as long as LOS conditions hold, thanks to its self-aligning capabilities which have been experimentally validated (e.g., cf. \cite{9677003}). Moreover, the transmitter's ability to generate simultaneous resonating beams also enables native multi-user support \cite{Xiong.2019}. However, broadcasting energy to multiple receivers is typically more inefficient, often motivating time division-multiple device-charging protocols \cite{8425720}. To overcome the LOS limitations, UAVs \cite{8618313} or supporting reflectors \cite{liu2022nlos} can relay energy transmissions to charge devices out of the coverage of the main transmitter.
\vspace{-6mm}
    \subsection{\uppercase{Acoustic WET}}
      This technology leverages acoustic waves to charge EH devices by typically equipping transmitters and receivers with piezoelectric transducers \cite{8944030}. Acoustic WET works in any medium capable of propagating pressure waves, e.g., metal, air, and human tissue, but it is particularly convenient in EM wave absorption-prone mediums, such as Faraday shielding structures \cite{9785780} and water. Besides, for the same operating frequency, acoustic WET transmitters/receivers have a more compact form factor and achieve a higher directivity than those implementations based on EM WET \cite{8101521}. 
    
    Similar to other WET technologies, the channel attenuation, which varies significantly according to the medium, affects severely the system performance. Notably, one cannot increase deliberately the intensity of the acoustic waves as it may cause hearing impairments, body heating, and other unpleasant effects in humans and animals depending on the operating frequency \cite{Kim.2021}. Notice that the attenuation increases with the impedance mismatch of the medium between the transmitter and the receiver since traveling waves may encounter different materials in their path. This has motivated the use of glue or electromagnets to ensure a firm connection of both the transmitter and receiver to the matching layers in the path (when solid) and therefore reduce the propagation losses \cite{9785780}. Moreover, when receivers are deployed inside a Faraday shielding structure but not connected to it, one can rely on hybrid WET approaches in which the last section of the transmission path relies on a different WET technology, such as inductive coupling \cite{10007605}.

      \begin{figure}[t!]
        \centering
        \includegraphics[width=\columnwidth]{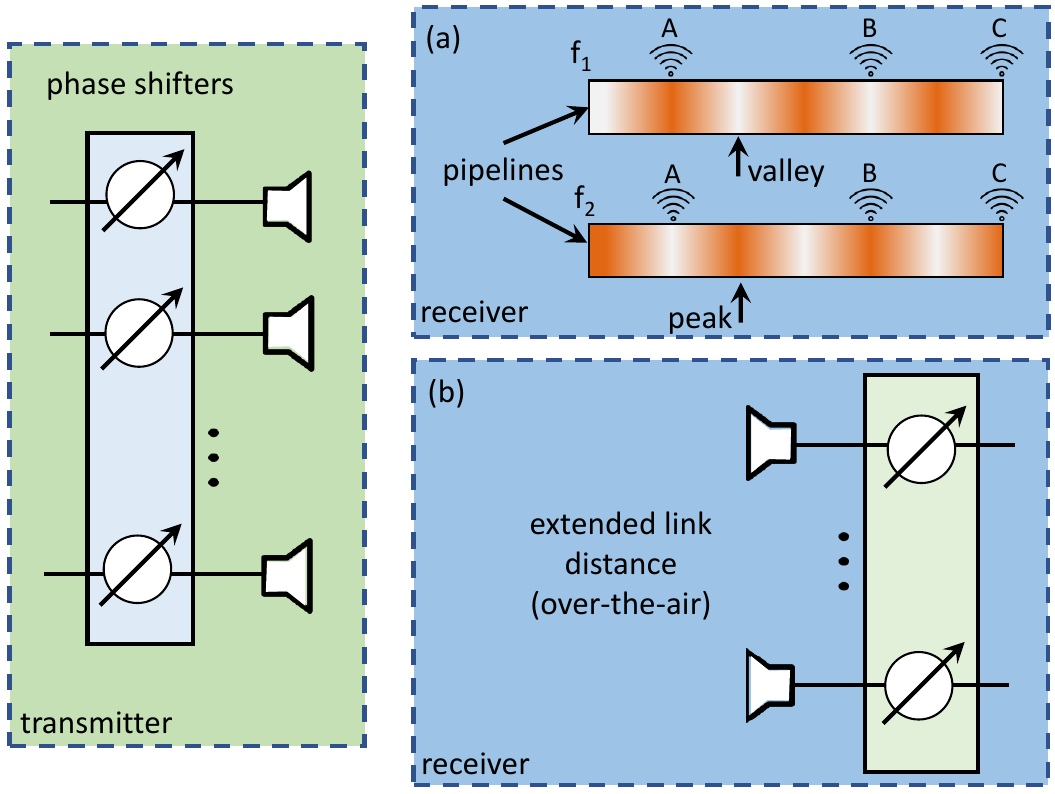}
        \caption{Acoustic phased-arrays architecture and use cases: (a) multiple piezoelectric transducers at the transmitter to power a sensor network embedded in pipelines and (b) phased-array at the receiver to extend the operating RG.}
        \label{fig:acousticWET}
        \vspace{-6mm}
    \end{figure}

    Phased acoustic arrays, illustrated in Figure~\ref{fig:acousticWET}, which are composed of multiple piezoelectric transducers, also aid in overcoming medium attenuation by focusing the energy toward the receiver direction using sound beams. In fact, the performance of transmit phased acoustic arrays increases with the number of transducers given a form factor constraint \cite{8101521,8589762}. Further improvements can be achieved by also equipping the receiver with phased acoustic arrays and hence allowing different combining techniques to improve the PTE, similar to a traditional multiple-input multiple-output (MIMO) wireless link \cite{7953842}. Notice that the achievable performance of a phased acoustic array also depends on the transducers' geometry and arrangement, and the array's aperture and diameter \cite{9670653}.

    Finally, acoustic WET can also provide heterogeneous quality of service (QoS) to the end users. For instance, acoustic transmitters can use Lamb waves to create a pattern of peaks and valleys in solid structures. Therefore, by changing the operating frequency, the resulting vibrations charge the devices deployed in different locations of the same structure with a different intensity \cite{8882371}. Besides, adaptive acoustic beamforming also serves to discriminate which devices to charge depending on their locations and energy demands \cite{9958008}. This might be especially useful in sensor networks embedded in structures such as buildings or bridges.
    \vspace{-2mm}
    \subsection{\uppercase{Inductive coupling-based WET}} 
    This technology exploits the inductive coupling phenomenon. The simplest setup consists of two wire coils coupled by a magnetic field such that the oscillating magnetic field in the transmitter's coil passes through the receiver's coil inducing an alternating current. This forms a highly efficient air-gap transformer whose performance depends on the operating frequency and the mutual inductance between both coils. However, this basic setup performs poorly when both coils are misaligned or too separated. 

    One can extend the charging coverage by tuning both coils to resonate at the same frequency. This is commercially known as magnetic resonant coupling, which compared to the basic inductive coupling, is more resilient to coils' misalignment. The most common magnetic resonant coupling methods include an external capacitor to compensate for the internal inductive reactance and extra coils for tuning and impedance matching. The first implementation is easier to realize, but the second one achieves a higher CE given that there are no power losses in external resonators \cite{8052089}.

    Magnetic resonant coupling allows a transmitter to charge multiple nodes simultaneously. Here, the internal resistance of the resonators can be adjusted to control the mutual inductance among coils and hence boost the PTE. For instance, increasing the internal resistance of the nearest receivers may increase the harvested energy at the most distant receivers, although at a higher transmit power cost \cite{7347451}. 
    
    Deploying multiple transmitters also improves the system performance. On the one hand, the optimal transmitter deployment guarantees uniform power coverage and ensures a minimum available energy at the receivers regardless of their locations \cite{7890356}. On the other hand, coordinated transmissions from multiple transmitters can result in a constructive combination of the magnetic fields at the receivers' locations, which is known as distributed magnetic beamforming \cite{9531554}.
\vspace{-3mm}
    \subsection{\uppercase{Capacitive coupling-based WET}} This technology exploits the capacitance coupling phenomenon. The basic architecture consists of two pairs of transmit and receive plates each coupled by an electric field. One pair forwards the displacement current while the other provides the return path to close the circuit. Capacitive coupling systems come in two different flavors depending on how the plates are arranged. The horizontal capacitive coupler has the transmitter's (and receiver's) plates placed side by side in the same plane, while the vertical capacitive coupler has both the forward and return paths overlapped in the direction of the electric field. Although the former architecture has the highest coupling, it comes at the cost of being bulky and less reliable due to the large number of required components. For a fixed device size, one can enhance the coupling of the system by combining the geometry of the basic architectures and adding more plates \cite{9447235}.
    
    Some critical challenges in capacitive coupling systems include the high excitation frequency/voltage due to the large capacitive reactance of the coupling and the control of the fringing electric fields, i.e., the non-uniform field at the edge of the plates, within safe limits. To cope with these issues, resonant matching networks can be added at both the transmit and receive sides. They compensate for the reactive losses in the system, hence reducing the required excitation voltage. Further, they can be designed as a transmit voltage gain stage to drop the displacement current of the plates, hence reducing the fringing field to the safety limits. Herein, the size of the inductors in the matching network is key to realizing compact designs. As an example, one can exploit the successive impedance transformation that provides multistage matching networks \cite{9194001}. The required amount of gain and compensation of such a matching network depends on the ratio between the load and excitation currents, and the misalignment/distance between couplers \cite{9457989}.

    Different from inductive coupling, capacitive coupling systems are more tolerant to the plates' misalignment, reduce the risk of interference to neighboring networks, and do not induce eddy currents in nearby metallic objects. Moreover, they are lighter, easy to integrate, and more mechanically robust. That is why they are regarded as promising for powering medical implants, vehicles, consumer electronics, and rotary electric machines \cite{9666847, 9855879}.
\vspace{-7mm}
    \subsection{\uppercase{RF-WET}}
    RF-WET technology relies on intentional RF transmissions to charge RF-EH devices. The main challenge of this technology is its limited coverage due to the channel attenuation and the regulations on the maximum radiating power of the RF energy transmitters, hereinafter referred to as power beacons (PBs). For this reason, directive antennas are preferred to increase the incident RF power in the desired direction without increasing the transmitter's radiating power. Notice that when serving multiple users,  the PB can utilize an omnidirectional antenna to provide basic service guarantees and a directional antenna (for out-band) energy transmissions to meet more specific user equipment (UE) requirements \cite{9723184}. However, in case the PB is equipped with non-reconfigurable antenna(s), a mechanical sweeping of the service area may be needed to charge those devices otherwise located in the minimum of the radiation pattern \cite{lopez2021csi}. 
    
    Another potential technique to extend RF-WET coverage is channel state information (CSI)-based energy beamforming. The main challenge here resides in the often unavoidable cost of instantaneous CSI acquisition, especially when powering a massive number of devices with strict HW constraints, which may null or even reverse the gains from accurate CSI-based transmit strategies. This has motivated alternative strategies relying on statistical CSI \cite{9184149}, received energy feedback \cite{Shen.2023}, and positioning information \cite{lopez2021csi}.

     \begin{figure}[t!]
        \centering
        \includegraphics[width=\columnwidth]{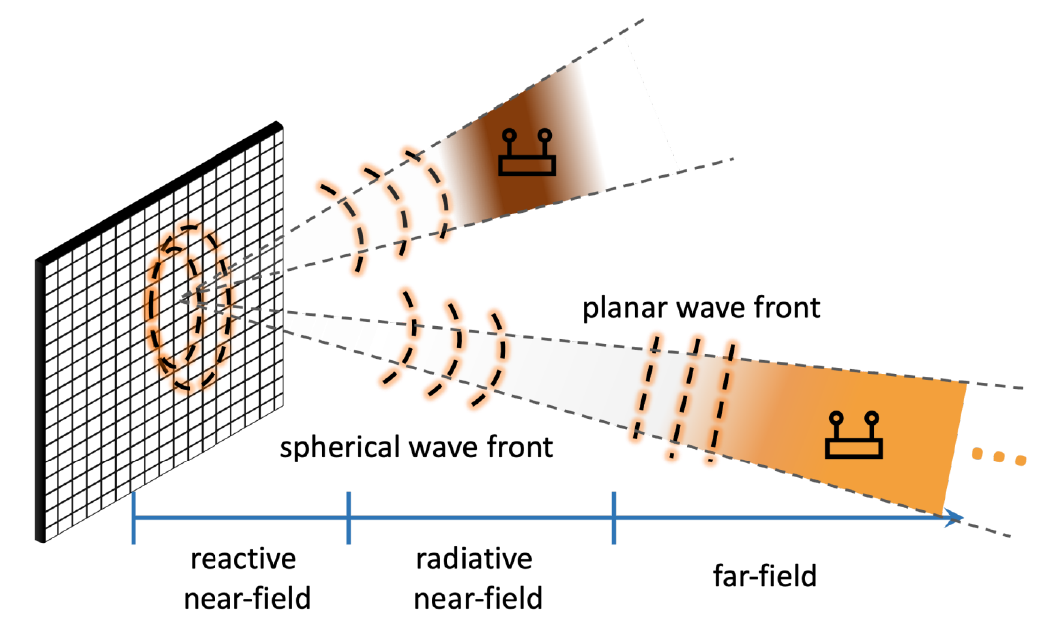}
        \caption{Near-field vs far-field beamforming.}
        \label{fig:nearFieldBeamforming}
        \vspace{-7mm}
    \end{figure}

    Conveniently deploying the PBs is also key to overcoming channel attenuation, banning blind spots, and homogenizing the incident RF power according to the network requirements \cite{Rosabal.2021,9310236,Lopez.2021}. Notably, distributed single-antenna PBs offer better service than a single PB equipped with the same total number of antennas due to the reduced charging distance \cite{9310236} and the reduced number of EH devices to be charged per PB \cite{Lopez.2021}. Robotic WET, in which the PBs move \cite{9209664} or fly \cite{yuan2022joint}, becomes also appealing not only to reduce the link distance but also to power the devices deployed in hard-to-reach places or to meet temporary service requirements during emergencies. Other potential technologies to boost the coverage of RF-WET are also discussed in Sections~\ref{EE}-\ref{IRS} and \ref{lowPMIMO}.

    The next generation of wireless systems envisions extremely large antenna arrays to compensate for the attenuation at high-frequency bands, increase reliability, and reduce interference. Consequently, many future IoT deployments may operate in the radiating near-field (or Fresnel region) of the transmit antennas, which contrasts with the traditional assumption of far-field operation. As Figure~\ref{fig:nearFieldBeamforming} illustrates, under such operating conditions, a PB can focus the energy in a particular location as opposed to what happens in far-field conditions, where the energy is steered towards a certain angular direction. As a consequence, near-field RF-WET will generate less RF pollution, thus interference, in both angle and distance domains and a reduced human RF exposure \cite{9743350}. Moreover, the CE achievable in the near-field can be significantly high even when the energy receiver is not in the focal point of the transmitter's antenna \cite{Zhang.2022}. 
    
    Due to the above, RF-WET is usually regarded as a short-distance solution to charge low-power devices. However, some experiments have shown energy transmissions over distances greater than $1~$km distance with a peak incident power of several Watts at the rectenna \cite{9318744}. Long-distance RF-WET could also help bypass the complicated infrastructure of the electric network to power hard-to-reach locations or to provide an energy supply during emergency situations.\footnote{Please refer to \url{https://emrod.energy/} for more details.}
    \vspace{-1mm}
	\section{\uppercase{Energy Efficient Communication Technologies}}\label{EE}
        \vspace{-2mm}
        Once the energy is available for operation through EP/ET mechanisms as those outlined in Sections~\ref{EP} and \ref{ETr}, the device(s) and/or the network(s) must ensure its efficient usage. Indeed, every technique and technology conceived for performance improvements in terms of coverage, throughput (THP), dependability, and other KPIs, can be in principle leveraged for EC reduction as well, hence, EE. That is the case of massive MIMO \cite{Bjornson.2014,Prasad.2017}, UAV-based connectivity \cite{Mozaffari.2016,Lin.2019,Martinez.2022}, satellite-assisted communications \cite{Kapovits.2018,Tondo.2023}, cooperation and diversity mechanisms \cite{You.2018,Lopez.2019,Lopez.2020}. Nevertheless, there are some technologies/techniques that are natively conceived for energy-limited/efficient operation, and these constitute the scope of this section.

       Notice that the primary EC sources in a device or network are i) the utilization of active circuit components such as transistors and power amplifiers, which require a power source to function, and ii) the running applications. Therefore, to enable energy-efficient and/or low-power operation, the use of active techniques and devices must be limited when possible while relying on efficient passive architectures. This comes with several challenges as active devices, although consume energy, can facilitate high-performance computing and communication, while passive devices (e.g., resistors, capacitors, inductors) are much less flexible and may dissipate energy in the form of heat as they interact with electric signals. This motivates the use of semi-passive/active architectures, which include some active techniques and components to efficiently support the application demands for which they are deployed, and/or tunable operational models as enabled by WuR technology. These approaches are discussed next through key technologies. Specifically, backscatter communication (BC), metasurface-aided communication, radio stripes, and WuR technologies are overviewed in Sections~\ref{EE}-\ref{BC}, \ref{IRS}, \ref{lowPMIMO}, and \ref{wakeUp}, respectively. Meanwhile, the ML approach, focused on intelligently reducing the EC burden of computation/communication application tasks at the devices and networks, is discussed in Section~\ref{ML}.\footnote{Please note that our focus in this work is on technologies (other than wireless communication standards/protocols) that we believe are most relevant for enabling low-power IoT connectivity in the near future.  The interested reader can refer to other potential technologies such as molecular communication \cite{Akyildiz.2019}, printable electronics \cite{Wiklund.2021}, and passive radar \cite{Griffiths.2022}.}
\vspace{-3mm}
        \subsection{\uppercase{Backscatter communications}}\label{BC}
\begin{figure}[t!]
        \centering
        \includegraphics[width=1\columnwidth]{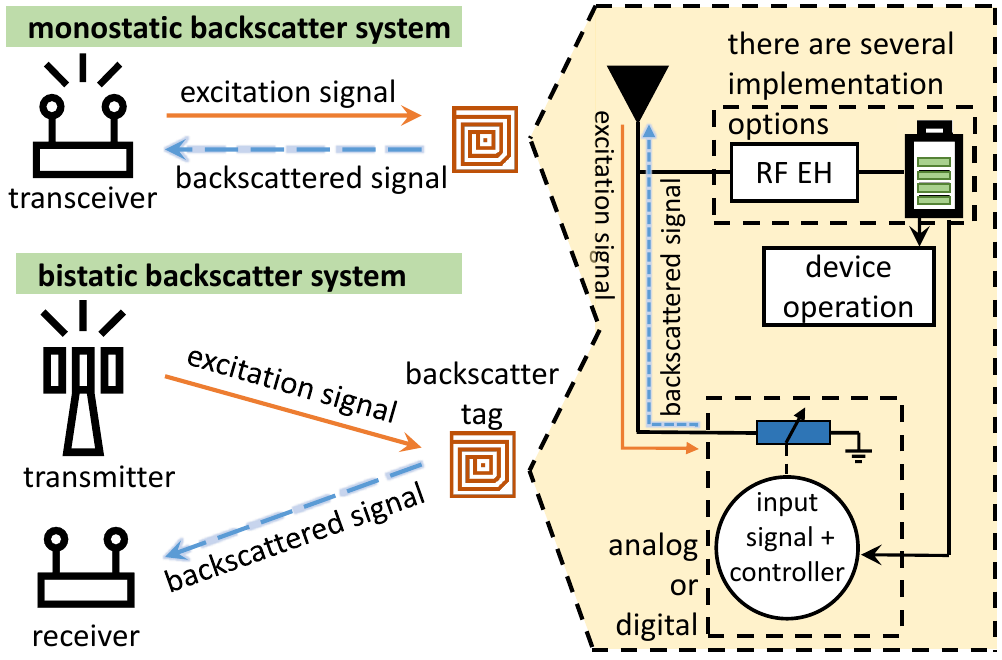}
        \caption{BC systems and main components.}
        \label{BCfig}
        \vspace{-5mm}
    \end{figure}

\vspace{-1.5mm}
        BCs are passive and involve backscatter tag(s) reflecting the signal from nearby transmitter(s) and modulating it by adjusting its amplitude, frequency, and/or phase via impedance mismatching tuning. At the receiver, the backscattered signal is processed to extract the information added by the backscatter tag. The typical components and types of BC systems are illustrated in Figure~\ref{BCfig}. Notice that the transmitter(s) may be dedicated or non-dedicated, the latter leading to the so-called ambient BC systems.
        
       There are multiple implementation options for powering the backscatter tags, e.g., relying on EH from ambient/dedicated RF signals \cite{Lin.2022,TallaKellogg.2017,Reed.2021}, 
        solar energy, vibration, and/or thermal gradient \cite{TallaKellogg.2017,Phan.2022}. 
       In general, the energy buffer can be as small as a capacitor of several nF, or as large as a digitally-controlled supercapacitor or rechargeable microbattery. The nature of the energy buffer depends on the application. A capacitor may suffice for ultra-low power applications, and it is not often regarded as a buffer but simply as another stage element within the circuitry. In other cases, a large and intelligently controlled energy buffer is required to enable substantial gains in RG and operating time that could not have been attained otherwise. Such energy storage topology influences also the type of modulation/demodulation of the backscatter signals, which can be digital or analog as in conventional communication systems, and may be activated by a wake-up code detector \cite{Soyata.2016}. 

       Table~\ref{tab2} lists representative state-of-the-art BC systems along with their distinctive features and performance figures. 

 \begin{table*}[p!]
            \caption{Some representative state-of-the-art (2017-2022) BC systems}
            \vspace{-1mm}
            \centering            
            \begin{tabular}{p{1.9cm}p{1.2cm}p{1.27cm}p{0.83cm}p{0.83cm}p{0.83cm}p{2.3cm}p{5cm}}
                \toprule
                 & & \multicolumn{2}{c}{Maximum Bit Rate} & \multicolumn{2}{c}{RG} & \ \ Deployment &  \\
                 \cmidrule(lr){3-4}\cmidrule(lr){5-6}\cmidrule(lr){7-7}
                  \textbf{BC System} &  \textbf{Power} $^a$ & \textbf{Bit Rate} & \textbf{Dist.} & \textbf{Tx-Tg} & \textbf{Tg-Rx} &\ \ \ \ \ \ \textbf{Tx, Rx} & \textbf{Distinctive Features}\\
                \midrule
              $\bullet$ LoRea \cite{Varshney.2017} & 70 $\mu$W & 197 kbps & 175 m & 1 m & 3.4 km &  patched commodity, customized HW & Compatible with properly-tuned WiFi, ZigBee, and BLE radios at 2.4 GHz.\\
              $\bullet$ Battery-free phone \cite{TallaKellogg.2017} & 3.5 $\mu$W & N/A & N/A & 15.2 m & 15.2 m & customized HW & Powered by RF \& light. Speech/data transmission and UE input via capacitive touch.\\
              $\bullet$ LoRa back- scatter \cite{Talla.2017} & 9.3 $\mu$W & 37.5 kbps & N/A & 238 m \hphantom{o} 5 m & 238 m 2.8 km & customized HW, commodity LoRa & Sensitivity of $-149$ dBm  even under strong in(out-of)-band interference.\\ 
              $\bullet$ FM back- scatter \cite{Wang.2017} & 11.1 $\mu$W & 3.2 kbps & 4.9 m& N/A & 18.3 m & ambient FM, commodity FM & Applications: i) posters communicating to smartphones/cars, ii)  smart fabric.\\
            \rowcolor{red!20} $\bullet$ Morse back- scatter \cite{Daskalakis.2018} & 20 $\mu$W  & 250 kbps & 2 m & N/A & N/A & customized HW, SDR & Morse code and on-off-keying modulation for low-complexity/power/cost.  \\
              $\bullet$ MOXcatter \cite{Zhao.2018} & N/A  & 50 kbps & 6 m & 0.3 m & 14 m & commodity WiFi & 
              Sensing data embedded in ambient spatial-stream packets. \\
              $\bullet$ BARNET \cite{Ryoo.2018} & 3 $\mu$W &  N/A & N/A & 2 m & 2 m & customized HW & Applications: i) identification of tagged objects, ii) `device-free' activity recognition.\\ 
              $\bullet$ PLoRa \cite{Peng.2018} &  2.6 mW & 6.25 kbps & 300 m & 1 m & 1.1 km & commodity LoRa, SDR & Backscatter signal decoder \& MAC protocol for coexisting with active LoRa nodes.              \\
              $\bullet$ TunnelTag \cite{Varshney.2019} & 57 $\mu$W & 2.9 kbps & 1 m & N/A & N/A & customized HW, SDR &
               Tunnel diode-based RF oscillator for transmission or amplifying reflection.
              \\
              $\bullet$ RF-Mehndi \cite{Zhao.2019} & N/A &  N/A & N/A & 0.3 m & 0.3 m & customized HW &
              User's fingertip profiled passive RFID tag -based identifier. \\
             \rowcolor{red!20} $\bullet$ ZeroScatter \cite{Dadkhah.2019} & 850 $\mu$W & 9.4 kbps & 3 m & 3.5 m & 3.5 m & customized HW &
              Unmodified digital I/O pins of FPGAs and microcontrollers (MCUs) for BPSK BC. \\
              $\bullet$ WiFi back- scatter \cite{Wang.2020} & 30.8 $\mu$W & 2 Mbps & N/A & 30 m & 90 m & commodity WiFi & First backscatter IC communicating directly with commodity WiFi transceivers.\\ 
              $\bullet$ BLE back- scatter \cite{Katanbaf.2020} & 50 $\mu$W & N/A & N/A & 1 m \hphantom{o}  \linebreak 2 m  & 42 m \hphantom{o}   \linebreak 37 m & commodity BLE & Closed-loop BC system exploiting frequency and spatial diversity. \\
              $\bullet$ DigiScatter \cite{Zhu.2020} & 54.5 $\mu$W & 620 kbps & 7 m & 7 m & 7 m & customized HW & OFDMA backscatter realizing digital frequency synthesis. \\
              $\bullet$ RF-powered tag \cite{Reed.2021} & 840 $\mu$W & 96 Mbps & N/A & N/A & N/A & customized HW & Compact tag powered by RF EH at 2.45 GHz and backscattering at 1.76 GHz. \\
              $\bullet$ SyncScatter \cite{Dunna.2021} & 30 $\mu$W & 450 kbps & 32.3 m & 12 m  & 30 m & commodity WiFi  & IC achieving synchronized BC with ambient signals.
              \\
              $\bullet$ RapidRider \cite{Wang.2021} & N/A & 200 kbps & 14.5 m &  & 14 m & commodity WiFi &
              Exploitation of single-symbol decoding via forward and backward deinterleavers. \\
              $\bullet$ IBLE \cite{Zhang.2021} & N/A & 8 kbps & 2.5 m & 1 m & 18 m &   commodity BLE &
              Reliable PHY and full BLE compatibility. IPS/GFSK modulation and FEC coding. \\
              $\bullet$ Long-RG FM tag \cite{Hu.2022} & 150 $\mu$W & 1 kbps & 19.6 m&  N/A & 20 m & dedicated FM carrier, SDR &
              Tunnel diode reflection amplifier with high gain and wide bandwidth. \\
              $\bullet$ CD-ZED \cite{Phan.2022} & 54 $\mu$W & N/A & N/A & 7.2 km & 3 m &  SDR & Mobile communication-based backscattering powered by indoor light. \\
              $\bullet$  CAB \cite{Yang.2022} & 271.2 $\mu$W & 341 Mbps & 0.5 m & N/A & N/A & commodity OFDM-WiFi &
              Content-agnostic backscatter that can demodulate both tag and ambient WiFi data. \\
              $\bullet$  SubScatter \cite{Yuan.2022} & 613 $\mu$W & 11 Mbps & 2 m  & N/A & 20 m &  customized, commodity WiFi & Subsymbol backscatter modulation for high THP. \\
              $\bullet$   STScatter \cite{YangGong.2022} & N/A & 500 kbps & 14 m & N/A & 20 m & commodity WiFi, SDR & Universal space-time stream backscatter. \\
                \bottomrule
            \end{tabular}
            \label{tab2}
            \vspace{1mm}
            \raggedright The two rows colored in light red correspond to mono-static BC systems, while the remaining are bistatic.\\
            \raggedright $^a$ \footnotesize This field corresponds to the minimum power consumption (PC) of the backscattering tag. 
            \end{table*}
            %
\subsubsection{Key BC Topologies}
As shown in Figure~\ref{BCfig}, there are two key BC topologies: monostatic and bistatic. In traditional or monostatic backscatter systems, the transmitter and receiver are integrated, e.g., in RFID, while they are separate in bistatic systems.\footnote{Note that the term ``backscatter'' may not be accurate for the bistatic architecture since the signal does not necessarily scatters back, but toward the receiver.} The latter implementation offers \cite{Xu.2018}:
        \begin{itemize}
            \item Temporal flexibility: The tag has additional time slots to transmit data rather than being restricted to waiting for a single reader's protocol-bound inquiry. This allows the backscatter tag (functioning as a sensor node) to transmit the sensory data immediately upon availability, which might be crucial in many sensing scenarios.
            \item Spatial flexibility:  The position of the receivers can be optimized. Moreover, the spatial deployment of backscatter tags can be quite flexible in urban and metropolitan areas with high ambient RF APD, while in other scenarios, dedicated transmitter(s) can be strategically placed in optimal locations to balance the scalability and performance of backscatter tags.  
            \item Technology flexibility: A variety of excitation signals can be exploited, e.g., from ambient RF sources such as TV or frequency modulation (FM) radio towers, cellular BSs, and WiFi access points, while several modulation schemes may be supported.
        \end{itemize}
  Due to the above, bistatic BC technology has become increasingly popular. Notice that the transmitter(s) and receiver(s) may not be independent entities in bistatic systems, but they may cooperate. For instance, the authors in \cite{Katanbaf.2020} propose for them to share link settings and metrics to improve the BC link performance. Nevertheless, cooperation is not possible in the case of ambient BC systems.
 %
 \subsubsection{Use cases}
\begin{figure}[t!]
        \centering
        \includegraphics[width=1\columnwidth]{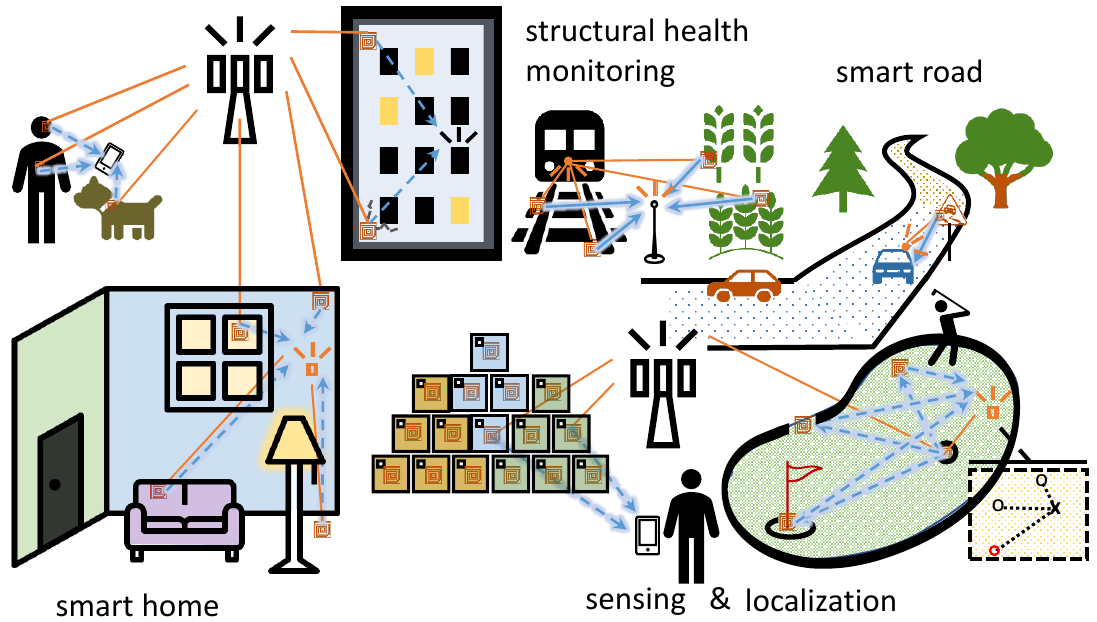}
        \caption{Illustration of some BC applications.}
        \label{BCapp}
        \vspace{-5mm}
    \end{figure}

 The applications of BC technology are numerous as illustrated in Figure~\ref{BCapp} and are rapidly increasing. The three main use cases are related to
 \begin{itemize}     
     \item Invasive monitoring:  BC technology may be useful in scenarios where sensors must be instrumented in an invasive manner because it can potentially eliminate the need for replacing batteries and the corresponding high cost. That is the case of applications such as i) structural monitoring, e.g., to ensure safe and reliable operation of railways, pipelines, dams, bridges, and aircraft, and ii) implantable health-care monitoring \cite{Xu.2018}.     
     \item Ubiquitous localization and sensing: The astonishing spatial diversity offered by the massive proliferation/deployment of backscatter (battery-free) devices (e.g., attached to everyday objects) can be exploited to extend networks' localization and sensing capabilities as envisioned by the 6G paradigm \cite{Lima.2020}.        
     \item Smart world: By incorporating backscatter tags into everyday objects, not only network sensing capabilities are naturally extended, but also each of these objects is digitized and becomes a source of information and/or control unit with added value. 
 \end{itemize}

 An appropriate choice between (non-passive) radios and BC for an application depends on the operating distance, data rate requirement, and power budget \cite{Talla.2021}. There is no one-size-fits-all solution. In general, BC is more energy-efficient but may not cope with stringent requirements in terms of coverage and rate. Indeed, BLE radio is likely the best option currently to support 1 Mbps at 20-40 m, while BC fits better for applications requiring around 100 kbps at 5-10 m. However, this is also a matter of deployment topology, cost, and other KPIs such as reliability, latency, and security as required by the target application. Please refer to Section~\ref{integration} for further discussions on this.
%
\subsubsection{Challenges and Research Directions} 
 Many state-of-the-art BC designs as those implementing LoRa and BLE backscatter in Table~\ref{tab2} are quite mature and deployable. However, there are still a number of challenges that require attention in the coming years to fully realize the potential of BC technology and enable its seamless integration with 5G and beyond generations, thus paving the way for the most advanced and futuristic applications. A compilation of such key challenges and associated research directions is presented below.
\begin{itemize}
    \item Imperceptible integration to everyday objects: Fabricating BC miniaturized electronic components and integrating them into small, and diverse, form factors is challenging, and so is the imperceptible integration into everyday objects. Printed electronics may be appealing here and also the exploitation of meta-materials with favorable electrical properties (see Section~\ref{EE}-\ref{IRS}) to achieve small computational materials that can communicate using backscatter.
\item High-frequency operation: Most current BC systems operate at well-established microwave frequencies. However, expanding the operation to higher frequencies (e.g., mm-wave and THz) may bring substantial benefits since more antennas per unit area can be packed, producing very directional transmit, receive, and backscattered beams while achieving longer communication RGs. Unfortunately, ambient RF energy availability is more limited in such a high-frequency operation regime, which may prevent exploiting ambient BC. Instead, dedicated RF signals/sources are required together with novel low-complex protocols that efficiently handle the issues related to beam search, especially for flexible bistatic BC systems\footnote{In monostatic systems, the tags can be designed to be retrodirective (i.e., capable of reflecting an incident signal toward the source direction without prior knowledge of its direction of arrival) to close a directional link with the reader, e.g. by using phase conjugate arrays, Van Atta arrays, and leaky-wave antennas \cite{Rahmani.2023}.}. Notice that current high-frequency BC prototypes are monostatic, e.g., \cite{Freidl.2017,Guerra.2017,Matos.2021,Simonjan.2022}, or non-flexible bistatic, e.g., \cite{Adibelli.2022}, thus, much further research is needed to realize flexible high-frequency BC systems suited to real-life applications.
\item Wide-band and frequency-agnostic designs: Wide (ultra-wide)-band BC designs (including advanced modulation schemes) are needed for supporting high data-rate applications, e.g., sensing and interaction related to augmented reality, which constitutes a challenging research direction. Also, frequency-agnostic BC systems able to operate across different protocols, locations, and applications, would be extremely appealing. In this regard, HW innovations, as well as low-power algorithms, are required to dynamically identify which frequency bands have the strongest signal. 
\item Security: Due to the limited power/complexity of BC systems, guaranteeing secure communications is extremely difficult. Notice that the i) accurate identification of a fake ambient BC tag, and ii) mitigation of interference generated by a tag maliciously backscattering ambient signals to a nearby reader, remain open problems in the literature. A promising direction lies in designing quantum backscattering mechanisms \cite{Candia.2018}.
\item Enhanced RF-EH sensitivity: RF EH, as the charging source, facilitates small form-factor and battery-free BC implementations. Unfortunately, RF-EH sensitivity is orders of magnitude worse than that of a BC receiver, which limits the connectivity RG. Improving RF-EH sensitivity, preferably up to two orders of magnitude, is a fundamental research and engineering challenge.\footnote{Under free-space propagation conditions, every 6 dB improvement in sensitivity roughly doubles the operating distance  \cite{Talla.2021}.} Techniques exploiting leakage power reduction, technology scaling, and sub-threshold operating using voltage scaling may be key to achieving this goal.
\item BC networks: Most of BC research and prototyping is focused on the physical layer (PHY) with piecemeal evaluation. A full-layer design constitutes a challenging next step for maturing the technology and realizing scalable, integrated, and practical BC networks with the capabilities for realizing, e.g., carrier sense, network management, and polling of devices. 
\end{itemize}
   \begin{figure}[t!]
        \centering
        \includegraphics[width=1\columnwidth]{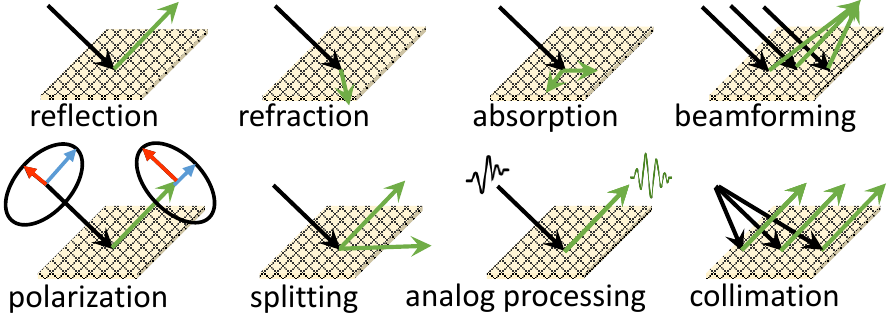}
        \caption{EM-based elementary metasurface functions.}
        \label{metaF}
        \vspace{-7mm}
    \end{figure}
    \vspace{-4mm}
\subsection{\uppercase{Metasurface-aided Communications}}\label{IRS}
Metasurfaces are surfaces composed of metamaterials with sub-wavelength thickness. The so-called metamaterials have special properties when interacting with EM radiation, and thus metasurfaces may support several functions as shown in Figure~\ref{metaF} \cite{Renzo.2020,Long.2021}:
\begin{itemize}
    \item \textbf{Reflection/refraction} of the incident RF waves to a given reflecting/refracting direction;
    \item \textbf{Absorption}, by which the reflected/refracted signals corresponding to an incident RF wave are nulled;
    \item \textbf{Beamforming}, by which the incident RF waves are focused toward a given direction/location. \textbf{Collimation} is the complementary operation;
    \item \textbf{Polarization change} of the reflected RF waves with respect to the incident ones.  For example, incident RF waves are transverse electric polarized, and reflected RF waves are transverse magnetic polarized;
    \item \textbf{Splitting}, by which multiple reflected or refracted RF waves are created from the incident RF waves;
    \item \textbf{Analog processing} includes mathematical operations at the EM level, e.g., the RF waves refracted by a metasurface can be the first-order derivative or the integral of incident RF waves.    
\end{itemize}
These functions motivate the use of metasurfaces for improving wireless systems performance, e.g., for passive signal or energy relaying \cite{Wu.2020} and wireless sensing \cite{Elbir.2022}.

\begin{table*}[h!]
            \caption{Some Representative State-of-the-Art (2017-2022) Experiments and Prototypes of Reconfigurable Metasurfaces}
            \vspace{-1mm}
            \centering          
            \begin{tabular}{p{0.7cm}p{1.3cm}p{1.6cm}p{2cm}p{2cm}p{7.5cm}}
                \toprule
                \textbf{Ref.}  & \textbf{Elements} & \textbf{Controller} & \textbf{Control Units} & \textbf{Operation Freq.} & \textbf{Functions and Other Features} \\
                \midrule
                \cite{Huang.2017} & $22\times 22$  & DC voltage \hphantom{0} source & PIN and varactor diodes & $11.5-13.5$ GHz & Dynamical beam deflection, splitting, and polarization change. $180^\circ$ reflection phase difference. \\
                \cite{Tian.2017} & $20\times 20$  & Bias network & 1 varactor diode & $3-4$ GHz & Discretely-tuned broadband beam reflection/ scanning. \\
                \cite{Zhou.2018} & $40\times 40$  & N/A & N/A & $8.7-11.3$ GHz & Generating several radiation patterns. Even distribution of reflection phases from 0 to $2\pi$. \\
                \cite{Zhang.2018} & $8\times 8$  & FPGA & Binary PIN \hphantom{hhh} diodes & $8-10$ GHz & Space-time harmonic beam steering and beam shaping.\\
                \cite{Tan.2018} & $14\times 16$  & MCU & Binary relay \hphantom{hhh} switches & 60 GHz & Beam reflection/steering. A two-stage beam-searching algorithm is implemented. \\
                \cite{Luo.2019} & 75 & N/A & N/A & 8.6, 10.1 GHz & Beam splitting. Two beams with different circular polarizations at each frequency. \\
                \cite{Tran.2019} & $16\times 16$  & Control board & Binary PIN \hphantom{hhh} diodes & 5.8 GHz & Beam steering/focusing for enhancing PTE of RF-WET systems.  \\
                \cite{Dai.2020} & $16\times 16$  & FPGA & PIN diodes \hphantom{000} (2-bit setups) & 2.3, 28.5 GHz & Active wide-angle beam-scanning/transmission. 21.7/19.1 dBi gain achieved at 2.3/28.5 GHz. \\
                \cite{Liu.2020} & $12\times 12$  & MCU & PIN diodes & $6.1-6.6$ GHz & Polarization selection/reconfiguration and frequency reconfigurable polarization conversion. \\         
                \cite{Zhang.2020} & $24\times 24$  & FPGA & 4 varactor diodes per element & 5.67 to 6.15 GHz &  i) spin-modulating circularly polarized waves, ii) dual-beam scanning, iii) dual-polarized shared-aperture antenna.\\
                \cite{Pei.2021} & $55\times 20$ \hphantom{00} ($0.25$ m$^2$) & FPGA & Binary varactor \hphantom{0} diodes & $5.7-5.9$ GHz &   Self-adaptive beamforming based on real-time feedback.  \\
                \cite{Vardakis.2021} & $5\times 20$ & SDR reader & RFID tag & $866$, 870 MHz & Backscattering (virtual IRS composed of an array of commodity RF-powered RFID tags). \\
                \cite{Xiao.2021} & $16\times 16$ & FPGA & N/A & $5.79-5.81$ GHz &  Beamforming exploiting low-complexity neural beam alignment. \\
                \cite{Staat.2021} & $8\times 16$ & MCU with \hphantom{00} shift registers  & Binary PIN \hphantom{0} diodes & 5.3 GHz & Channel randomization to generate temporal variation and provide UE-independent entropy source. \\
                \cite{Liang.2022} & $10\times 10$ \hphantom{00} ($0.07$ m$^2$) & N/A & Varactor diodes & $2.9-3.35$ GHz & Beam reflection with \hphantom{00} stable phase and amplitude responses at oblique incidences. \\
                \cite{Trichopoulos.2022} & $16\times 10$ \hphantom{00} $0.1$ cm$^2$ & MCU with \hphantom{00} shift registers & Binary PIN \hphantom{00} diodes & 5.8 GHz   &  Beam reflection/scanning. $60^\circ$ angle resolution for maintaining a single main lobe.   \\                
                \cite{Fara.2022} & $14\times 14$ & N/A & 4 varactor diodes per element & $5.15-5.75$ GHz & Beam steering/reflection with $40^\circ$ angular resolution.  \\
                \cite{Ouyang.2022} & $20\times 20$ & FPGA & Binary PIN \hphantom{00} diodes & 5.4 GHz & Beam steering/reflection assisted by computer vision methods leveraging a camera attached to the IRS. \\
                \cite{Rossanese.2022} & $10\times 10$ & MCU & 3-bit RF switches \& delay lines & 5.3 GHz & Beam steering/reflection with high spatial resolution.                \\                
                \bottomrule 
            \end{tabular}           
         \label{IRSprot} \vspace{-4mm}        
            \end{table*}

Metasurfaces are low-cost fully-passive devices with zero EC, which can be engineered to statically perform one or several of the above functions. 
However, more recently, the research community and industries are pursuing a more dynamic approach, where the operation of the metasurfaces is SW-controlled in real-time, thus, leading to intelligent metasurfaces. Next, we briefly discuss the main configurations of intelligent metasurfaces in the state of the art: i.e., intelligent reflective surface (IRS) and large intelligent surface (LIS), associated research challenges, and relevant research directions. A list of representative prototypes and experiments with metasurfaces in the last years is presented in Table~\ref{IRSprot} along with their main distinctive features.
%
\subsubsection{IRS}
\begin{figure}[t!]
        \centering
        \includegraphics[width=1\columnwidth]{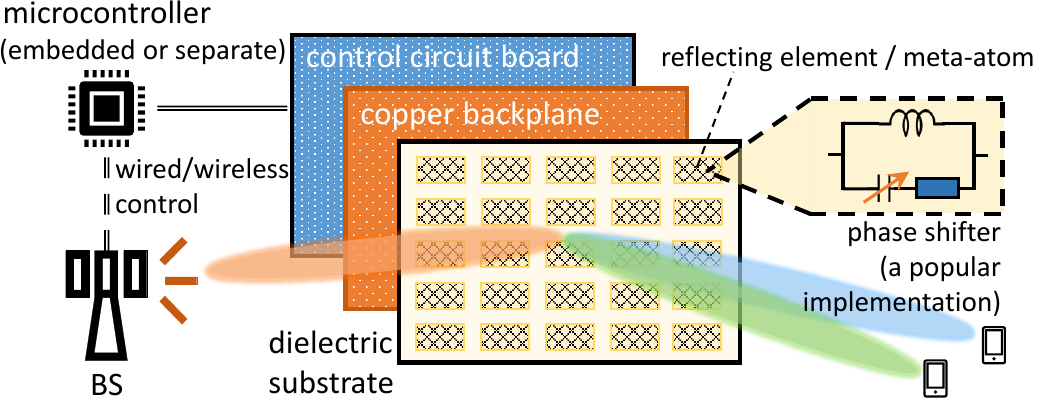}
        \vspace{-1mm}
        \caption{IRS architecture and assisted communication.}
        \label{IRShw}
        \vspace{-6mm}
    \end{figure}

As illustrated in Figure~\ref{IRShw}, IRS (also known as reconfigurable intelligent surface - RIS) is a metasurface composed of a large number of $N$ subwavelength-spaced passive scattering metamaterial elements. Each scattering element can be controlled in an SW-defined manner by the so-called IRS controller, which can be embedded or separated,  to properly tune the EM properties of the output signals given a set of incident RF signals impinging the scattering elements. Therefore, IRS  is a promising technology to dynamically control the radio propagation environment and improve the wireless system performance in a cost/energy-effective manner. Specifically, an IRS can reflect the incident signals so they are added constructively in the desired direction to increase the signal power (so-called passive beamforming) or destructively for mitigating undesired interference, either for high data rate, dependable, secure, NLOS, or wide-coverage communications, and RF-WET (including joint communication and RF-WET).

Conventional IRSs, so-called passive IRSs, e.g., \cite{Huang.2017,Tian.2017,Zhou.2018,Zhang.2018,Tan.2018,Dai.2020,Liu.2020,Zhang.2020,Pei.2021,Xiao.2021,Staat.2021,Liang.2022,Trichopoulos.2022,Fara.2022,Ouyang.2022,Rossanese.2022}, are composed of fully passive reflecting elements and the only active EC comes from the controlling HW (IRS controller and related active circuitry). This allows low-power/cost energy-efficient implementations that do not incorporate additional RF radiation into the environment, which is undoubtedly appealing in terms of sustainability. 

Unfortunately, the effective coverage of passive IRS may be seriously limited since the reflected signal suffers high \textit{product-distance path loss}, which critically constrains the signal power at the receiver(s). This can be addressed by either equipping the IRS with an increasingly massive number of passive elements and/or conveniently deploying the IRS closer to the transmitter/receiver to reduce the cascaded channel path loss. Unfortunately, these options are not always viable due to, e.g., deployment difficulty (limited space and unavailable site) and enormous training overhead for the CSI acquisition required to optimize the operation of the passive IRS elements in real time. 
This has motivated the introduction of active IRS \cite{LongRuizhe.2021,Kang.2023}.

\begin{table*}[h!]
            \caption{Main Features of IRS and  Competing Relaying Technologies}
            \vspace{-1mm}
            \centering          
            \begin{tabular}{C{1.5cm}C{0.95cm}C{1.7cm}C{1.7cm}C{2.0cm}C{1.3cm}C{2.4cm}C{2.5cm}}                \toprule
                \textbf{Technology}  & \textbf{Duplex Mode} & \hphantom{00} \textbf{Power} \hphantom{000} \textbf{Consumption} & \textbf{Amplification Noise} & \textbf{EE} & \textbf{HW Cost} & \textbf{Network Spectral Efficiency}$^b$ & \textbf{Operation Mechanism} \\
                \midrule
                Passive IRS & FD & Very Low & No & High (large $N$)  & Very Low & Low & Reflect \\
                Active IRS & FD & Low & Yes & High (small $N$)  & Low & Moderate & Reflect \&  Amplify \\ 
                AF Relay & HD/FD & Moderate & Yes & Low & High & Moderate-High & Amplify \& Forward  \\
                DF Relay & HD/FD & High & No & Low & High & High & Decode \& Forward \\               
                \bottomrule
            \end{tabular}            \begin{flushleft}\footnotesize{\hspace{2mm} $^b$ Based on the ability to simultaneously serve multiple users and to act in other roles (e.g., as sources).}\end{flushleft} \vspace{-5mm}
         \label{IRSfeatures}         
            \end{table*}

In active IRS, the reflecting elements incorporate reflection-type amplifiers, which are implemented without high-cost and power-hungry RF chains and that can simultaneously alter the signal's phase and amplitude to enhance the signal power at the receiver. This requires modestly higher EC and HW cost compared to conventional IRS.

Interestingly, the achievable signal-to-noise ratio scales as $\mathcal{O}(N)$ and $\mathcal{O}(N^2)$ when using active and conventional IRS, respectively \cite{Kang.2023}. The limited gain when using active IRSs is due to the noise power amplification introduced by the active design. Nevertheless, an active IRS implementation still provides superior rate performance (subject to a given total power budget constraint) in comparison to a conventional IRS when $N$ is moderate thanks to the additional signal power amplification gain \cite{Kang.2023}. Table~\ref{IRSfeatures} illustrates  the main features of conventional and active IRSs with competing relaying technologies. All in all, there is no \textit{one-size-fits-all} solution (as usually in engineering), and the use of one technology over another depends on the specific network's characteristics, constraints, and performance requirements.
%
\subsubsection{LIS}
LIS is a metasurface equipped with RF circuits and signal processing units and composed of a virtually infinite number of elements to form a spatially continuous transceiver aperture. This structure can be used to transmit/receive communication signals across the entire surface by leveraging the hologram principle \cite{Huang.2020}.\footnote{Alternatively, a LIS can be implemented as an IRS-aided antenna by deploying an external communication and antenna module to wirelessly feed the IRS for active communication \cite{Jamali.2021}, e.g., \cite{Luo.2019,Tran.2019}. Another LIS implementation can be based on discrete photonic antenna arrays integrating active optical-electrical detectors, converters, and modulators for transmission, reception, and conversion of optical or RF signals \cite{Huang.2020}.}

\begin{figure}[t!]
        \centering
        \includegraphics[width=1\columnwidth]{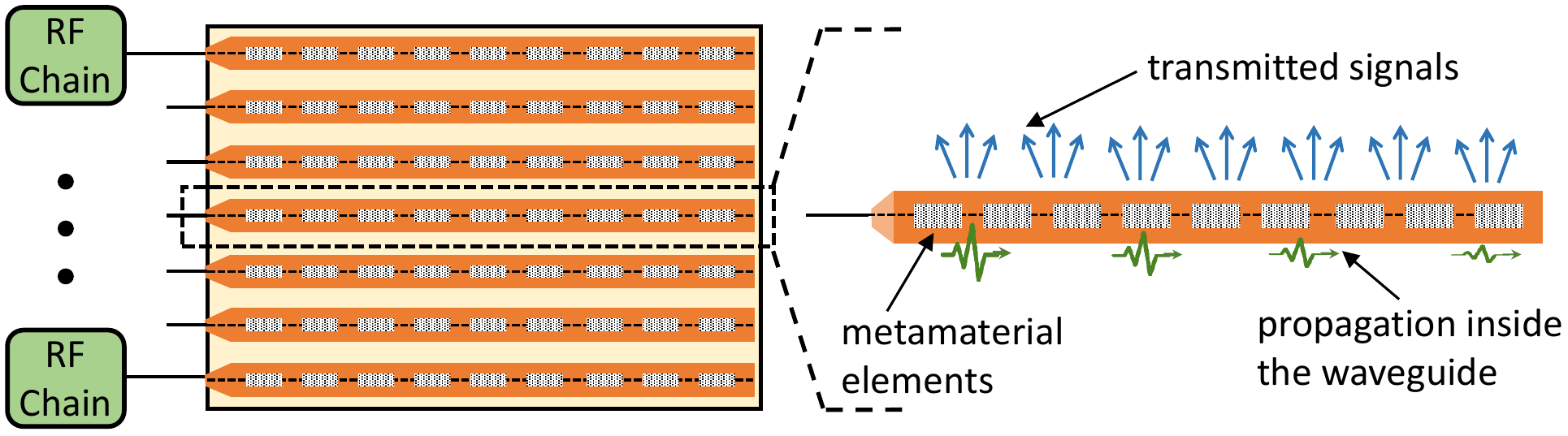}
        \caption{LIS implemented with multiple microstrips. Each microstrip is connected to a single RF chain.}
        \label{DMA}
        \vspace{-8mm}
    \end{figure}

For instance, an LIS may be comprised of multiple waveguides, e.g. microstrip. Each waveguide may embed a large set of radiating metamaterial elements whose frequency response can be externally and individually adjusted by varying the local electrical properties. Each microstrip is fed by one RF chain, and the input signal is radiated by all the elements located on the microstrip, as shown in Figure~\ref{DMA} \cite{Zhang.2022}. The figure also illustrates an example of transmitting a signal using a single microstrip with multiple elements. 

  \begin{figure*}[t!]
        \centering
        \includegraphics[width=2\columnwidth]{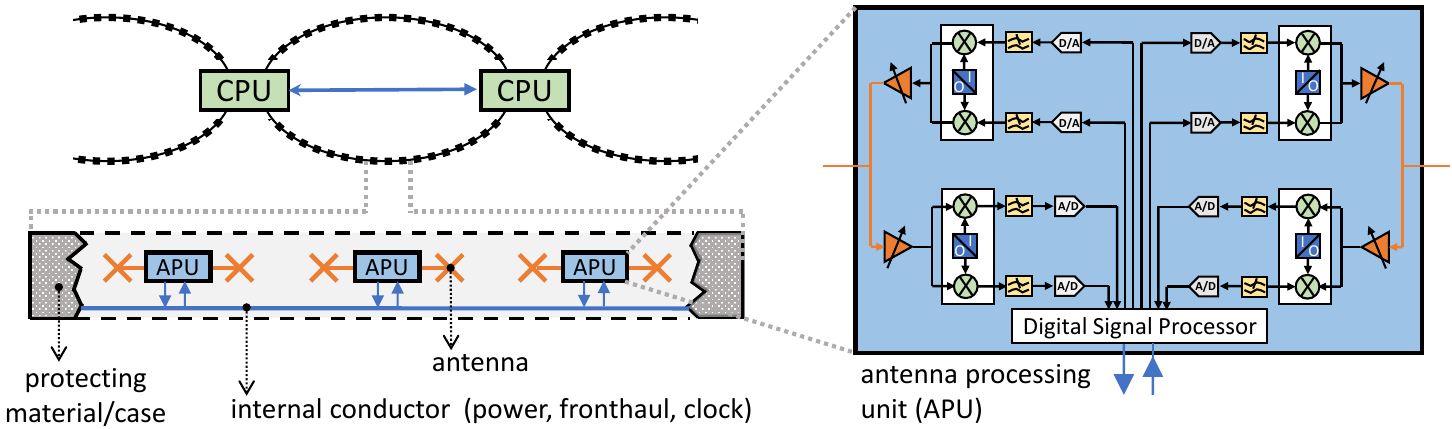}
        \caption{Structure of a radio stripe system. Each radio stripe can be connected to one or multiple central processing units (CPUs).}
        \label{radioStripe}
        \vspace{-5mm}
    \end{figure*}
    
\subsubsection{Challenges and Research Directions}
Some key challenges and associated research directions for maturing metasurface-assisted communication technology and making it a reality in future  sustainable networks are briefly discussed in the following:
\begin{itemize}
\vspace{-1mm}
    \item Low-cost/energy control: The use of tunable reflecting elements with discrete amplitude/phase shift levels favors cost/energy-effective implementations. However, this may significantly limit the beamforming/reflection capabilities of the metasurface, especially in the case of limited form-factor implementations. Therefore, further studies are required to unveil such underlying trade-offs. Moreover, further research is required on the metasurface controller circuitry, which interfaces with all the tunable reflecting elements and constitutes the only EC source in passive IRS implementations. Specifically, low-cost/energy-effective control mechanisms must be developed to connect and communicate with massive tunable elements, and thus agilely and jointly control their EM behaviors on demand.  
    \item CSI acquisition: A tunable passive beamforming/reflection typically requires accurate CSI, which is challenging to acquire in practice. The two main approaches proposed in the literature for passive IRS (without any active RF chain or reflection-type amplifier), but that still require further research in terms of performance trade-offs, are \cite{Wu.2020}: i) estimate the concatenated (TX-IRS and IRS-RX) channel with some known IRS reflection patterns, or ii) exploit feedback from the TX/RX pertaining to their received signals that are reflected by the IRS (no explicit channel estimation). Meanwhile, CSI acquisition for active IRS is generally more challenging because explicit CSI of the separate TX-IRS and IRS-RX links, instead of the cascade CSI, is needed due to the amplification noise. The research addressing this issue is still incipient, e.g., cf. \cite{Kang.2023}.  Notably, approaches exploiting limited CSI as those relying, e.g., on positioning information \cite{LopezMahmood.2020}, may be appealing.  
    \item Advanced metasurface implementations: The most commonly investigated/prototyped metasurface applications are those related to passive beamforming/reflection. However, as illustrated in Figure~\ref{metaF}, there are several other metamaterial functions, which are so far only incorporated into non-flexible/configurable implementations. In the next years, SW-controlled engineering solutions must be developed exploiting these metamaterial functions for real-life applications. For this, accurate physics- and EM-compliant models are needed.
    \item Data-driven optimization: Accurate modeling/analysis and optimum design/implementation/deployment of IRSs/LISs are challenging due to the inherent complexity of such systems. This calls for efficient data-driven methods, e.g., based on deep learning (DL), reinforcement learning (RL),  and transfer learning \cite{Zappone.2019}. Moreover, low-complexity/energy ML mechanisms, the so-called TinyML (see Section~\ref{ML}-\ref{tinyMLsec}), seem appealing for incorporation into the metasurface controller circuitry for more cost/energy-effective implementations.
\end{itemize}
\vspace{-4mm}
\subsection{\uppercase{Radio Stripes}} \label{lowPMIMO}
    Massive MIMO technology addresses challenging 5G performance requirements, especially in terms of network coverage, capacity, and THP. However, the required high manufacturing and operating expenses, as well as the increased EC, make the development and deployment of truly large-scale antenna arrays extremely difficult. This motivates the research on more affordable and low-power MIMO architectures that can scale with the number of antennas more sustainably. The metasurface-aided communication architectures overviewed in Section~\ref{EE}-\ref{IRS} constitute one active research front in this direction. Indeed, IRSs can be deployed to efficiently assist already-deployed MIMO networks (thus, avoiding the need to install new active HW); while LIS are basically low-power/cost massive MIMO that can replace traditional MIMO HW in many scenarios. However, the deployment of IRS/LIS alone may not fulfill all the use cases' constraints/requirements, at least in the near future, for which alternative low-power/cost MIMO architectures are still needed. That is the case of the radio stripes system discussed in the following. 

Radio stripe systems enable cost/energy-effective distributed massive MIMO \cite{Interdonato.2019}. In such a system, with architecture depicted in Figure~\ref{radioStripe}, antenna elements and circuit-mounted chips (including power amplifiers, phase shifters, filters, modulators, and A/D and D/A converters)  are serially located inside the protective casing of a cable or a stripe, which also provides synchronization, data transfer, and power supply via a shared bus. Unlike traditional massive MIMO BSs, radio stripes \cite{Interdonato.2019,Lopez.2021}: i) allow imperceptible/flexible installation in existing construction elements and alleviate the deployment permissions problem, ii) support native system resiliency to failures, and iii) facilitate low EC due to the inherent low-complex and distributed architecture functionality. Also, additional HW, including temperature/vibration sensors and microphones/speakers, can be deployed in the radio stripes to provide additional features/services, e.g., fire/burglar alarms, earthquake warnings, indoor positioning, and climate monitoring/control. 

Due to the above, radio stripes technology is attractive for supporting energy-efficient (and sustainable) networks. Applications are numerous, e.g., to: i) facilitate high spatial multiplexing and low EC in indoor communications \cite{Ganesan.2020}, ii) increase the coverage and end-to-end PTE of RF-WET  \cite{Lopez.2021,Lopez.2022}, and iii)  support ultra-reliable low-latency communications in industrial IoT networks \cite{Tominaga.2022}. 
In general, although the last few years have witnessed significant advances in this technology (e.g., see \cite{He.2021,Shaik.2021,ShaikZakir.2021} and references therein), more advanced/efficient resource allocation schemes, circuit implementations and prototypes, and distributed processing architectures to avoid costly signaling between the antenna elements, still pose open research challenges.

  \begin{figure}[t!]
        \centering
            \includegraphics[width=0.85\linewidth]{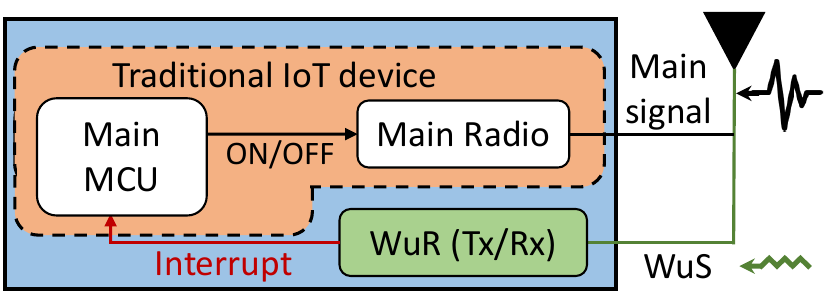}
        \caption{IoT device with WuR architecture for the same-band operation of WuR and main radio. 
        Note that different architectures may or may not include all the components, such as a WuR with Tx capabilities. }
        \label{arq_wur}
        \vspace{-6mm}
    \end{figure}

 \vspace{-4mm}
    \subsection{\uppercase{W\lowercase{u}R}}\label{wakeUp}
\vspace{-1mm}
    Duty cycling, consisting of turning off the radio component of a device, pausing its MCU, and using a timer to activate the device periodically, extends the lifetime of devices with limited battery capacity (BCAP). However, the device with the radio off cannot exchange data, which may result in excessive communication delays~\cite{ghribi2020survey}.  This motivates the adoption of WuR, which allows activating the devices' main radio on-demand when there is data to communicate.     Since its EC is several orders of magnitude lower than that of the traditional main radio, e.g., about 1000 times lower~\cite{oller2015has}, the WuR can be kept always on, in contrast to the duty cycling operation. Furthermore, although the EC for operating a clock is relatively low, it is still non-zero, while this EC may be eliminated entirely by employing WuRs. 
    
\begin{figure}[t!]
        \centering
            \includegraphics[width=0.9\linewidth]{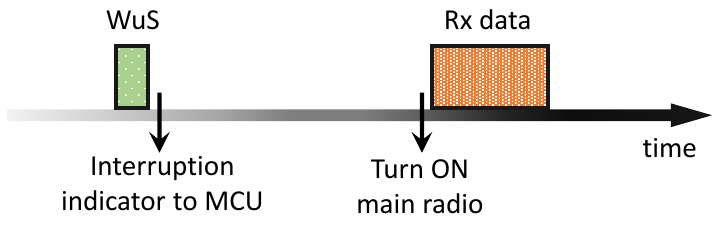}\\
            \vspace{3mm}
            \includegraphics[width=0.9\linewidth]{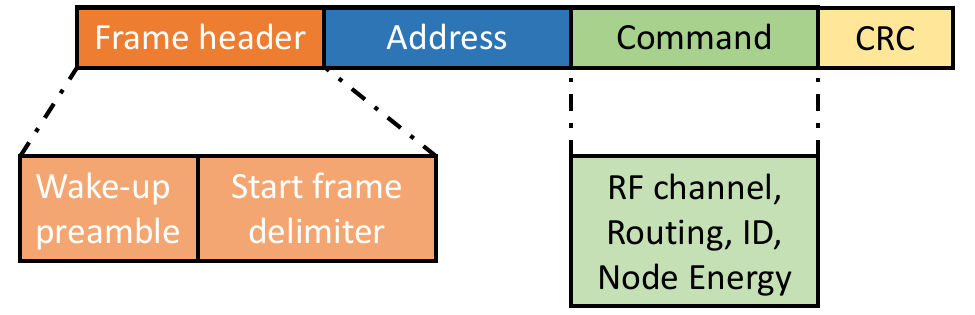}
        \caption{Typical triggering using WuR scheme (top) and the WuS packet structure (bottom).}
        \label{arq_wur1}
        \vspace{-5mm}
    \end{figure}

    Figure~\ref{arq_wur} illustrates an exemplary architecture of a WuR-capable device, which includes a low-power radio that receives and detects the WuR signal to activate the main radio for communication. The specific components and their inclusion in different architectures can vary based on the design requirements and specific use cases. A typical WuR setting is illustrated in Figure~\ref{arq_wur1}. The main radio of the device remains deactivated (OFF) until it is required for communication, or until a special packet known as the Wake-up signal (WuS) is received by the WuR, which generates an interrupt signal to the main MCU to switch it ON. Subsequently, the main radio can exchange data packets with the other node in a conventional manner~\cite{wus_survey}. Figure~\ref{arq_wur1} also illustrates the composition of a WuS packet. Note that the frame header consists of the wake-up preamble and start frame delimiter, a standard byte pattern agreed upon between the transmitter and the receiver. The preamble is used for synchronization whereas the start frame delimiter indicates the start of the frame that contains relevant information. Meanwhile, the address field contains the destination node identifier and the payload contains application data, commands, or extra instructions specified by the UE or application. Finally, there is an error detection frame, using cyclic redundancy check, aka CRC, to check data integrity~\cite{wus_survey}. 
	
    The main benefits of using WuR are: 
    \begin{itemize}
       \vspace{-2mm}
	      \item EE i) avoiding unnecessary idle listening, ii) avoiding energy wasting related to start-up/power-down, and iii) combining with uplink reference signaling such that even high-speed UEs can reside in the sleep state for long periods, while not increasing the handover failure rate~\cite{FWuS,rostami2020wake}. 
	      \item Short buffering delay~\cite{rostami2020wake}.
	      \item Synchronization assistance~\cite{FWuS,rostami2020wake}.
	      \item Low wake-up reconfiguration need for many realistic unsaturated traffic scenarios~\cite{FWuS,rostami2020wake}.
	      \item Less signaling overhead~\cite{rostami2020wake}.
	\end{itemize}

    \subsubsection{WuS standarization}   
\begin{figure*}[t!]
        \centering
        \includegraphics[width=0.9\linewidth]{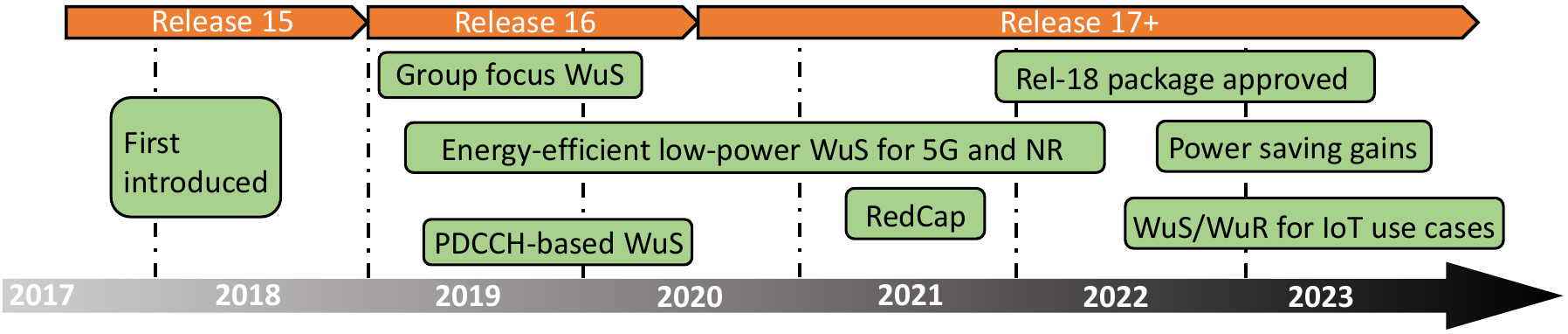}
        \caption{3GPP Release evolution related to WuS.}
        \label{timeline}
        \vspace{-5mm}
    \end{figure*}

    Figure~\ref{timeline} depicts a timeline overview of the key aspects and features associated with WuS in each 3rd Generation Partnership Project (3GPP) Release. 
    Within the 3GPP standardization process, WuS was initially introduced in Release 15~\cite{3gpp2019ts38,3gpp} as a paging signal sent over the PHY downlink shared channel that ``wakes up'' a UE from an idle state so that it can prepare to receive data. Concerns towards enabling energy-efficient techniques resulted in a feature for 5G called low-power WuS in Release 16~\cite{3gpp16,3gpp162}. Next, we primarily focus on the standardization of WuR within 3GPP. 

            \begin{table*}[t!]
            \caption{WuR taxonomy}
            \vspace{-1mm}
            \centering
            \label{taxonomy}
            \begin{tabular}{p{1.8cm}p{6.5cm}p{6.85cm}c}
                \toprule
                   &\multicolumn{2}{c}{\textbf{Classification}} & \textbf{Ref.}\\
                \midrule
                WuR-based MAC protocols
                &\textbf{Only MAC} (access decision is done just in MAC layer)   &\textbf{Cross layer} (interaction between layers to optimize MAC layer performance)
                &\cite{ghribi2020survey}\\ \hdashline
                Wake-up circuit
                &\textbf{Duty cycled} (``duty-cycled'' WuR, which is activated periodically to sense the radio channel and send WuS)    &\textbf{Continues cycled} (WuR's circuit is always ON)    &\cite{djiroun2016mac}\\ \hdashline
                Communication initiator
                &\textbf{Transmitter initiated} (the transmitter first sends a WuS to activate the desired node and then sends data)      &\textbf{Receiver initiated} (receiver sends a WuS asking for data, \textbf{bidirectional} WuS is also supported)    &\cite{ghribi2020survey}\\ \hdashline
                Addressing scheme                &\textbf{Broadcast} (the WuS is received by all the neighboring nodes of the sender)   &\textbf{Identity-based} (only the desired next hop node is wakening up)     &\cite{ghribi2020survey}\\ \hdashline
                Frequency band &\textbf{In-band with the main radio} (main radio has predictable dormant periods, cheaper as there is no need for a separate antenna)
                &\textbf{Out-of-band} (unpredictable dormant periods, it reduces interference, increases signal capacity, but may increase cost and complexity)
                &\cite{wus_survey} \\ \hdashline
                Power source &\textbf{Active} (battery-powered, immediate responsiveness but consumes extra power, ideal for real-time monitoring, rapid event detection, and instant data transmission)    &\textbf{Fully-passive} (no internal power supply is needed, while \textbf{hybrid} combines EH with a small backup battery for responsiveness, limited operational RG, and minimal EC)
                &\cite{rinne2017viability} \\ 
                \bottomrule
            \end{tabular}
            \vspace{-2mm}
            \end{table*}  
    
    Release 16 and Release 17 incorporate improved cross-slot scheduling~\cite{white,3gpp161}. Specifically, the network can inform a device when there is a guaranteed minimum time interval between downlink packet transmissions, leading to a significant reduction of unnecessary RF operations. In addition, group focus WuS is defined, allowing the network to wake up a configurable group of UEs (instead of all UEs) by configuring the WuR of each UE to listen for a specific pattern or sequence in the WuS that is unique to their assigned group, thereby reducing EC~\cite{white}. Another feature is that the network transmits a PHY downlink control channel-based WuS before active on-duration within discontinuous reception cycles. This allows for UEs to avoid PHY downlink control channel monitoring during on-durations within which the network is not transmitting any data to the UE, and may enable between 15\% and 30\% energy savings. Alternatively, WuS may allow for shorter discontinuous reception cycles, i.e., faster response time, with similar EC~\cite{white}. 

   Release 17 specified power-saving enhancements for reduced-capability devices, referred to as RedCap, while Release 18 promises to provide further EE gains~\cite{lin2022overview}. The current set of work and study items for 3GPP Release 18 radio access network (RAN) features is split into four RAN working groups. Specifically, the first group (RAN1), is responsible for the PHY specification including PHY channels and modulation, PHY multiplexing and channel coding, PHY procedures and measurements, as well as PHY-related UE capabilities. Within the 5G-Advanced Release 18 scope, low-power WuS and WuR constitute a new study item in RAN1, which aims at studying power-saving schemes that do not require existing signals to be used as WuS and includes i) evaluation methodologies for low-power WuS/WuR for power-sensitive small form-factor devices; ii) evaluation of low-power wake-up receiver architectures and WuS designs to support wake-up receivers; iii) PHY procedures and higher layer protocols to support WuS; and iv) UE EE gains compared to the existing Rel-15/16/17 UE power saving mechanisms and their coverage availability, and latency impact~\cite{3gpp18}. Finally, WUR standardization is also being carried out by IEEE, e.g., as a part of IEEE 802.11ba standard~\cite{10017392}.
%
    \subsubsection{State-of-the-art}
    WuR constitutes a promising technique for achieving IoT devices' lifespan superior to 10 years~\cite{froytlog2019ultra,oller2015has,rinne2017viability}, and thus has recently received significant attention from the scientific community. For example, the authors in ~\cite{froytlog2019ultra} presented a WuR-enabled BLE device targeting an IoT scenario where such battery-powered massive IoT devices do not support direct 3GPP connections. Real-life results show that the system meets the over-10-year lifetime target while satisfying the latency requirements for 5G IoT devices. Similarly, the authors in~\cite{mikhaylov2020wake} shed some light on BLE-compatible sensor devices enriched with a WUR, and their results demonstrate EE gains in low-latency applications (under 2s). In \cite{rostamitrans}, the authors optimize operational parameters, determined by BS at the beginning of the session, to save energy.     Meanwhile, an accurate traffic forecasting model is proposed in \cite{FWuS} to optimize the wake-up parameters, achieving up to 35\% EC reduction. Additionally, \cite{petajajarvi2016human} proposes a super-regenerative WuR solution to improve EE in human-body communication. Herein, WuR operates at a very low data rate, e.g., 1.25 kbps, for higher sensitivity while consuming $\sim$40 $\mu$W. Likewise, the authors in~\cite{petajajarvi2015wban} show the EE from using WuR in wireless body area network applications with event-driven traffic and propose a WuR capable of receiving small control commands besides WuS. All in all, the different WuR proposals vary depending on the particular protocols used, the type of circuitry, and the application. Here, we summarize the main classification of WuR research in Table~\ref{taxonomy}. 
    %
 \subsubsection{Challenges and trade-offs}
Using a WuR brings two main impacts on the devices' performance. On the one side, there is the problem of miss-detection of the WuS, which happens when the device does not receive the page occasion scheduled within the WuS for information exchange. As a result, the device misses the chance to wake up and has to wait until the next page occasion, increasing the latency of packets/information. 
    On the other hand, due to the inherent simplicity of the WuS, the problem of false alarms needs to be addressed. In a false alarm event, the device/WuR receives a page occasion needlessly when no information is intended to be transmitted/received. Therefore, special attention to these problems is needed when using WuR~\cite{FWuS}.      

Some key challenges and research directions are:
        \begin{itemize}
            \item WuS may complicate radio resource management and device scheduling in the network due to sleep patterns, reducing potential EE gains. In this context, employing advanced ML-based scheduling algorithms considering the sleep patterns of devices may be appealing~\cite{FWuS}.
            \item A design where the WuR utilizes a different frequency band than the main radio increases the complexity and cost of the devices. Therefore, in-band operation and RF integrated circuit-embedded WuR implementation is desired. However, this approach complicates resource management and reduces the available spectral resources for transferring application data.
            \item Beamformed WuS at mm-waves and mobility management is still an open challenge since beam sweeping for WuS is required to reach a desired device. The network should be able to optimize the number of beams in a single WuS burst utilized for waking up the device.
            \item Applying WuR brings trade-offs between EE and other KPIs, depending on the application scenario, like latency, reliability, and robustness~\cite{rostami2020wake}. Therefore, more research should be directed in this direction, especially when dealing with massive low-power IoT scenarios.
        \end{itemize}     
     \vspace{-2mm}
    \section{\uppercase{Energy-efficient ML}}\label{ML}
    
    Native ML support is essential for dealing with the increasing complexity and automation of 6G networks while improving their performance~\cite{loven2019edgeai,ahmad2020machine,peltonen20206g,LopezMahmood.2022}. ML techniques can help address issues such as increasing traffic demands, real-time QoS requirements, and resource allocation. However, these benefits usually come at high computational and memory requirements. Therefore, energy-efficient ML algorithms are paramount for network sustainability~\cite{ML_IoT}.
    
    Several ML features affect EC. In general, larger models, i.e., with more parameters, require more energy to be trained and exploited. Also, training time (TRNT) and inference time (INFT) directly affect the EC at the training and inference phases, respectively.  Therefore, the trade-offs between power cost and performance reward require special attention.

\begin{figure}[t!]
        \centering
        \includegraphics[width=\linewidth]{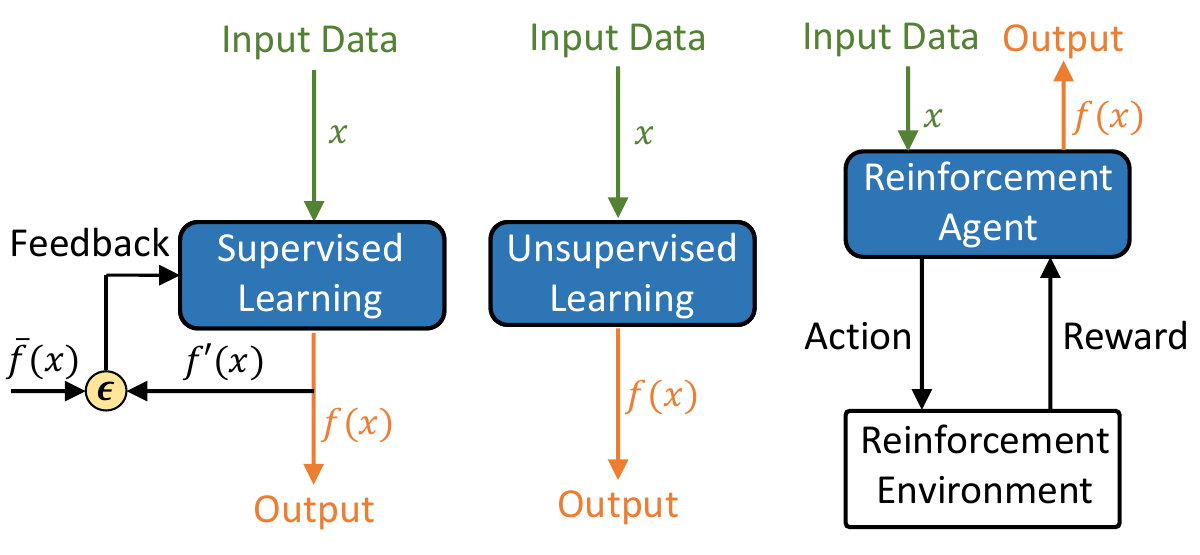}
        \caption{High-level representation of SL, UL, and RL working principles.}
        \label{mainML}
        \vspace{-8mm}
    \end{figure}

       As shown in Figure~\ref{mainML}, ML approaches are broadly classified as supervised learning (SL), unsupervised learning (UL), and RL. In SL, the model is trained on a labeled dataset that contains the correct outcome for the corresponding input. SL can be used for tasks such as predicting energy usage based on historical data or identifying energy-efficient products based on specific features~\cite{morocho2019machine}. Meanwhile, UL does not require labeled data and thus can be used, e.g., to identify patterns in energy usage data, detect anomalies that may indicate energy waste, or reduce the dimensionality of the data~\cite{fourati2021survey}. In RL, effective solutions are learned over time given constraints imposed by the inputs and without attempting to find hidden categories or structures. RL is suitable for solving problems with multiple optimal solutions\cite{klaine2017survey}, such as optimizing energy usage in buildings or predicting optimal times to charge electric vehicles. Notice that RL algorithms can be computationally intensive, but they have the potential to improve EE over time by learning from experiences and making adjustments accordingly~\cite{van2016deep}.
    
     \begin{figure*}[t!]
        \centering
        \includegraphics[width=\linewidth]{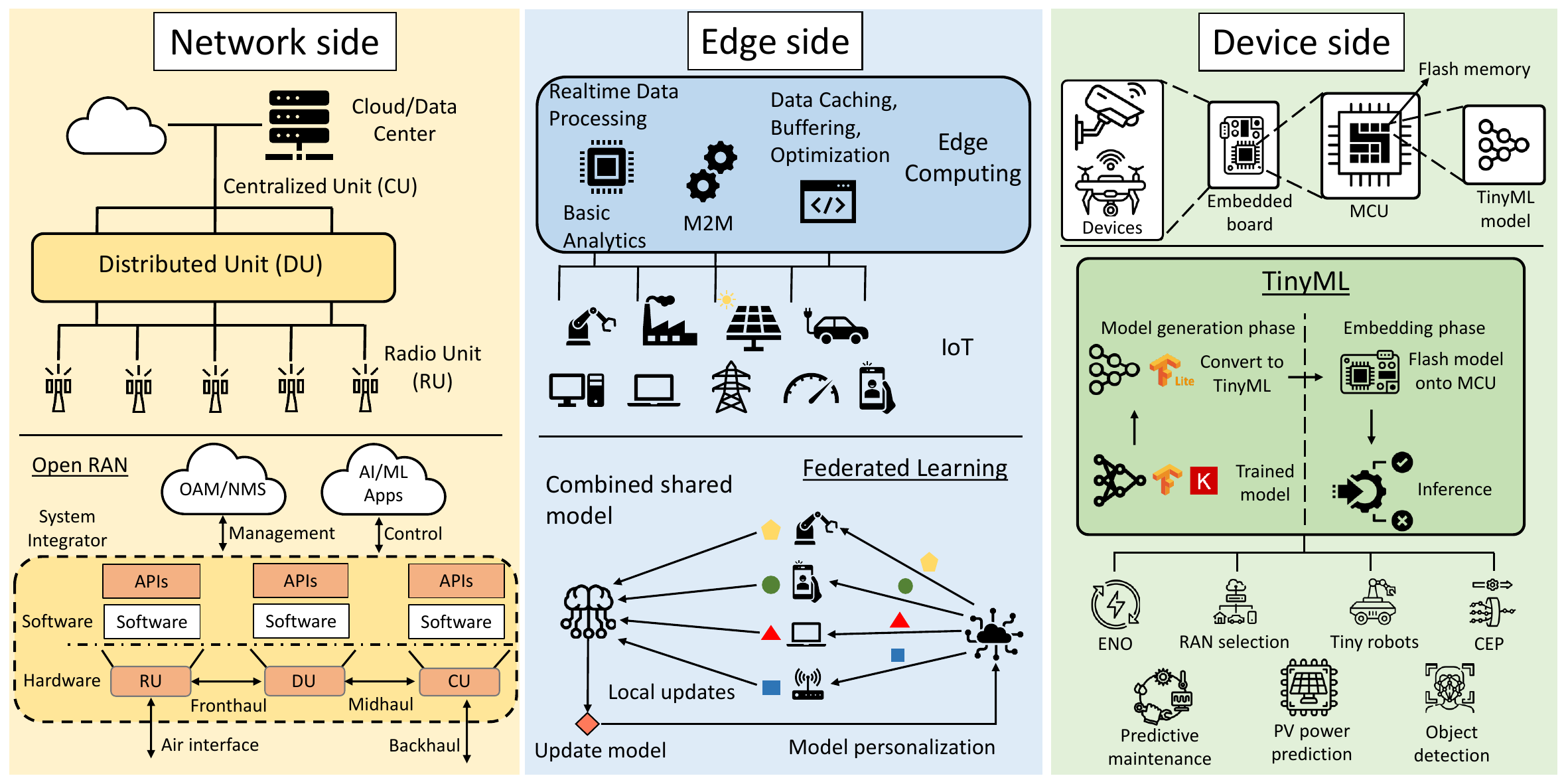}
        \caption{ML at the network, at the edge, and at the device side.}
        \label{EEML}
        \vspace{-6mm}
    \end{figure*}

   ML in wireless communications can be approached from three distinct perspectives: i) the network side, ii) the edge, and iii) the device side, as illustrated in Figure~\ref{EEML}. Note that network-based computing treats mobile devices as data collectors sending data to cloud servers for processing. The drawback of this scenario is the introduced overhead and potentially severe latency~\cite{zhang2019deep}. Also, ML algorithms can be complex, data-hungry, and computationally costly on the network side. Contrary, ML at the device side must be simpler (i.e., of lower power/cost), while a balance/trade-off between these two extremes is achievable by implementing ML at the edge. Indeed, ML at the edge takes advantage of local processing and data storage capabilities while communicating with the network and devices. Specifically, pre-trained models are offloaded from the cloud to individual devices, such that they can make inferences locally. This allows for more efficient and effective use of ML in wireless communications. However, it can only support tasks that require lightweight computations~\cite{zhang2019deep}. 
    
    Next, we discuss each of the aforementioned approaches, Indeed, our purpose here is to explore the application of ML algorithms in specific local aspects of IoT connectivity, such as computation task optimization and offloading decision-making, which provide EE enhancements.   
    \vspace{-3mm}
    \subsection{\uppercase{ML at the Network Side}}\label{networkML}
    
    Currently, RANs account for $73\%$ of the total EC in modern cellular networks~\cite{lopez2021survey}. Therefore, intelligent resource management is imperative to maximize EE, and thus minimize EC. Specifically, the current trend is to replace rule-based heuristics with optimal parameters configured through the knowledge acquired by data-driven approaches~\cite{lopez2021survey}. 
    In this regard, ML models play a crucial role in enabling intelligent networks to characterize their environment, predict system changes in real-time, and react accordingly~\cite{ML_IoT}.    

    \subsubsection{ML algorithms}

    Choosing the most appropriate ML algorithm is critical for solving EE problems on the network side. The algorithm choice should be based on the specific problem, network architecture, and available data. A comprehensive/holistic approach that considers all aspects of the network, including HW, SW, and communication protocols, is needed to design energy-efficient solutions~\cite{IoT_ML_Network,kokkonen2022autonomy,Guirola.2023}. Moreover, the specific EE problem, the architecture and the characteristics of the network, the available data, and the strengths and weaknesses of each ML algorithm must be considered~\cite{IoT_ML_Network,kokkonen2022autonomy,Guirola.2023}. For instance, decision trees are suitable for routing optimization, while support vector machines are effective for network anomaly detection and prediction. Genetic and clustering algorithms are well-suited for optimizing the placement of network and edge resources~\cite{lahderanta2021edge,ruha2020capacitated,IoT_ML_Network2}, while clustering algorithms are also appealing for grouping network devices, e.g., based on their EC patterns~\cite{IoT_ML_Network2}. 
                
    DL is one of the most popular ML algorithms at the network side~\cite{fourati2021survey}. Specifically, deep SL (DSL) and deep RL (DRL) are commonly used on the network side due to the availability of labeled data and the effectiveness and flexibility of the models.
    Both can be used to optimize EC by adjusting the network resources based on the network traffic load.     
    By combining feature extraction with prediction, DL models classify, predict, and make accurate decisions more effectively than traditional ML algorithms~\cite{fourati2021survey}. This is because DL architectures such as multi-layer perceptrons, convolutional neural networks (NN)~\cite{lecun2015deep}, and transformers~\cite{vaswani2017attention} can estimate complex mappings between input and labels in the training data~\cite{lecun2015deep}, all while efficiently utilizing HW-based accelerators such as graphical processing units (GPU) for training and inference~\cite{maimo2017performance}. Other benefits include the distribution of processing, avoiding redundant capacity in hotspots, and the efficient marshaling of big data, generated in-network or at user devices. 
    
    Notably, DL has its own drawbacks, e.g., i) DL requires large amounts of training data, whose curation and labeling may be costly and face privacy concerns; ii) DL algorithms are largely black boxes with low interpretability and explainability; and iii) DL may require dedicated ML accelerators for efficient operation~\cite{lopez2021survey,reuther2021ai}. All in all, the DL benefits may be outweighed by the costs in many cases, calling for careful and informed designs/implementations.
                
    \subsubsection{Use cases and applications}

    In 6G networks, ubiquitous intelligence is key for providing efficient and personalized services. However, this poses a challenge in terms of data management and EC~\cite{liu2020federated}. Indeed, EC can increase considerably unless energy-efficient approaches are used \cite{abrol2016power}. Thus, there is a pressing need for developing lightweight, flexible, and adaptive solutions that effectively tame the dynamics of the environment and minimize EC in practical IoT networks~\cite{lopez2021survey}. This has motivated the adoption of DRL for solving a variety of wireless communication problems \cite{guo2022energy}, making it a valuable tool for data-driven optimization in new-generation systems~\cite{lopez2021survey}. 

    ML forecasting network environment can help DRL algorithms converge faster to proper operational policies, resulting in faster adaptation to changing conditions and greater EE in network operations\cite{lopez2021survey}.  Indeed, they can trigger well-informed decisions to address integration issues with IoT devices. For instance, ML models can analyze historical data from IoT devices and predict their traffic patterns and energy requirements. This information can then be used to optimize resource allocation, dynamically adjust power management settings, and improve EE. In~\cite{ML_K1}, ML is used to estimate empirical path loss and shadowing. These estimations guide transmission power adjustments, leading to significant energy savings of up to 43\% compared to conventional protocols. Moreover, DRL is exploited in~\cite{ML_K2} to design an adaptive LoRa strategy for improved reliability in industrial applications, while~\cite{ML_K3} uses a combination of decentralized and centralized RL for allocating spreading factor and transmission power to the devices, showing significant improvement in both network level good-put and EC, especially for large and highly congested networks. Additionally, ML models can help identify IoT behavior patterns, such as peak usage times and idle periods of devices, enabling the implementation of energy-saving strategies like sleep modes and adaptive transmission power control. Ultimately, ML-driven insights can contribute to achieving a coordinated balance between network performance and EE in IoT environments.
                
    Notably, novel DRL schemes have been proposed to manage advanced sleep modes in BSs. Specifically, in \cite{salem2019traffic}, the sleeping level length is set by the BS in a sequential manner. When the cell becomes idle, the BS departs from the deepest level of sleep and gradually switches to higher activity levels. At each stage, the BS decides the number of slots for the current sleep mode. Most relevant components must thus only be active (consuming energy) when handling actual data~\cite{lopez2021survey}. In this scheme, the BS is the agent, and the network is the environment, including the traffic load, available energy, and the state of other network devices. The BS takes actions (i.e., sets the sleeping level length) based on its observations of the environment (i.e., traffic load and energy budget), and the RL algorithm provides feedback (i.e., rewards or penalties) based on the EC and network performance. The RL algorithm aims to maximize the reward by finding the optimal sleeping level and duration for the BS while maintaining the required network performance. 

    On the other hand, demand forecasting, i.e., predicting how much time and resources will be spent on applications, is a key problem in data center management.  Notice that good forecasting techniques can lead to minimizing EC by scheduling jobs efficiently~\cite{demirci2015survey}. Network-level data usually exhibit significant spatiotemporal variations, which can be utilized for network diagnosis and management, UE mobility analysis, and public transportation planning~\cite{zhang2019deep}. In this context, DL has the potential to improve EE in a variety of settings by improving demand forecasting, optimizing resource allocation, and identifying patterns and trends in data that can be used to reduce EC. 
                
    Establishing a data collection path model is another solution for minimizing EC. Specifically, data collection can avoid visiting needless nodes and collecting unreliable/correlated data, resulting in outperforming traditional data collection methods in both energy and delay. In that direction, proactive caching can also contribute to EC minimization using forecasted lookup patterns to jointly optimize computation offloading policies and caching decisions~\cite{shafik2020network}. These approaches may minimize radio interface usage, which is a significant EC source in IoT devices.
    \subsubsection{Open-RAN (O-RAN) and virtualization as key technological enablers}

    Open interfaces are appealing for the operators to swiftly introduce novel services and tailor the network to their own requirements. Motivated by this, RAN is evolving towards the O-RAN concept, which focuses on openness and ntelligence~\cite{theodoropoulos2022cloud}. O-RAN brings new business opportunities and encourages local 5G innovations. O-RAN aims at decoupling the RAN components from their underlying SW and HW components, enabling operators to cover more users in a cost-effective, secure, and energy-efficient manner~\cite{update_nodate}.

    In O-RAN, the RAN is disaggregated into three main building blocks: i) radio unit (RU), ii) distributed unit (DU), and iii) centralized unit (CU)~\cite{theodoropoulos2022cloud} as shown in Figure~\ref{EEML}. Also, the O-RAN ALLIANCE has defined different interfaces within the RAN including those for i) fronthaul between RU and DU and ii) midhaul between DU and CU. Another feature of O-RAN is the RAN intelligent controller, which adds programmability to the RAN and the ability to introduce new services and features~\cite{update_nodate, theodoropoulos2022cloud,Karbalaee.2023}. 

    The open interfaces/protocols of O-RAN enable a seamless integration of ML models to optimize radio resource management, interference mitigation, and network planning. In addition, O-RAN supports real-time data processing and analysis, which is essential for ML algorithms that require fast and accurate decision-making. This enables intelligent and adaptive management of the RAN, resulting in better network performance, improved UE experience, and lower operating costs. These all may be crucial to extend the battery lifetime (BL) of IoT devices~\cite{10008676,Karbalaee.2023}, and support energy-sustainable IoT ecosystems in general.

      Notice that network functions virtualization (NFV) enables network functions to be implemented as SW applications running on virtualized infrastructure. This improves EE in several ways, such as reducing the required number of physical network devices, dynamically scaling network functions based on demand to avoid over-provisioning, and optimizing energy usage across the network through centralized management and orchestration~\cite{NFV2}. Furthermore, NFV allows for the implementation of EE features like sleep modes and power management in network devices, leading to even further EC reduction while maintaining network performance. In general, O-RAN and NFV are two complementary technologies that can work together to improve the EE and flexibility of wireless networks.

       O-RAN initiative is still in its infancy, with a lot of work in progress. Hence, it is important that future research activities specifically focus on practical and real-world trials with respect to the virtualized RAN concepts~\cite{gavrilovska2020cloud}, which heavily rely on self-organization and other ML-based approaches. 
         \vspace{-3mm}
    \subsection{\uppercase{ML at the Edge}}\label{edgeML}
        
    ML at the edge (popularly known as edge ML~\cite{loven2019edgeai} or edge intelligence~\cite{xu2021edge}) refers to the training and use of ML models across the computing continuum: on UEs, edge nodes, and cloud servers, rather than only in cloud-based centralized setups. his is a rapidly evolving field with prominent innovation opportunities and impact~\cite{park2019wireless,peltonen2022many}. 
    
    This topic is often studied from two viewpoints. First, \textit{ML on edge} refers to adapting ML methods for the distributed edge environment, while \textit{ML for edge} is the use of ML methods for the benefit of the edge environment~\cite{loven2019edgeai, deng2020edge}. As an example of ML on edge, distributed learning and inference allow the efficient distribution of ML computations across the computing continuum. This allows optimizing the overall processing time, resulting in lower EC of in/near-sensor devices \cite{ML_IoT}. Moreover, ML on edge can reduce latency, allow localized filtering of unwanted data, and increase system uptime as data is locally processed. On the other hand, edge ML can introduce various benefits, including the application of a predictive approach in troubleshooting. With edge ML, real-time data can be analyzed by ML models to proactively identify potential issues or anomalies before they escalate into critical problems. This can provide EE gains and minimize the downtime of IoT systems. Moreover, operational efficiency and EE can be promoted in IoT environments by reducing unnecessary offloads and tasks and optimizing resource utilization~\cite{loven2021weathering,loven2022dark}.
    
    Shortcomings that need to be addressed include: i) complexity due to coordination issues related to IoT constraints like processing power, memory, and delay in real-time applications; ii) heterogeneity, opportunism, and geographical distribution of computing resources; iii) fluctuating or intermittent connectivity; iv) security and privacy; v) standardization-related concerns like interoperability of IoT with ML integration; and vi) accuracy and latency issues in real-time applications \cite{peltonen20206g, xu2021edge}.

    \subsubsection{ML algorithms}\label{MLalgs}
                
    ML can be used for multiple purposes in the context of edge computing. For example, edge computing can filter data, with only relevant data getting transmitted between the user devices, the edge nodes, and the cloud. This results in substantial savings in terms of bandwidth and cost of data transmission~\cite{Albreem.2021}. Moreover, advanced ML techniques can be utilized, for example, to optimize computation tasks, make offloading decisions on a wireless device, and identify the best scheduling solutions for working and sleeping time, thus lowering EC and enhancing EE \cite{liu2020federated}. 
                
    There are several ML approaches that can be applied at the edge depending on the application requirements and constraints. These include distributed learning, referring to a family of methods such as federated learning (FL) for distributing the learning and training data/process across various nodes~\cite{kairouz2021advances}, and transfer learning, a method for transferring knowledge between ML models in different domains~\cite{Albreem.2021}. For example, FL and other variants of distributed learning have emerged as possible solutions for solving complex operational decisions at the edge side~\cite{liu2020federated} (Figure~\ref{EEML}). Although undoubtedly beneficial, FL faces some challenges~\cite{liu2020federated}: 
    \begin{itemize}
      \item \textbf{Expensive communication and synchronization.} FL communication overhead may be limited by reducing the number of i) communication rounds and ii) gradients in each communication round.
      \item \textbf{Security/privacy/robustness issues.} FL must: i) provide protection against malicious attacks, ii) support privacy, iii) tolerate heterogeneous HW, and iv) support robust aggregation algorithms.                            
      \item \textbf{Model size (MDS).} A large FL MDS might be unsuitable for an IoT device, especially given stringent real-time constraints. Thus, efficient training and inference are necessary for massive and heterogeneous networks. 
    \end{itemize}
    \subsubsection{Use cases and applications}
    IoT places significant demands on three main areas: transmission, storage, and computation. Indeed, IoT devices generate a large volume of data and have limited storage and computational capabilities, thus edge-based storage and computing are appealing. Such an edge integration effectively accelerates data uploads/computation, reduces response and device active time, and improves EE~\cite{yu2017survey}.

    Edge computing is a rapidly evolving field, and the trend is towards moving cloud functions to the network edges~\cite{maray2022computation}. Edge computing benefits the advancement and implementation of 5G and beyond networks with applications such as augmented/virtual reality~\cite{qiao2018new}, low-complexity IoT~\cite{yu2017survey}, Internet of vehicles~\cite{zhu2020multiagent}, and video stream analysis~\cite{feng2022computation}. Delay-sensitive augmented/virtual reality applications can be migrated to edge servers to guarantee a high-quality user experience with a timely response. For low-complexity IoT tasks, edge servers can reduce the HW complexity and increase device lifespan by processing tasks that were originally done locally. Additionally, edge cloud networks close to vehicles can improve transportation safety, reduce traffic congestion, and provide value-added services. By utilizing the capabilities of MEC networks, video playback can be expedited, and user experience can be enhanced. Moreover, smart meters can be used to detect electricity consumption and gather data into a central controller to facilitate real-time power control, leading to improved systematic EE.
    
    The spatial and temporal correlation of IoT traffic data is exploited in dual prediction and data compression techniques. Specifically, dual prediction techniques use the correlation between the current and previous data to predict future data values. On the other hand, data compression techniques use the correlation between the different sensor nodes to compress the data before transmission. Both techniques reduce the amount of data that needs to be transmitted and hence the EC and bandwidth requirements~\cite{mao2017survey}.
    
    At the edge, communication-efficient FL and efficient training aims at reducing the required communication overhead of the training process while maintaining or improving the model's performance. Federated parallelization is an efficient training FL technique for parallelizing the training process across multiple devices and thus accelerating the training. Efficient training can also be conducted via distillation which consists of training a small model that mimics the behavior of the original model; therefore, significantly reducing the amount of required training data~\cite{liu2020federated}.
    \subsubsection{MEC - A key technological enabler}\label{MEC}

    MEC, or multi-access edge computing, is an architecture for mobile networks. MEC was introduced to address the latency issue during mobile cloud computing offloading, pushing computing and storage resources to the edge with the aim of bringing those resources as well as applications and services near the end-users. MEC is characterized by two key features, namely, low latency and high workload capacity, stemming from proximity to users and their devices~\cite{gasmi2022survey}. However, MEC has unique design considerations, such as complex wireless environments and MEC servers' inherently limited computational capabilities~\cite{mao2017survey}.    
                
\begin{figure}[t!]
      \centering
      \includegraphics[width=0.95\linewidth]{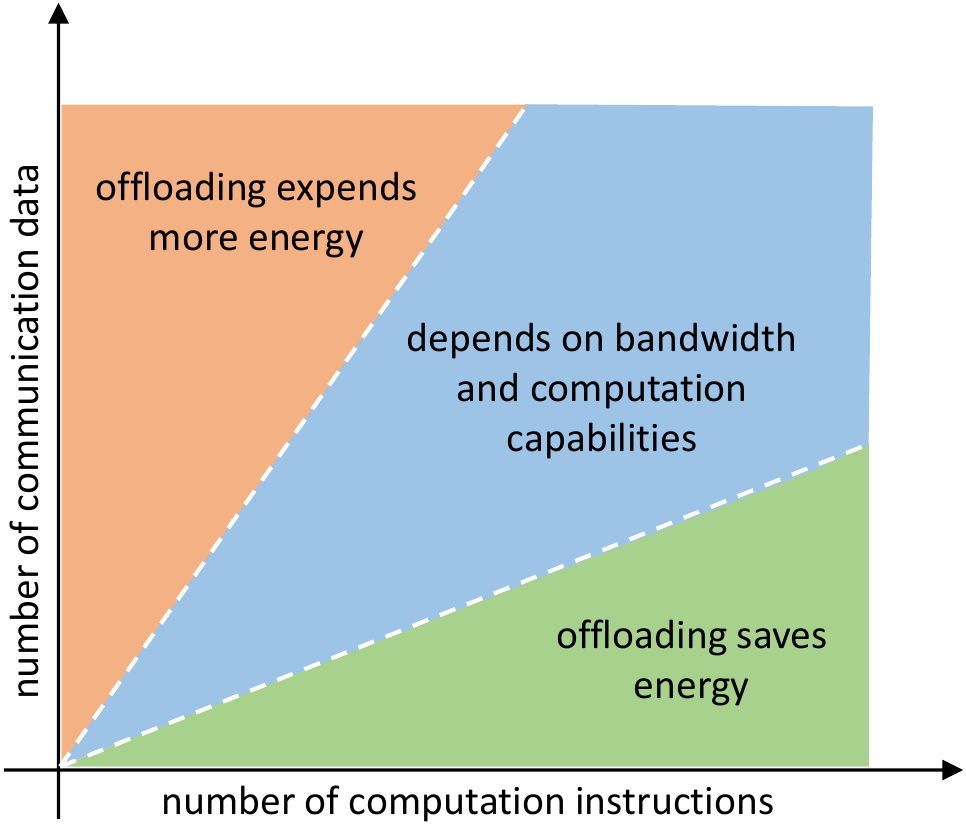}
      \caption{EE when using computation offloading \cite{MEC2}.}
      \label{mec_offloading}
       \vspace{-5mm}
    \end{figure}

    Computation offloading is a significant MEC feature as it may prolong the lifespan of IoT devices by delegating computation tasks to edge devices as long as the communication overhead remains reasonable~\cite{IoT_ML_Network2}. Specifically, the amount of energy that a mobile node can save by offloading an application depends on the number of computation instructions and communication data. If the computation instructions are much larger than the communication data, it is more energy-efficient to offload the computation-intensive application to the server. However, if communication is expensive, it is better to carry out the application at the mobile node itself. The condition also depends on the bandwidth available for communication, where a large bandwidth can save communication time and improve EE~\cite{IoT_ML_Network2}. Figure~\ref{mec_offloading} illustrates when offloading can save energy~\cite{MEC2}.

    There are various energy-efficient offloading methods: i) computation-based methods, which involve partitioning the offloading application program and offloading computation-intensive parts; ii) communication-based methods, which involve reducing the amount of communication by aggregating or compressing data~\cite{MEC2}; and iii) hybrid/joint optimizations methods, which involve collaboratively executing tasks on the mobile node and the cloud to minimize EC while ensuring the total execution of tasks~\cite{mathur2021machine}. The study in~\cite{mathur2021machine} demonstrated that computation offloading can reduce EC and increase battery life up to 50\% for practical applications.

   Notice that establishing a MEC infrastructure, especially server locations, is the first step toward constructing a MEC system. For this, the system planners and administrators must consider the deployment cost, demand for computation, and availability of renewable energy sources for EP in the case of green MEC~\cite{lahderanta2021edge,ruha2020capacitated, gasmi2022survey,sun2017green}. Indeed, green MEC is an emerging technology that combines the benefits of MEC with energy-efficient computing and EP from renewable sources to create a more sustainable and eco-friendly approach to mobile computing~\cite{sun2017green}. The need for green MEC arises from the growing awareness of the environmental impact of mobile computing and the increasing demand for sustainable solutions. As mobile networks and devices become more ubiquitous and essential to our daily lives, their EC and carbon footprint also increase. Green MEC may reduce this impact by optimizing the use of resources and energy-efficient computing techniques. One method for designing green MEC systems is dynamic right-sizing, where the EC of a MEC server is dependent on its utilization, and a server should consume energy proportional to its workload~\cite{barroso2007case}.\footnote{In a typical MEC system, the EC of a server is often fixed, regardless of its utilization or workload. This means that even if the server is underutilized, it still consumes a significant amount of energy, leading to wastage and higher energy costs. Refer to the EPC metric in Section~\ref{KPI}-\ref{others}-\ref{other1}.} To achieve energy-efficient servers, the processing speeds of lightly-loaded edge servers can be reduced. Another method is geographical load balancing (GLB)~\cite{sun2017green}, which utilizes spatial variations in workload patterns, temperatures, and electricity costs to direct the flow of workload between different data centers. These methods can help to reduce the EC of MEC systems and provide a carbon-neutral energy supply, leading to more sustainable and environmentally friendly network operations~\cite{sun2017green}.
    \vspace{-2mm}
    \subsection{\uppercase{ML at the Device: TinyML}}\label{tinyMLsec}
 
	TinyML focuses on computationally efficient and HW-constrained ML algorithms for low-power IoT devices \cite{Kallimani.2023}. TinyML-equipped IoT devices can energy-efficiently execute tasks and make autonomous decisions without continuously relying on cloud/network services. This reduces network traffic and latency in decision-making and increases privacy. Due to this and its on-device data processing capability, TinyML may assist with intermittent connectivity/computing.  
	    
    \subsubsection{Use cases and applications}
    In the following, several use cases and applications motivating the deployment of ML on the device side are discussed.

     Energy Management is a promising solution to enable i) energy-neutral operation (ENO) in always-on IoT devices utilizing EH \cite{basaklar2022tinyman} and ii) efficient allocation of PV energy from a single-panel or off-grid system to multiple tasks \cite{9954033}. In this regard, TinyML models can forecast future EH values to help devise a proactive ENO strategy for IoT devices. Besides, when the number of EH samples becomes insufficient to elaborate a forecast on the incoming ambient energy, energy management strategies can still benefit from TinyML by using the current battery state and previous EH measurements \cite{basaklar2022tinyman}. Again, PV power prediction is necessary for proper energy management/distribution among tasks. The study in \cite{9954033} employs TinyML to perform PV power prediction and suggests that it can also serve as an indicator for measuring the effects of aging on the power-generating capacity of solar panels.  
      
     In addition, TinyML may jointly ensure EE and the QoS of the data transmitted by an IoT device with multiple radio access technologies (RATs) by enabling a proper RAT selection. This can be implemented by considering situational characteristics, such as the available energy on the device and the size/urgency level of data to be transmitted \cite{9166461}. Also, TinyML may support complex event processing (CEP), which aims to identify complex event patterns in real-time data streams from multiple sources using predefined logic rules.\footnote{For instance, CEP can be used in a manufacturing plant to detect a sequence of events that occur before a machine breaks down by monitoring, e.g., changes in pressure, temperature, and vibration, and thus triggering an alert for maintenance.} Note that large IoT deployments pose a challenge for CEP in performing sequence matching over raw data due to unexpected events or outliers not covered by predefined rules, motivating the use of ML-based CEP. Nowadays, the data privacy and latency issues faced by centralized IoT are empowering the idea of performing ML-based on-device CEP in IoT. Interestingly, the authors in \cite{ren2021synergy} designed a framework that puts together TinyML and CEP for machine safety monitoring in a distributed IoT network. 

 \begin{figure*}[t!]
        \centering
        \includegraphics[width=0.95\linewidth]{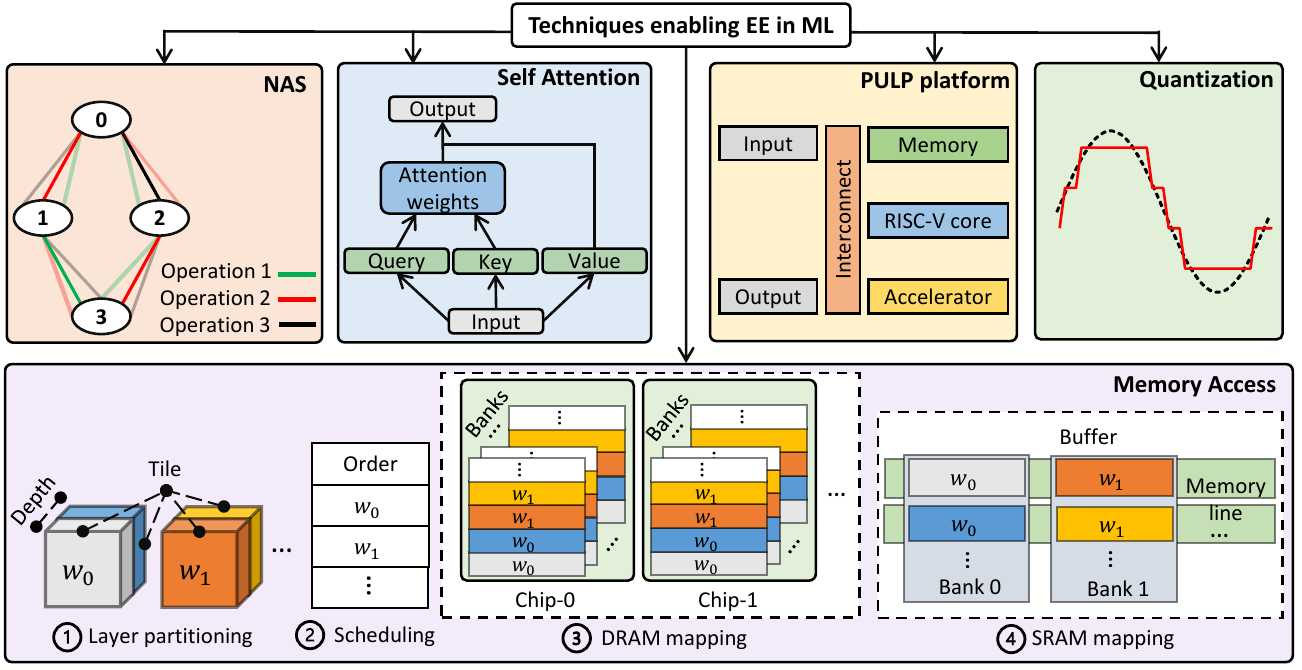}
        \caption{Techniques enabling EE in ML \cite{putra2019romanet, videoselfattention, PULPplatform}.}
        \label{tech_tinyml}
        \vspace{-5mm}
    \end{figure*}
             
    Object detection running in EH IoT devices constitutes another application opportunity for TinyML. Indeed, data privacy and latency requirements in various object detection applications, such as intelligent video surveillance and number-plate recognition, are encouraging the idea of on-device processing in IoT, thus calling for efficient TinyML implementations.  Due to the dynamic nature of the EH environment, on-device processing must be aware of the available energy on the device, and also of the time, energy that must be collected, and the deadline to complete a task. Interestingly, TinyML is used in \cite{Sabovic.2023} to detect a person over a battery-less IoT device. This study also suggests the use of TinyML for applications where battery-less devices require a long lifetime and are difficult to reach.           
	            
   TinyML allows performing predictive maintenance procedures, e.g., for detecting and preemptively solving the impending failures that a system might face, over an MCU-based sensor device rather than on the cloud. For instance, authors in \cite{10.1007/978-3-031-08337-2_6} implemented TinyML to perform anomalous behavior detection tasks over the sound recordings of the ToyADMOS dataset \cite{koizumi2019toyadmos}, where spectrograms (images generated from audio) are used as input. It is shown that optimizing the audio sampling rate used to form a spectrogram can lead to a decrement in the spectrogram dimensions, which further leads to a reduced INFT, required memory, and EC. In addition, \cite{bn2022prema} introduces TinyML for detecting various kinds of faults in a solenoid valve, which is an electro-mechanical actuator. Here, the transient response of the drive current of the solenoid valve is used as input to the TinyML model. This encapsulates information about the solenoid valve's electro-mechanical action, from which predictions about the present working condition can be made.   
	                          
   Finally, TinyML is appealing for realizing tiny robots. Note that legged robots use imitation learning \cite{peng2020learning} and RL to learn their walking gaits. However, such procedures are not functional for tiny robots, which are low-cost resource-constrained robots that can potentially be used in search $\&$ rescue operations, military reconnaissance, space robotics, and routine equipment monitoring \cite{neuman2022tiny}. To address this, authors in \cite{jabbour2022closing} proposed exploiting TinyML techniques, such as graph freezing and float$16$ quantization, to shrink the trained walking gaits of the NN MDS by $8 \times$. Graph freezing made all NN variables constant and float$16$ quantization changed the floating point weights from 32-bit to 16-bit.   
   %
 \subsubsection{Enabling techniques}
	Next, we discuss the techniques to support energy-efficient ML models. They are illustrated in Figure~\ref{tech_tinyml}, while representative state-of-the-art works using them are listed in Table~\ref{table_2} together with key features.
	\begin{itemize}
	  \item \textbf{Neural architecture search (NAS)} aims to find the best NN architecture that can fit in the available MCU resources. For instance, the authors in\cite{lin2020mcunet} proposed MCUNet, a framework to optimally handle/utilize the MCU resources. MCUNet consists of TinyNAS, a low-complexity NAS, and TinyEngine, an inference library that reduces the required runtime memory. The latter utilizes a memory scheduling method to reduce the SRAM required by an ML model during its inference, thus reducing the EC per inference. Meanwhile, a drop in the required SRAM frees up the MCU memory, thus allowing TinyNAS to select a larger ML model to achieve higher accuracy. As shown in Table \ref{table_2}, MCUNet achieves higher accuracy compared to MobileNetV2 \cite{sandler2018mobilenetv2} and ResNet-18 \cite{he2016deep}, while utilizing relatively less SRAM and Flash. All in all, MCUNet may provide energy-cum-memory efficient NN architectures that can be implemented on an MCU.  
	                    
	 \item \textbf{Parallel ultra-low power (PULP)} is an IoT processor architecture providing SW-level acceleration for TinyML. PULP adopts both data and thread-level parallelism to deliver a steady performance irrespective of the operating voltages and mW-level computations/inference. As shown in Table~\ref{table_2}, PULP uses 3.5$\times$ less energy than ARM Cortex-M4 processor and also exhibits a significantly lower INFT. For instance, the authors in \cite{tabanelli2021dnn} adopted PULP-based MCU to facilitate the parallel run in non-neural ML algorithms, while a fast NN (FANN)-on-MCU framework supporting both ARM Cortex-M series MCUs and PULP-based MCUs is used in \cite{wang2020fann}. The latter framework can be used with both fixed-point and floating-point NN models.   

    \item \textbf{Model compression} allows the creation of models that are compact enough to fit on limited-capability devices and efficient enough to run with limited power. Specifically, model compression is a collection of techniques aimed at reducing the ML MDS, thereby enhancing its efficiency in terms of memory usage and computational requirements~\cite{cheng2017survey}. Several model compression techniques are commonly employed in TinyML: i) quantization, which reduces the precision of the numbers used in the model, significantly decreasing MDS and computational requirements~\cite{jacob2018quantization}. For example, coarse quantization drastically reduces the precision of an NN model's parameters to less than $8$ bits, making the model suitable for an MCU. Moreover, the quantized NN model enjoys less EC and faster computation than its full-precision counterpart. However, these advantages come at the cost of lower NN accuracy~\cite{lu2022improving}; ii) pruning, which eliminates parts of the NN that contribute little to the output, such as weights that are close to zero~\cite{han2015learning}; iii) knowledge distillation, which trains a smaller model (the student) to mimic the behavior of a larger model (the teacher). The smaller model is then used in place of the larger model, resulting in significant computational savings~\cite{hinton2015distilling}; and iv) weight sharing, which involves using the same weights for multiple neurons, reducing the total number of weights that need to be stored~\cite{chen2015compressing}. By applying these and other techniques, it is possible to create small, efficient models that can run on low-power devices, thereby expanding the ML applications.

    \begin{table*}[h!]
        \caption{Representative state-of-the-art works on energy-efficient TinyML (2020-2022)}
        \vspace{-1mm}
        \centering
        \label{table_2}
        \begin{tabular}{ p{0.68cm} p{3cm} p{2cm} p{10cm} }
          \toprule
          \textbf{Ref.} & \textbf{Application} & \textbf{EE strategy} & \textbf{Additional information}\\
          \midrule
          \cite{basaklar2022tinyman} & ENO & tinyMAN & EC = $27.75$ $\mu$J per inference, memory footprint (SRAM + Flash) = $91$ KB\\  
          \hhline{~---}
          \cite{9223156} & Face recognition & Quantization & MNIST dataset. 
          Chainer ML framework (prediction latency $= 0.005$ ms, training time $= 3.54$ s) and Keras/Tensorflow (prediction latency $= 5.3$ ms, training time $= 243$ s)\\   
          \hhline{~---}
          \!\cite{putra2019romanet, shafique2021towards} & Smart health-care, image classification, object detection & ROMANet & Compared to the baseline scheme SmartShuttle \cite{8342033}, the number of DRAM accesses is $12 \%$, $36 \%$ and $45 \%$ smaller for the AlexNet, VGG-16, and MobileNet, respectively. Moreover, the DRAM access energy is $12 \%$, $36 \%$, and $46 \%$ smaller,  while the numbers are $12 \%$, $34 \%$, and $45 \%$ smaller in terms of  DRAM operations\\    
          \hhline{~---}
          \cite{lin2020mcunet} & Image classification, object detection, keyword spotting in an audio & MCUNet, linear quantization & ImageNet dataset. 
          MCUNet ($4$-bit Quantization): SRAM $= 0.49$ MB, Flash usage $= 1.9$ MB, accuracy $= 70.7 \%$. 
          MobileNetV2 ($8$-bit Quantization): SRAM $= 1.7$ MB, Flash usage $= 2.5$ MB, accuracy $= 69.8 \%$, 
          ResNet-18 ($8$-bit Quantization): SRAM $= 0.9$ MB, Flash usage $= 11.2$ MB, accuracy $= 69.8 \%$ \\     
          \hhline{~---}
          \cite{wang2020fann} & Hand gesture recognition, fall detection, human activity classification
          & PULP & Applied to a NN with $103800$ multiply-add operations. 
          FANN-on-MCU with ARM Cortex-M4 processor: EC $= 183$ $\mu$J per inference, INFT = $17.6$ ms. FANN-on-MCU with PULP processor: EC $= 50$ $\mu$J per inference, INFT $\le 1$ ms \\  
          \hhline{~---}
          \cite{wong2020attendnets} & Object recognition & AttendNets &  ImageNet$_{50}$ dataset. 
          MobileNetV2: accuracy $= 68.7 \%$, number of parameters $= 2290$ K, number of multiply-add operations $= 299.7$ M.                AttendNets: accuracy $= 71.7 \%$, number of parameters $= 782$ K, number of multiply-add operations $= 191.3$ M \\     
          \hhline{~---}
          \cite{bn2022prema} & Fault detection & PreMa & PC = $140$ mW per inference, INFT = $76.2$ ms\\  
          \bottomrule
      \end{tabular}
      \vspace{-4mm}
  \end{table*}  
    
	\item \textbf{Self-attention} allows each input in a sequence to weigh the importance of all other inputs, thereby enabling each element to ``attend to'' or ``focus on'' all other elements in the sequence~\cite{vaswani2017attention}. This enables high levels of parallelization and increased efficiency, and it is a significant departure from traditional recurrent NNs or long short-term memory networks, which typically focus on preceding words in a sequence. Note that the self-attention mechanism is a core component of the Transformer architecture~\cite{vaswani2017attention}, which is revolutionizing natural language processing. In the context of TinyML, the benefits of self-attention and Transformer architectures can be manifold. Firstly, the ability to process sequence data in a non-temporal manner allows for more efficient computation, which is crucial for low-power edge devices~\cite{zhang2019fixup}. Secondly, the parallelizability of Transformer architectures makes them well-suited to the resource-constrained environments typical of TinyML applications. By processing all elements of a sequence simultaneously rather than sequentially, Transformers can deliver faster INFT, which is crucial for real-time applications on edge devices~\cite{howard2019searching}. However, the original Transformer models are often too large and computationally intensive for TinyML applications. Therefore, further research on model compression techniques and efficient Transformer variants, such as TinyBERT~\cite{turc2019well} and DistilBERT~\cite{sanh2019distilbert} is needed. For example, AttendNets, a TinyML model, combines self-attention with a machine-driven design exploration, resulting in a compact deep NN with low-precision parameters. As shown in Table~\ref{table_2}, AttendNets achieves better performance in terms of accuracy, EC, and memory consumption than MobileNet-V2 for pictorial recognition with ImageNet$_{50}$ dataset \cite{fang2018nestdnn}. 
    
	\item \textbf{Memory access techniques} aim to reduce the average DRAM access energy in an inference. Note that HW accelerators, consisting of DRAM (off-chip part), and SRAM and compute engine (on-chip parts), are required to implement complex NNs, such as CNN, on a device. The portion of the NN layer that an accelerator can process at a single time instance depends on the data storage capacity of SRAM. Also, the same input data could be used in multiple NN operations, which leads to multiple DRAM accesses for the same data. Notably, the energy required for DRAM access is relatively higher than that for a NN operation, which means that a significant amount of energy can be saved by exploiting memory access techniques. Indeed, authors in \cite{shafique2021towards} recommend using ROMANet \cite{putra2019romanet}, a memory access technique to cut down the average energy-per-DRAM-access, on-chip buffer access energy, and the number of DRAM accesses. Specifically, ROMANet partitions an NN layer and then schedules the processing of the portions to minimize the number of DRAM accesses. Based on the data storage capacity of SRAM, data is partitioned into various blocks. Then, ROMANet maps them to the available DRAM and SRAM resources to minimize the row buffer conflicts and maximize the bank-level parallelism, respectively. The advantages of using ROMANet with various NN architectures, such as AlexNet \cite{10.1145/3065386}, VGG-16 \cite{simonyan2014very}, and MobileNet \cite{howard2017mobilenets}, are clearly visible in Table \ref{table_2}.     
  \end{itemize}   
	    
  The reader interested in TinyML can refer to the python package hls4ml \cite{fahim2021hls4ml}, which allows designing ML algorithms for low-power FPGA or ASIC devices. Through hls4ml, it is possible to optimize the HW implementation of TinyML models by leveraging the parallelism and pipelining capabilities of FPGA. This can lead to more efficient use of HW resources and reduced EC.   
  %
  \subsubsection{Future Directions}\label{futureD}
        
  Bio-inspired optimization \cite{gupta2023bio} may brush aside unnecessary ML computations, providing potential benefits for TinyML. Furthermore, alternate NN models, such as spiking NN \cite{Donati2019DiscriminationOE} and analog NN \cite{7551379}, should be investigated for TinyML design. Also, new computing models for TinyML systems are needed, while the existing computing models, such as in-memory computing, require further research in the context of TinyML systems. Note that MobileNet is a baseline deep NN model for edge computing, however, no baseline TinyML model is currently available for the end devices. Indeed, a baseline TinyML model serving as a reference point for future TinyML models is appealing. Finally, a ferroelectric RAM \cite{FRAM66} based MCU is a viable option for tinyML applications where some data must be retained on the device along with the tinyML code. Ferroelectric RAM offers non-volatile storage and allows flexible memory partitioning for tinyML code and data storage. Also, it allows writing a single data word to memory in $\sim 100$ ns, whereas flash memory demands several ms to perform the same operation. However, these non-volatile RAMs are currently bulky and have a significantly limited number of write cycles, so further research and technological advances are still needed to make them really appealing.
\vspace{-2mm}
\section{\uppercase{Energy-related Performance Metrics}} \label{KPI}
The potential of a technique/technology, especially when compared to a competitor, can only be assessed via relevant metrics quantifying performance. Herein, we discuss several performance metrics related to energy that serve this purpose for the discussed IoT techniques and technologies at a component/device and/or system/network level. The key features of the discussed performance metrics in terms of related energy processes, application level (at the component/device or system/network level), and relevance/applications are summarized in Table~\ref{MetricsT}. Meanwhile, Table~\ref{KPItable} lists relevant performance targets (i.e., KPIs) for the current and next generation of wireless systems. The numeric values are indicative and were extracted from the vast literature consulted for this work, together with data extrapolation and trend analysis in some cases.

\begin{table*}[t!]
            \caption{Performance metrics and related key features}
            \vspace{-1mm}  
            \begin{tabular}{C{0.3cm}L{1.7cm}C{0.385cm}C{0.387cm}C{0.387cm}C{0.76cm}C{0.6cm}L{9.7cm}}
                \toprule
                 & \textbf{Metric} & \multicolumn{3}{c}{\textbf{Energy Process}} & \multicolumn{2}{c}{\textbf{Level}}  & \textbf{Relevance/Applications}  \\  
                 & & EP & ET & EE & Comp. & Syst. &  \\
                \midrule
                \multirow{9}{*}{\rotatebox[origin=c]{90}{Conversion \& Transfer}} & CEE  &  &  & X & X & & EE assessment of power supplies/amplifiers, motors, filters, phase shifters \\ \hhline{~-------}  
                & sensitivity$^c$  & X & X &  & X & & KPI of EH circuits specifying minimum input power for operation  \\ \hhline{~-------} 
                & saturation  & X & X &  & X & & KPI of EH circuits specifying maximum input power for operation \\ \hhline{~-------}  
                & max CE  & X & X & X & X & & KPI of EH circuits specifying the maximum attainable CE \\ \hhline{~-------}  
                & APD  & X & X & X & X &  & KPI of EH transducers useful for determining miniaturization levels \\ \hhline{~-------}   
                & PTE & & X & X & & X & EE assessment of a WET technology/system \\ \hhline{~-------}   
                & AHE & X & X & & & X &  Assessment of the EH capabilities of device(s) in a network/deployment  \\  \hhline{~-------}   
                & EOP  & X & X  & & & X &  Operational availability of battery-less EH device(s) in a network/deployment \\  \hhline{~-------}   
                & MDHE & X &  X & & & X & Fine-grained theoretical performance analysis of battery-less EH networks  \\   
                \midrule
                \multirow{12}{*}{\rotatebox[origin=c]{90}{Storage \& Consumption}} & BCAP & X & & & X & & Assessment of the devices' lifetime in a network/deployment \\ \hhline{~-------}
                 & ED & X &  &  & X & & Form-factor assessment for given devices' energy requirements  \\ \hhline{~-------} 
                & lifetime & X & X & X & X & X & Operational assessment and deployment/maintenance planning     \\ \hhline{~-------}   
                &  EC & & & X & X & X & Determination of the energy demands. Critical for energy-limited systems \\ \hhline{~-------} 
                & FLOPS & & & X & X & X & EC KPI for digital signal processors   \\ \hhline{~-------}
                & MIPS & & & X & X & X & EC KPI for computer systems  \\ \hhline{~-------} 
                & NHE & X & X & X & X &  & Assessment of energy availability after losses from EH-related protocols   \\ \hhline{~-------} 
                & RPS & & & X & X & X & Assessment/comparison of techniques in terms of triggered EC reduction  \\ \hhline{~-------} 
                & PKD & & & X & X & X & Assessment of maximum energy demands. Critical for energy-limited systems  \\ \hhline{~-------} 
                & GECS & X &  & X & X & X & Assessment of the penetration of green energy  \\  \hhline{~-------} 
                & LCOE & X &  & X &  & X & Cost assessment for energy generation. Critical for feasibility studies  \\  \hhline{~-------} 
                & GRS & X & X &  &  & X &  Reliability/stability assessment of energy generation/transfer systems \\  
                \midrule
                \multirow{7}{*}{\rotatebox[origin=c]{90}{EER}} & -data transfer  &  & & X &  & X & Typical KPI for wireless communication networks with/without QoS constraints  \\  \hhline{~-------} 
                & -RG,CA,CV &  &  & X &  X &  X & Energy-efficient coverage assessment of a wireless communication technology   \\ \hhline{~-------}   
                & -bandwidth & &  & X & X & & EE assessment of the utilization of the bandwidth by a technology/device   \\ \hhline{~-------}    
                & -AroI (-BoI) & &  & X & X &   &  Energy-efficient coverage (bandwidth utilization) assessment of an IRS  \\ \hhline{~-------}    
                & -MLA & &  & X & X & X   &  EE assessment of the accuracy of an ML model  \\ \hhline{~-------}
                & -LA  & &  & X & X & X  & EE assessment of the accuracy of a positioning/localization system   \\ \hhline{~-------}
                & -WUA  & &  & X & X & X  &  EE assessment of a WuR system responsiveness \\ 
                \midrule
                \multirow{5}{*}{\rotatebox[origin=c]{90}{Others}} & EPC,EPI, IPR,LDR  &  & & X & X & X & Evaluation of system's EE according to the load. Useful for proper operation planning and resource allocation/scheduling  \\ \hhline{~-------} 
                & WET exposure level & & X & & & X & Assessment of the environmental disturbances of deploying a WET technology \\ \hhline{~-------}
                & WuT &  & & X & X &  &  Assessment of a WuR responsiveness and corresponding EC \\ \hhline{~-------}
                & INFT &  & & X & X & X &  Assessment of the INFT (being proportional to EC) of an ML model  \\ \hhline{~-------}
                & MDS,SRAM &  & & X & X &  & KPI for the deployability of an ML model in a device/platform  \\ 
                \bottomrule
            \end{tabular}
            \begin{flushleft}{\footnotesize{$^c$Please note that sensitivity is not solely related to EH but also applies to radios in general, including WuRs. Specifically in WuR, sensitivity is taken into account in WUA, since the sensitivity of the WuR directly contributes to achieving higher WUA values.}}\end{flushleft}
            \label{MetricsT}  \vspace{-7mm}          
            \end{table*}
\vspace{-4mm}
\subsection{\uppercase{Energy Conversion and Transfer Metrics}}\label{ECTM}
Every energy conversion/transfer process introduces losses and other potential non-linearities affecting device/network EE. Some related metrics are discussed below.

\begin{table*}[t!]
            \caption{Some performance metrics and associated target values for current and next generation of wireless systems}
            \vspace{-1mm}
            \centering          
            \begin{tabular}{L{1.8cm}L{2.5cm}L{3.2cm}L{2.6cm}L{5.2cm}}
                \toprule
                \textbf{KPI}  & \textbf{Technology} & \textbf{Current Values} & \textbf{Target Values, 2030} & \textbf{Enabling techniques} \\
                \midrule
                max CE, APD & PV & $20\%$, $175~\text{mW/cm}^2$ & 50\% & Multijunction cells/CPV\\               
                    & TEG & $10\%$, $20~\text{mW/cm}^2$ & $30\%$, $70~\text{mW/cm}^2$ & Highly efficient thermoelectric materials \\
                    & MFC & NE, $5~\text{mW/cm}^3$ & NE, $500~\text{mW/m}^3$ & Electrode engineering \\
                    & Piezoelectric & NE, $<1~\text{mW/cm}^3$ & NE, $5~\text{W/cm}^3$& Structural optimization \\
                    & Triboelectric & NE, $50~\text{mW/cm}^2$& NE, $300~\text{mW/cm}^2$& Structural optimization \\
                    & Electrostatic & NE, $20~\text{mW/cm}^3$ & NE, $200~\text{mW/cm}^3$ & Structural optimization \\
                    & Vibration EM & NE, $<1~\text{mW/cm}^3$& NE, $3~\text{mW/cm}^3$ & Structural optimization \\
                    & Wind & $35-50\%$, NE & $60\%$, NE & Blade geometry optimization \\
                    & RF-EH & $30-70\%$, NE & $90\%$, NE & Adaptive receiver design \\ \hdashline
                PTE & RF-WET & $< 50\%$ & $70\%$ & Joint end-to-end efficiency optimization \\
                     & Inductive WET & $80-95\%$ & $99\%$ & Joint end-to-end efficiency optimization \\
                     & Capacity WET & $70-90\%$ & $ 99\%$ & Joint end-to-end efficiency optimization \\
                     & Laser WET & $< 10\%$ & $30\%$ & Joint end-to-end efficiency optimization \\
                     & Acoustic WET & $< 5\%$ & $50\%$ & Joint end-to-end efficiency optimization \\ \hdashline
                AHE, EOP & Ambient RF-EH & $0.01-1$ $\mu$W, $10^{-1}$ &  $\ge 0.1$ $\mu$W, $10^{-3}$ & Dynamic (online) hybrid combing \\
                     & RF-WET & $\ge 1$ $\mu$W, $10^{-3}$ & $\ge 50$ $\mu$W, $10^{-5}$  & Low-power massive MIMO and IRS \\ \hdashline
                ED, BL & Lithium-ion & $100-260~\text{Wh/kg}$, $300-500$ cycles & $500~\text{Wh/kg}$, $-$ & High-capacity electrode materials \\
                     & Supercapacitor & $10~\text{Wh/kg}$, $10^5-10^6$ cycles & $300~\text{Wh/kg}$, $-$ & Novel carbon-based composite materials \\
                     & Solid-state battery & $300~\text{Wh/kg}$, $5000$ cycles & $500~\text{Wh/kg}$, $30\times$ current value & High-capacity electrode materials \\   \hdashline
                EC & BC & $10-1000$ $\mu$W & $\le 10$ $\mu$W & Meta-materials and printed electronics \\                   
                   & Passive IRS & $0.1-0.5$ mW/element & $\sim 10$ $\mu$W/element & Low-resolution reflecting elements \& \\
                   & & & &  novel meta-materials\\
                   & Active IRS &  $\ge 0.5$ mW/element & $\sim 0.1$ mW/element & Advanced negative resistance components,\\
                   & & & & e.g., tunnel diodes \\
                   & WuR & $1 - 50~\mu$W & $< 1~\mu$W & Optimization of WuS detection \\
                   & & & & Edge traffic prediction for smart wake up\\
                   & TinyML & $0.1 - 50$ $\mu$J/inference & $0.05 - 5$ $\mu$J/inference
                   & PULP-based MCU \& coarse quantization \\ \hdashline 
                THP-EER & BC & $0.1-1$ nJ/bit & 1 pJ/bit & Passive frequency tuning and low-power \\
                & & & & amplifiers using, e.g., tunnel diodes    \\
                 & LIS & $0.01-1$ $\mu$J/bit & $< 10$ nJ/bit & Effective deployment  \& low-power HW \\ \hdashline
                RG-EER & BC & 100 $\mu$W/m & 10 $\mu$W/m & Passive frequency tuning and low-power\\
                & & & & amplifiers using, e.g., tunnel diodes \\
                 & Radio stripes & $10-100$ $\mu$W/m/element & $1-50$\ $\mu$W/element & Effective/dynamic deployment \\ \hdashline
                 MLA-EER & ML at network & $90-99\%$ &$>99.99\%$ & Specialized HW \& high-quality datasets\\
                  & ML at edge & $70-99\%$ &$>99.9\%$ & Specialized infrastructure \& ensembling\\
                  & TinyML & $65 - 98\%$ & $>99 \%$ & Advanced model search space and NAS   \\ \hdashline %
                INFT, MDS & ML at network &$10-30$ ms, $<250$ MB &$<0.1$ ms, $<1$ MB & Model compression \& optimization \\
                  & ML at edge &$10-100$ ms, $<15$ MB &$<1$ ms, $<1$ MB & Knowledge distillation\\
                  & TinyML & $10 - 1000$ ms, $<2$ MB & $<10$ ms, $<50$ KB  & Coarse quantization and NAS\\ \hdashline
                WuT & WuR & $10-200$ ms &$\sim 1-24$ ms & Fusion of WuR and main radio \\
                & & & &Shorter transmission time intervals\\
                \bottomrule
            \end{tabular}
         \label{KPItable}  \vspace{-4mm}       
            \end{table*}
\subsubsection{Component EE (CEE)} 
CEE refers to the ratio of the output to the total energy at the input of a certain electronic component. CEE considers losses due to heat, friction, and other inefficiencies, and describes the effectiveness of electronic components such as power supplies, power amplifiers, motors, filter circuits, phase shifters, etc. The ideal CEE value is 100\% or 1.
%
\subsubsection{EH Input/Output Relationship}
The main parameters impacting the performance of EH circuits are sensitivity, saturation, and CE. Sensitivity refers to the minimum magnitude or change in the input signal required for the EH circuit to produce a usable electric signal. Saturation, on the other hand, occurs when the circuit reaches maximum output and thus the harvested power is independent of any input power increase. Both sensitivity and saturation metrics correspond to absolute values of power or energy, while the EH CE is defined as the ratio of the output to the total input power. Notably, CE is a non-linear function of the input power, and EH datasheets commonly include input-output power transfer curves under different settings and emphasize the maximum achievable CE (hereinafter referred to as max CE). Notice that in WET systems, one can exploit the fact that CE depends on the operating conditions, such as the operating frequency, incident energy, and distance to the source, to drive the EH circuits to their maximum CE.

\subsubsection{APD of EH transducers} 
This metric refers to the ratio of the EH transducer's peak output power to its size under specific operating conditions. It is given in W per unit of area/volume and characterizes the miniaturization level achievable by an EH transducer.
\subsubsection{PTE} 
This metric, given as a percentage or a dimensionless quantity, refers to the ratio of the power captured by EH receiver(s) to the power consumed by the corresponding transmitter(s) and captures the joint energy CE of the transmitter(s), the medium, and the EH circuit(s). Several factors can affect PTE, including the distance between the ET and EH receiver and the specific WET technology.
\subsubsection{EH Coverage} 
This comprises a set of metrics characterizing the EH effectiveness in a given IoT network. For example, 
\begin{itemize}
\vspace{-1mm}
    \item Average harvested energy (AHE) characterizes the average energy harvested by (either a specific or random) device in the network in a given time period. It can be given in J, or W if time is normalized.
    \item Energy outage probability (EOP) denotes the probability that the energy harvested by a device in a certain period falls below an energy threshold $\xi$. Such a threshold may indicate the energy required for the device operation or the execution of a relevant task. EOP is the complement of the energy coverage probability.
    \item Meta distribution of the harvested energy (MDHE) is the distribution of the EOP conditioned on certain time/spatial-varying network parameters, thus, it is a finer-grained performance metric compared to the typical EOP. For instance, within the stochastic geometry framework, the authors in \cite{Deng.2019,Alves.2021} express MDHE  as the distribution of the EOP conditioned on the locations of the RF ETs. In this context, MDHE can be interpreted as the fraction of IoT devices in each realization of the point process (i.e., each possible network realization) that have harvested energy above $\xi$ with probability at least $\varepsilon$.
\end{itemize}

Notice that EOP and MDHE are relevant metrics for networks with battery-less IoT devices. Meanwhile, AHE can suffice in scenarios where devices are equipped with batteries, especially when AHE is measured per device.
All in all, these metrics measure the extent to which a region or network has access to EH services and/or can reliably maintain an energy supply over time. Therefore, they can be used to assess the level of energy availability/ubiquity.
\vspace{-4mm}
\subsection{\uppercase{Energy Storage and Consumption Metrics}}\label{subsec:energystorageKPIs}
\subsubsection{BCAP and energy density (ED)} 
BCAP, given in ampere-hour (Ah), refers to the amount of potential energy (usually chemical energy) that batteries can store. Over time, batteries' electrodes degrade due to continuous chemical reactions causing capacity fades and unexpected voltage drops. Notice that even when batteries remain disconnected from an external circuit, the internal chemical reactions will cause a self-discharge thus reducing the amount of energy stored over time. 

The ED of a battery measures the amount of power a battery can store per unit of volume (Wh/m$^3$) or weight (Wh/kg). High ED batteries can provide longer battery life while maintaining a compact and lightweight form factor. The factors that mainly impact batteries' ED are their chemistry and internal cell design.
%
\subsubsection{Lifetime}
In general terms, the lifetime of a device/component is a measure of its durability and 
refers to the period of time over which the device/component is expected to remain functional and perform its intended tasks without requiring significant repairs or maintenance. When the device/component is a battery, we refer to BL, which is influenced by many factors such as the chemistry and design of the battery, the conditions under which it is used and stored, the level of usage and discharge, and the charging and maintenance practices over time. The BL may be given by the number of charging cycles the battery can endure before its capacity decreases significantly, or by the time the battery can hold a charge before it needs to be recharged.

Finally, there is also the so-called network lifetime, which refers to the time period a network functions as intended. This metric is obviously related to the lifetime of the devices, but the relationship depends on the specific network's applications and performance requirements.
\subsubsection{EC}\label{EC} 
The EC of a component/device, specified in J, depends on the specific operation modes, i.e., active, idle, and sleep states, and the time spent in each. As briefly discussed in Section~\ref{EE}-\ref{wakeUp},  duty cycling takes care of properly scheduling these states to reduce EC subject to QoS requirements. Notably, tasks such as computation and communication, mostly executed in active modes, consume different amounts of energy. For instance,  transmitting typically consumes more energy than receiving data wirelessly, and the EC from computational tasks increases with the computation complexity. Obviously, the EC of a system/network is given by the sum of the EC of its components/devices.

The EC scales linearly with the floating-point operations per second (FLOPS)  and the million instructions per second (MIPS) in digital signal processors and computer systems, respectively \cite{Chen.2010}. The scaling factor depends on the amount of computational work, the number and type of functional units, the clock frequency, and the complexity of the instruction set architecture. Therefore, FLOPS and MIPS units are usually more useful than J units for EC performance comparisons in these systems.\footnote{However, comparing different systems in terms of FLOPS/MIPS may be sometimes misleading for EC assessment. For instance, there are processors with architectures specialized for specific tasks. That is the case of security accelerators, whose designs are optimized for encryption/decryption operations. Due to their specialized architecture, these processors require significantly less energy compared to conventional general-purpose processors when performing the required computations.} Nevertheless, when assessing network EC performance figures, one may inevitably rely on J units and averages over different states, workloads, components/devices, etc.

Other three key metrics related to EC are:
\begin{itemize}
    \item Net harvested energy (NHE), which represents the amount of energy that is available for use by an EH device after considering losses, i.e., due to energy conversion and storage, and EC related to the EH protocol, e.g., for CSI acquisition in RF-WET systems \cite{Zeng.2015}.
    \item Relative power saving (RPS), which quantifies the EC reduction driven by a certain approach compared to a specific benchmark \cite{FWuS}.
    \item Peak demand (PKD), which measures the highest level of energy demand during a given period.  It can be used to assess the capacity of a renewable energy system to meet maximum demand and identify opportunities for demand management and energy storage \cite{Stern.2017}.
\end{itemize}
%
\subsubsection{Green EC Share (GECS)} 
\vspace{-1mm}          
The exploitation of renewable sources is fundamental to support IoT sustainability and thus will continue expanding in the next few years. Consequently, quantifying their relative contribution to the total energy budget of IoT devices/networks/systems is increasingly relevant. For this, the GECS metric, which quantifies the portion of the energy consumed from renewable/green sources relative to the total EC, is undoubtedly attractive. Through GECS, one can get a clear understanding of the extent to which a given IoT device/network/system relies on green energy sources. GECS can even evolve to quantify the energy contribution from specific renewable sources, thus, supporting more granular insights. Notice that this metric can be used as a benchmark to set targets for increasing the green energy share(s) of a solution, as part of a broader sustainability strategy.
%
\subsubsection{Levelized cost of electricity (LCOE)}
This metric evaluates the economic feasibility of electricity generation systems. It is defined as the ratio of the estimated electricity cost and the estimated power plant's electricity generation during its lifetime, thus given in monetary units per kiloWatt-hour (e.g., \$/kWh). In the context of renewable energy, LCOE estimation is heavily determined by the ambient energy availability over the system lifetime. In fact, the number, size, and complexity of the EH transducer/circuits and the energy storage vary depending on the geographic location. Besides, the uncertainty in the amount of electricity generated by the system may cause unexpected expenses due to outages or energy trading with the grid to cope with sporadic high electricity demands. Fortunately, energy trading can also be a source of revenue if the network sells its surplus energy (cf. Section~\ref{integration}-\ref{techI}-\ref{Etrading}). One important consideration is the market dynamics that may occur over the anticipated lifespan of EH systems, which is typically around 30 years. This timeframe allows for the potential increase in renewable energy market penetration, as well as the development of technological breakthroughs. These advancements could lead to reduced LCOE due to lower costs for spare parts and more efficient electricity generation.

Finally, LCOE can be extrapolated in WET-enabled networks to estimate the ratio between the electricity cost to the harvested energy. In such cases, the system PTE, the deployment of the ETs, and CE of the EH circuits may be the main impacting factors when estimating LCOE.
%
\subsubsection{Grid reliability/stability (GRS)}
GRS captures the ability of the (micro)grid to maintain a stable energy supply, despite fluctuations in demand and supply, especially when considering contributions from renewable energy sources. It can be given in percentage and used to assess the resilience of the grid and identify opportunities for improving grid management and infrastructure.
\vspace{-4mm}          
\subsection{\uppercase{EER Metrics}}\label{EERM}
 An alternative (and popular) way to define EE is the ratio between the achievable QoS performance and the corresponding EC/PC that is required, or the ratio between the EC/PC and the corresponding achievable QoS performance. We define this category of EE metrics as EER and notice that they differ from one another in the type of QoS performance metric that is used.

Resource allocation based on optimized EER metrics has been a popular approach in wireless communication engineering, e.g., \cite{Bjornson.2014,Prasad.2017,Lopez.2019}. However, this may not always be the most effective way to balance performance and EC in all scenarios, unless the EER metrics already include some guarantee for QoS support and/or maximum EC. If not, minimizing EC while maintaining QoS requirements, e.g., \cite{Mozaffari.2016,Martinez.2022,You.2018,Lopez.2022,MEC3}, or maximizing performance given energy constraints, e.g., \cite{LongRuizhe.2021,Jamali.2021,Zhang.2022,basaklar2022tinyman,10.1145/1274858.1274870,8203801}, may be more important.

Although algorithmically/analytically optimizing EER metrics may not always make sense due to intrinsic operational constraints, they are certainly useful for comparison and drawing valuable performance insights. Next, we briefly discuss relevant QoS performance metrics and associated EER units.
%
 \subsubsection{Data transfer metrics}
 This is a set of highly related measures used to evaluate data transmission performance over wireless networks, e.g.,
 \begin{itemize}
 \vspace{-2mm}          
     \item spectral efficiency, in bps/Hz, measures the amount of data that can be transmitted per unit of time and frequency spectrum; 
     \item THP (capacity), in bps, measures the amount of data that can be transmitted over a wireless link or network (under ideal conditions) within a given period of time;
     \item $\epsilon-$capacity, in bps, constitutes the best upper-bound for the attainable THP  (capacity) supporting an outage probability that does not exceed $\epsilon$;
     \item goodput, in bps, measures the useful data rate (considering overhead and error correction) delivered to the end-user/application;
     \item effective capacity, given in bits per channel use (bpcu), constitutes the highest arrival rate that can be served by a network under a particular latency constraint, thus capturing PHY and link layer characteristics.
     \vspace{-2mm}          
 \end{itemize}
 The corresponding EER metrics are often given in bits/Hz/J, bits/J (or J/bit), or bpcu/J.
%
\subsubsection{Range (RG) or coverage area (CA)}
RG  refers to the maximum distance over which a specific device, or a generic technology device, can transmit/receive signals effectively, according to, e.g., target QoS guarantees. Similarly, CA refers to the geographical area within which wireless connections can be established and maintained with certain QoS guarantees.\footnote{Sometimes, RG/CA and associated QoS guarantees, e.g., expressed via data transfer metrics, appear directly intertwined in the form of new metrics such as area spectral efficiency and area THP/capacity.} The corresponding EER metrics are often given in W/m or W/$\mathrm{m}^2$, i.e., characterizing the required amount of power per distance/area unit. 

Related to CA, but specifically for IRS-assisted networks, the authors in \cite{D24} proposed the so-called area of influence (referred to as AroI) metric. AroI comprises the area of significantly improved wireless connectivity triggered by the IRS(s) when optimizing for the whole area under consideration instead of a single nominal receive position. Notably, the spatial resolution of IRSs in horizontal and vertical axes depends on the specific element array, thus, AroI specified in $\mathrm{m}^2$ might be insufficient. Similarly, in non-terrestrial networks, including those composed at least partially of UAVs, high-altitude platforms, and satellites, the network becomes inevitably three-dimensional. In such scenarios, the CA may evolve to coverage volume (CV) and the corresponding EER to be measured in W/$\mathrm{m}^3$.
%
\subsubsection{Bandwidth}  
This metric corresponds to the frequency region in which a technology or device operates, e.g., harvests sufficient energy in the case of RF-EH networks. Note that for some communication systems, especially those with frequency hopping (e.g., BLE or LoRa), there might be two different bandwidth notations: i) signal bandwidth referring to the bandwidth of one channel and ii) system bandwidth referring to the bandwidth of all channels used by the system. Another special case of interest is in IRS-assisted networks, where there is the so-called bandwidth of influence (BoI), which indicates the frequency region in which, any wave hitting the IRS, will be reflected \cite{D24}. The corresponding EER metric is given in Hz/W.
\subsubsection{Accuracy}\label{accura}
 This metric somewhat characterizes the proximity of a measured/estimated value to the true value. Both relative units, such as percentages and relative errors (e.g., probability of error, miss-detection (PMD), false-alarm (PFA)), and absolute units, such as mean absolute error (MAE), root mean square error (RMSE), and standard deviation (STD), can be used depending on the scenario/application.
 Next, three specific examples are briefly discussed. 
\begin{itemize}
\vspace{-2mm}          
    \item ML accuracy (MLA) is a crucial KPI for any ML model, especially TinyML, which is subject to tight HW and SW-related constraints. In general, the definition of learning accuracy depends on the task for which the ML model has been employed. For example, for anomalous behavior detection \cite{info13100450} and object detection \cite{Sabovic.2023} tasks, MLA may be the percentage of correct predictions an ML model makes. Meanwhile, for tasks such as PV power prediction \cite{9954033}, learning accuracy can either be defined by the RMSE or MAE among the predictions and actual observations. 
    
    Notice that the EER metric corresponding to an ML model may be expressed as the ratio between the relative/absolute accuracy that is achievable per EC at each inference step. Specifically, the corresponding EER metric can be expressed in  $\% /$J when using relative accuracy metrics. Finally, recall that FLOPs can be also used as a measure of energy as discussed in Section~\ref{KPI}-\ref{subsec:energystorageKPIs}-\ref{EC}, thus, resulting in $\% /$FLOP EER units.
    
    \item Localization accuracy (LA) is the precision with which a system can estimate the position of an object or feature in a given environment. LA can be given in angular or distance units, which may correspond to RMSE, MAE, or STD statistics. For instance, indoor localization systems supported by future 6G wireless communication systems may operate with sub-meter STD LA \cite{Lopez.2023}. The corresponding EER metric may be given in rad/W or m/W.
    
    \item Wake-up accuracy (WUA) refers to the ability of a WuR to reliably detect WuS and avoid false alarms. WUA is typically characterized by two other metrics: PFA and PMD~\cite{FWuS}. PFA refers to the probability that the WuR wakes up erroneously in the absence of WuS (false positive, FP). On the other hand, PMD refers to the probability that the WuR fails to wake up in the presence of WuS (false negative, FN). Specifically, PFA is given by FP$/$(FP+TN) while PMD is FN$/$(TP+FN), where TP (true positive) is the number of times the WuR correctly detects WuS, and TN (true negative) is the number of times the WuR correctly determines that WuS is not present. Moreover, WUA is calculated as (TP + TN)$/$(TP + TN + FP + FN). Both PFA and PMD depend on various factors such as the type of WuS used, the power level of the signal, the distance between the transmitter and receiver, and the RF conditions of the environment. The accuracy of WuR can be improved by optimizing these parameters and using more sophisticated WuS detection algorithms~\cite{rostami2020wake}.
\end{itemize}
%
\subsection{\uppercase{Other Metrics}}\label{others}
\subsubsection{Energy proportionality coefficient (EPC)}\label{other1}
This metric represents the PC of a device/system as a function of
the offered load. In general, the observed PC increases non-linearly with the load, and the PC in an idle state is often non-negligible, e.g., network switches consume up to 85\% of their peak PC when idle \cite{Fiandrino.2017}. EPC is defined in $[-1, 1]$, where $\text{EPC}=1$ ($-1$) means that each increase in load leads to an equal increase (decrease) in EC, while $\text{EPC}=0$ describes the case when the system EC is constant and does not depend on the load.

Three other metrics related to EPC are \cite{Fiandrino.2017}: 
\begin{itemize}
\vspace{-1mm}          
    \item Energy proportionality index (EPI), which captures the difference between the measured power and the ideal power, i.e., the power that the device should consume if it is fully energy proportional. EPI is expressed in the region between idle and peak PC only. $\text{EPI}=0$ (1) indicates that the EC is agnostic (fully energy proportional) to the workload. 
    \item Idle-to-peak power ratio (IPR), which measures the ratio between the idle and the peak PC. IPR values tending to zero indicate energy-proportional designs. 
    \item Linear deviation ratio (LDR), which captures the deviation of the observed PC from the fully proportional case, i.e., a straight line connecting idle and peak PC values.  $\text{LDR}=0$ corresponds to a linear system.
\end{itemize}
%
\subsubsection{WET exposure level} This metric describes how different WET energy-carrying signals disturb the surrounding environment. 

A common metric to characterize RF transmissions is the power spectral density (in W/Hz), which describes the power distribution of a signal across different frequency components. In the spatial domain, one can resort to the effective isotropic radiated power (in W) which is the hypothetical power that an isotropic antenna must radiate to yield the same signal strength as the actual RF transmission in the direction of the antenna's maximum gain. Besides, the EM field (EMF) radiated by RF sources may cause disturbances in nearby equipment operating in the same frequency band. For such a case, international organizations have resorted to EM compatibility, a boolean metric that takes on ``pass'' or ``fail'' and describes the ability of electronic equipment to successfully operate in a certain EM environment without being affected by (or affecting) other devices.

International EMF exposure limits are available, e.g., ICNIRP guidelines \cite{international2020guidelines}, which are set with large margins to protect against both short- and long-term health effects. Limits are provided for whole-body and local exposure scenarios while considering both general public and occupational environments (cf. \cite{LopezRosabal.2023} for their impact on the RF-WET performance). The exposure limit values relate to basic physical quantities that are known to induce adverse health effects. For instance, the specific absorption rate (SAR, in W/kg), specific absorption (SA, in kJ/kg), absorbed power density (in W/m$^2$), and absorbed ED (in kJ/m$^2$) are basic physical quantities measuring the absorption rate of EM energy in the human body. Other reference quantities that are more easily evaluated, e.g., incident power density (in W/m$^2$), incident ED (kJ/m$^2$), electric field strength (in V/m), and magnetic field strength (in A/m) are also provided. These have been derived from the basic restrictions to provide a more practical way to demonstrate compliance. Reference levels provide an equivalent degree of protection as the basic restrictions, and thus an exposure is taken to be compliant with the guidelines if it is shown to be below either the relevant basic restrictions or relevant reference levels. There are also EMF exposure assessment standards available for wireless communication and WET technologies.

The exposure to ultrasound signals is commonly evaluated by measuring the intensity of the pressure wave in dB, scaled to a frequency sensitivity response curve. However, the exposure duration is also relevant, as prolonged exposure can cause hearing impairments in animals whose hearing response falls within the operating frequency. In humans, the impact of ultrasound signals is primarily characterized by two indexes: the thermal index and the mechanical index. The thermal index is a unitless metric that measures the ratio between the acoustic power penetrating the skin and the amount of power required to raise the body temperature by one degree Celsius. On the other hand, the mechanical index indicates the ability of the acoustic signal to create tissue mechanical stress and damage.
%
\subsubsection{Wake-up time} 
Wake-up or start-up time refers to the time it takes for an IoT device to wake up from a low-power sleep state and become fully operational, thus it is given in time units. This is an important factor in determining EC and responsiveness of a network implementing duty cycling and/or WuR. Typically, wake-up time depends on several factors such as the type of device, the complexity of the wake-up process, and the power management scheme used. For instance, some devices may require a longer wake-up time to initialize, calibrate, or pre-heat sensors or to establish a wireless connection. Similarly, wake-up time can be longer if the device is in a deep sleep mode that requires more time to restore the device's state. To optimize EC and responsiveness, it is important to minimize the wake-up time. This can be achieved by using low-power HW components, optimizing the SW for fast wake-up time, and selecting appropriate power management schemes~\cite{ketata2022design}.

\begin{figure*}
    \centering
    \includegraphics[width=\linewidth]{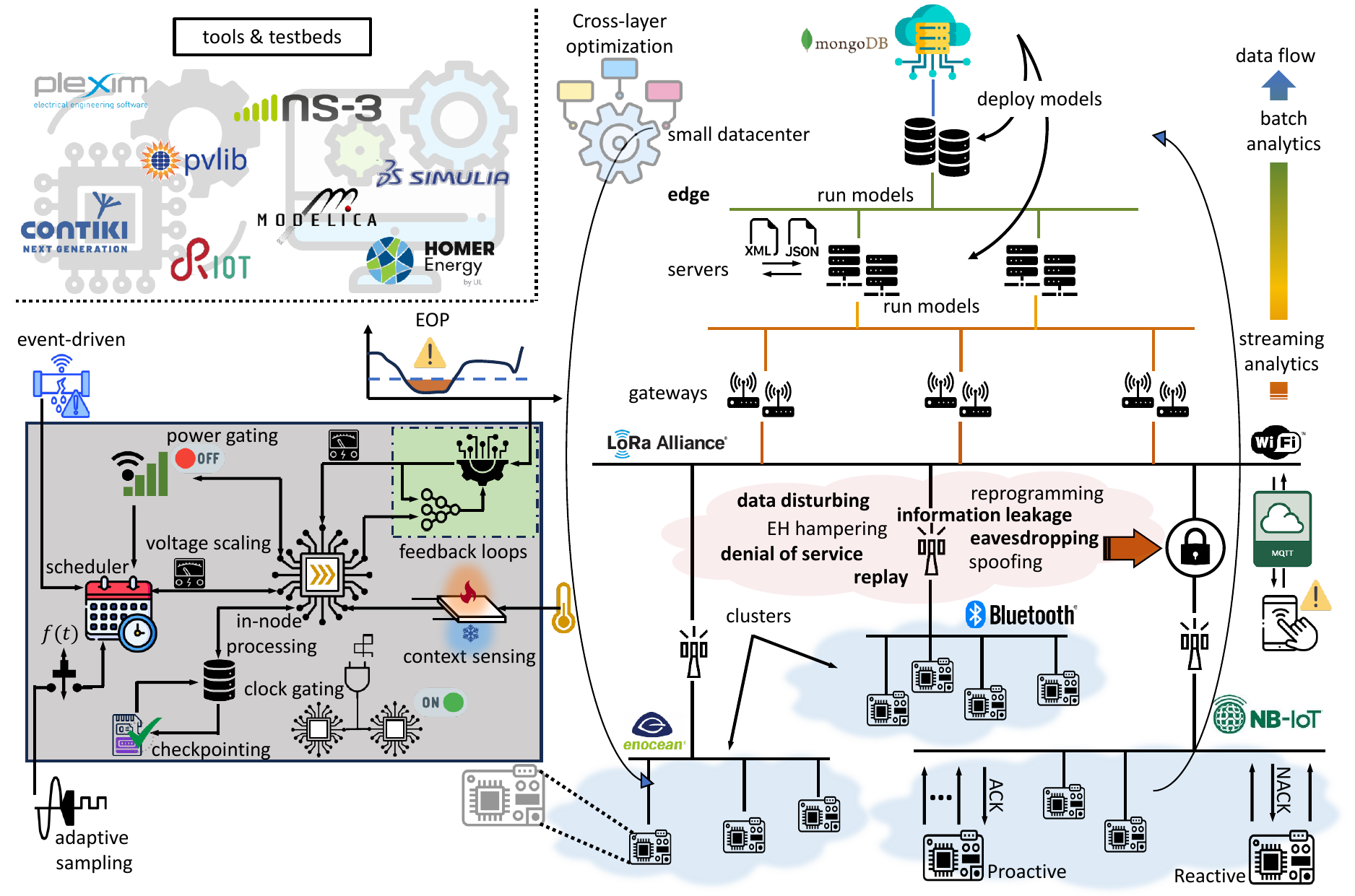}
    \caption{Integration, protocols, and deployments for energy-sustainable IoT. Herein, we also illustrate lightweight data-interchange formats, e.g., .xml and .json, and the database management tool MongoDB, which can be used for ML model storage and distribution, configuration files, and measurement formatting.}
    \label{fig:techIntegration}
    \vspace{-5mm}  
\end{figure*}
%
\subsubsection{INFT}\label{inft} 
INFT refers to the time it takes for an ML model to make a prediction (i.e, inference) on a new data sample. INFT can be affected by the complexity of the model, the size of the input data, the HW configuration, and the SW implementation. This metric is relevant for scenarios demanding real-time inference and/or low EC (since EC is proportional to active time) such as power management and sleep mode handling. 
%
\subsubsection{MDS}\label{mds} 
This metric refers to the amount of memory required to store the model's parameters and configuration, thus it is given in bytes (kB, MB, etc.). A small MDS makes the ML model easier to deploy on resource-constrained devices and/or can lead to fast INFT since the number of computation parameters during the inference phase is limited~\cite{kumar2017resource}. However, reducing the MDS can also have a negative impact on model performance, since smaller models may have limited capacity to capture complex patterns in the data. 
%
\subsubsection{Peak SRAM for inference} 
SRAM is an onboard memory space that accepts both read and write operations. Thus, an ML model's mutable parameters during its runtime are stored in SRAM. Notably, the peak memory required by a TinyML model during its inference, also called peak SRAM, becomes noteworthy in MCUs because of the limitations on onboard available memory. The standard units to express peak SRAM are kB and MB. Also, the peak SRAM depends on the memory scheduling procedure carried out by an inference library, while NAS decides MDS. The available SRAM on an MCU sets an upper bound on peak SRAM, while flash memory on an MCU constrains MDS \cite{lin2020mcunet}.
\vspace{-3mm}          
\section{Integration, Protocols, and Deployments}\label{integration}
Energy-sustainable IoT ecosystems require the seamless integration of the EP, ET, and EE-related technologies overviewed in this paper. This must be supported by proper multi/cross-layer protocol designs, energy-aware/efficient connectivity solutions, advanced simulation/testing tools, and validation procedures. This section delves into these critical aspects, namely integration, protocols, and real-world deployments as illustrated in Figure~\ref{fig:techIntegration}, highlighting their pivotal role in advancing sustainability KPIs.
\vspace{-3mm}  
\subsection{Technology Integration}\label{techI}
Herein, we discuss the integration of the surveyed technologies with IoT ecosystems while indicating the unique challenges and opportunities that arise.
\subsubsection{Hybrid EH}
\begin{figure}
    \centering
    \includegraphics[width=\linewidth]{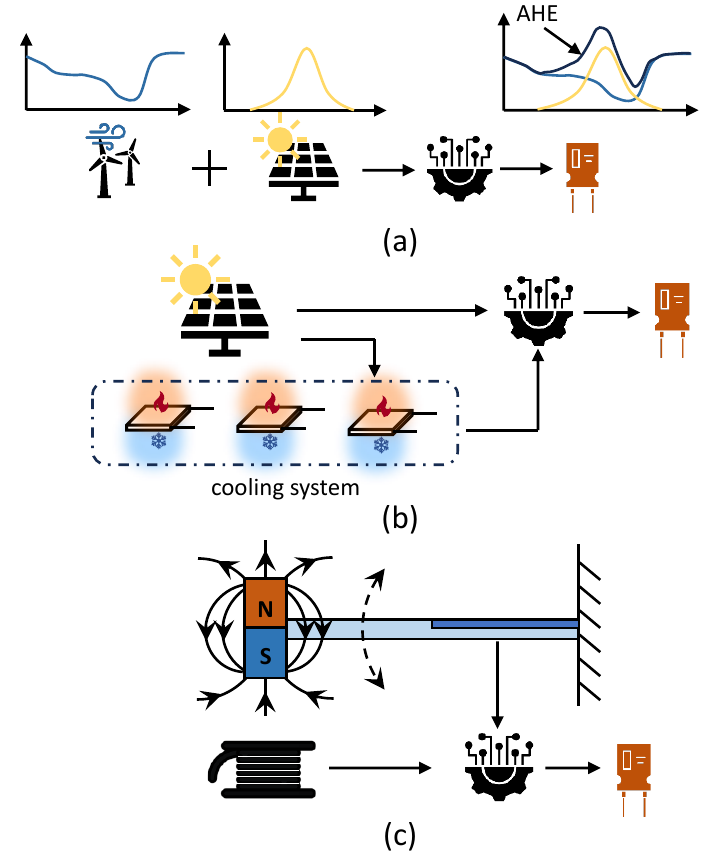}
    \vspace{-4mm}  
    \caption{Hybrid EH systems from (a) multiple sources using PVs and wind turbines, (b) a single source using a CPV and heat-based energy harvesters for heat waste recovery in the cooling system, and (c) a single source using different VEH transducers.}
    \label{fig:hybridEH}
    \vspace{-6mm}  
\end{figure}
The performance of ambient EH technologies is severely limited by the uncontrollability and unpredictability of ambient sources. This can be addressed by integrating multiple transducers into a single device. As shown in Figure ~\ref{fig:hybridEH}, a hybrid energy harvester can harness energy from either multiple or a single energy source. In the former case, the system can benefit from a higher AHE if all the sources are simultaneously available or can reduce the EOP of the energy supply as it is more likely that at least one of the sources is available \cite{Chandrarathna.2023}. As an example, Figure~\ref{fig:hybridEH}a illustrates a device harvesting energy from the wind and sunlight simultaneously. Observe that AHE is greater during sunny hours while harnessing energy from the wind compensates when sunlight is unavailable. In the case of single-source hybrid EH, the system incorporates additional transducers to harness the wasted energy from another EH process. For instance, Figure~\ref{fig:hybridEH}b shows that the excessive heat created in CPVs for light EH can constitute an additional EH source by using a heat-based transducer \cite{8861340}. Besides, multiple transducers designed for harnessing energy from the same source type can be combined to create a more versatile EH system. For instance, Figure~\ref{fig:hybridEH}c shows that one can increase the AHE, enable dynamic matching to the frequency of the source, or allow the transducer to resonate at multiple frequencies by combining different types of VEH transducers \cite{khalid2019review}. Finally, EH from ambient or dedicated sources can amicably coexist, e.g., dedicated ET may be turned off when ambient energy is sufficient for the device operation, thus saving energy at the energy transmitter \cite{wu.2019}.

Packing multiple transducers into a single device to realize hybrid EH is not exempt from challenges. First, it increases the manufacturing costs and compromises the aesthetics and form factor of the device. Hence, the trade-offs between extended device lifetime with the added costs require careful analysis. Second, it increases the HW's complexity in dealing with transducers' varying characteristics and optimal operating conditions. Indeed, the power management circuits must combine different output voltage levels and match the internal impedance of multiple transducers \cite{Abuellil.2020} to maximize the NHE. Finally, the environmental conditions can affect each transducer differently, which may affect the overall performance and lifetime of a hybrid EH system.
%
\subsubsection{Energy Management and Forecasting \label{energyF}}
\begin{figure}[t!]
    \centering
    \includegraphics[width=\linewidth]{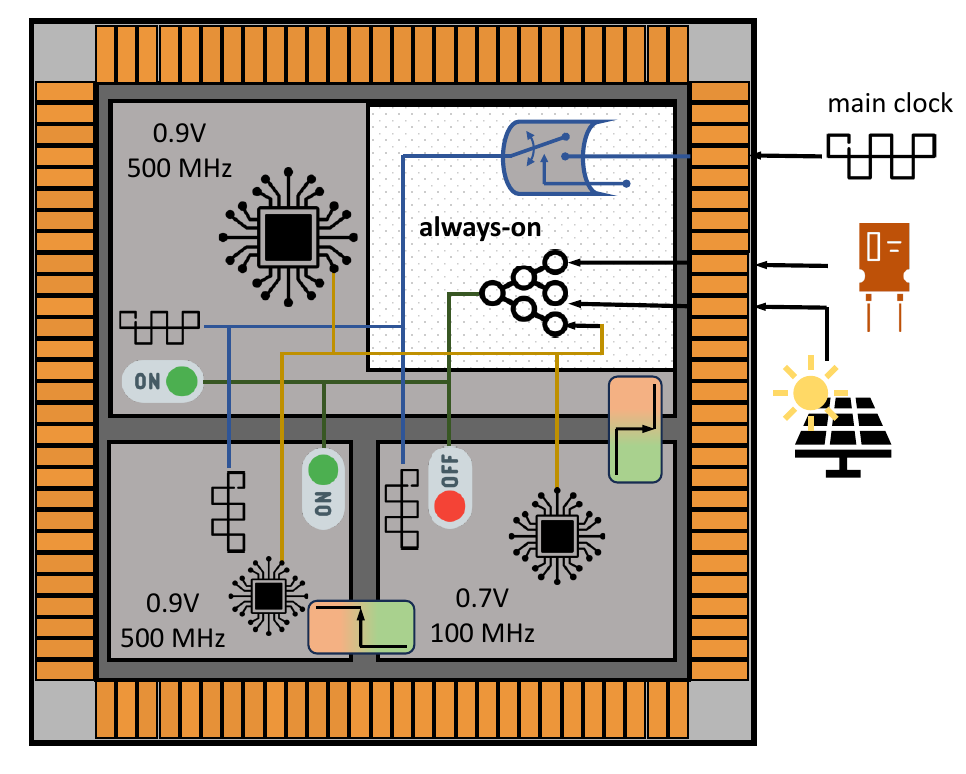}
    \caption{Energy management techniques for digital circuits.}
    \label{fig:energyManagement}
    \vspace{-5mm}  
\end{figure}

Harvesting energy from ambient sources and its judicious usage are equally essential to achieving IoT ENO. Figure~\ref{fig:energyManagement} shows the energy management techniques discussed next.

Digital circuits' EC comprises dynamic and static PC components \cite{Ali.2021}. The former results from the switching activity when load capacitances are charged and discharged and short-circuit currents flow during state transitions. Meanwhile, the static PC is caused by the leakage currents throughout reverse-biased semiconductors' junctions and sub-threshold conduction of inactive transistors. Notice that EC increases with the supply voltage following a square and linear law in the dynamic and static component cases, respectively. Hence, an effective technique to reduce the overall EC consists of dynamically scaling the supply voltage and the clock frequency to just meet the application demands  \cite{Hou.2023}. However, reducing the supply voltage/clock frequency leads to slower circuits, motivating modern digital circuits to be designed using multi-voltage architectures where each sub-system is operated at a different voltage/clock frequency. Such a design increases the complexity of the power distribution network and requires voltage level shifters to guarantee the correct transfer of logic levels between different power domains \cite{Hou.2023}. Finally, the supply/clock signal on inactive sub-systems in digital circuits may be cut off to prevent leakage using the so-called power/clock gating techniques. Under such a paradigm, conditional logic operates the circuits that provide power and clock to the sub-systems, pushing them into a deep sleep state whenever inactive. This technique was used in \cite{Moritz.2022} to design a heterogeneous integrated circuit composed of an always-on ultra-low-power MCU for power management and reduced wake-up time and a general-purpose MCU, with higher computational capabilities, for communication, processing, and sensing. Notice that the additional logic required for implementing power and clock gating techniques increases circuit complexity, cost, clock signal delays, and EC \cite{Gokhan.2021}. 

Combining the aforementioned approaches with proactive energy management techniques is key to boosting the reliability and lifetime of EH-enabled applications \cite{9154459}. As a first step, one can enable SW-reconfigurable power management circuits that cater to varying operating circuit conditions while adaptively pushing the system into $\mu$W PC levels when energy is scarce \cite{9226619}. This enables running ML algorithms to aid in the complex process of optimizing the previously discussed power management techniques \cite{Hou.2023}. Indeed, ML methods may help predict the availability and demand for energy through historical data extrapolation and thus facilitate preemptive scheduling maximizing EE.

In general, real-time data aggregation, advanced predictions, and adjustable algorithms are needed to ensure accurate predictions and efficient energy allocation~\cite{zhou2021deep}. For instance, at a device level, an RL-based TinyML algorithm using past samples from the EH process and the current battery level may judiciously allocate energy for future operations \cite{basaklar2022tinyman}. The main limitation of the approach in \cite{basaklar2022tinyman} is that the TinyML algorithm is trained offline which ignores that unpredictable changes in EH patterns can lead to improper energy management strategies. A potential solution to address this limitation could involve integrating retraining procedures at the edge or cloud level.

Leveraging predictive analytics and ML algorithms for proactive energy management shows promise for efficient IoT energy management~\cite{9154459}. Depending on the model's complexity, robustness, and delay considerations (cf. Section~\ref{KPI}-\ref{EERM}-\ref{accura} and Section~\ref{KPI}-\ref{others}-\ref{inft},\ref{mds}), implementation may be more feasible to occur on-device or on the edge/network side. These strategies together allow for the optimization of the EC of IoT devices in accordance with their real-time needs, leading to longer BL and contributing to the sustainable operation of IoT ecosystems.
%
\subsubsection{Energy Trading \& Microgrids}\label{Etrading}
The exploitation of renewable sources to power self-sufficient energy transmitters becomes an appealing solution on the road toward sustainable networks \cite{Rosabal.2023}. However, although energy management and forecasting techniques can help match energy generation and demand, this is not always possible, calling for energy trading mechanisms.
    
The above vision aligns with the emergence of microgrids, which are becoming increasingly popular to provide reliable and sustainable electricity in remote areas. A microgrid is a local energy network that comprises renewable electricity generators (cf. Section~\ref{EP}), energy storage, fuel-based generators, and controllers/inverters. Notice that fuel-based generators can be used as a backup source of electricity when renewable sources cannot meet the energy demand alone while the controllers/inverters ensure that energy flows efficiently between the various microgrid components.

Since energy trading with the main grid serves as a backup for renewable-powered WET systems \cite{8014289}, minimizing/maximizing LCOE/GECS while achieving a reliable network operation is necessary. This is not only more environmentally friendly but also reduces energy production and distribution costs. Moreover, direct energy trading among energy transmitters can also be possible using wired and wireless links that bypass the hierarchy of the power grid distribution network. In the case of wireless energy trading, static energy transmitters can exchange energy using, for instance, RF-WET and laser power beaming, to transfer energy over large distances. Moving/flying energy transmitters can benefit from these but also from near-field WET technologies, such as inductive coupling and acoustic WET, which provide high PTE, to trade energy with static transmitters. The idea here is that those transmitters cannot be equipped with large solar panels or wind turbines, so it is more beneficial for them to trade with other ground infrastructures with superior ambient EH capabilities \cite{9210076,Cetinkaya.2020}. Finally, energy transmitters can trade their surplus energy using the grid distribution network \cite{8753627}. This scenario becomes useful in the case when there is no direct connection between the parties, e.g., when the seller and the buyer transmitters are operated by different companies or deployed so distant from each other that a direct connection becomes infeasible.

Scalable algorithms are needed to support the increasing number of trading energy transmitters. Also, the latter may be managed by different operators, posing a challenge in terms of privacy and security. This motivates the use of DLT, which can provide transparency to the transactions and protect the ledger against possible forging by energy transmitters/receivers behaving maliciously \cite{8234700}. In other situations, the operators may not want to disclose their generation capacity and demand, hence potentially limiting the information exchange during the trading process \cite{Xia.2023}.
%
%
\subsubsection{Energy-aware Sensing and Computing}\label{sensComp}
Although accurate energy availability forecasting, efficient energy management, and energy trading techniques (cf. Sections~\ref{integration}-\ref{techI}-\ref{energyF} and \ref{integration}-\ref{techI}-\ref{Etrading}) can help prevent energy outages and task execution interruptions, they may not avoid them completely. These adverse events may still occur in those cases where the devices must uninterruptedly monitor/sense the environment, report, and/or execute high-complex tasks given limited energy availability. Sensing, computing, and communication protocols must consume extremely low energy and be resilient against these situations. Herein, we focus on the former two, while communication technologies and protocols are addressed in Section~\ref{integration}-\ref{protocols}.

In terms of sensing, the concept of ``context sensing from EH patterns'' is gaining traction. This is based on the observation that EH patterns often reflect the context in which energy is being harvested. For instance, kinetic-powered wearable IoTs can identify and tally the user's steps as the energy harvester produces distinctive peaks in its signal with each footstrike \cite{Kalantarian.2016}, while a thermoelectric energy harvester can detect changes in surface temperature from the variations in the EH signal \cite{Campbell.2014}. Note that by substituting dedicated sensors with a context detection algorithm based on the EH patterns, the device EC might reduce significantly.\footnote{Readers may refer to \cite{8944276} for further details on possible applications, e.g., human activity recognition, transportation mode detection, estimation of calorie expenditure,  gait recognition, hotword detection, HVAC airflow monitoring, and acoustic communication.} The two basic approaches for sensing from EH signals are \cite{8944276}: i) analyzing the patterns of the instantaneous power generated by the EH transducer, which allows the detection of a rich set of contexts at the expense of frequent sampling of the fluctuating power values; and ii) analyzing the amount of the total energy accumulated in the storage over a specific period of time, which consumes less energy by sampling the stored energy only once in a while at the expense of more coarse-grained context sensing. Very simple TinyML (cf. Section~\ref{ML}-\ref{tinyMLsec}) algorithms, e.g., for peak detection, moving average computation, etc, may be used for these tasks.

In terms of computation, efficient ``checkpointing'' techniques are necessary to periodically save volatile states in non-volatile memory so that the program execution can restart from a known state if the power supply fails. Notice that all variables and registers stored in the volatile memory during the program execution are completely lost in an energy-outage situation. The main challenges encountered by checkpointing techniques for EH IoT devices and main state-of-the-art approaches are \cite{8944276}: 
\begin{itemize}
    \item Non-negligible EC: the energy overhead of checkpointing increases with the number of variables to be copied/stored/reloaded. Therefore research on the area focuses on identifying how often to implement checkpointing and specific stages/locations for it. One appealing approach lies in inserting potential checkpoints to the program at compile-time and activating checkpointing processes at run-time.  Observe that checkpointing after a function call, instead of during the call, is desirable since it requires copying less volatile variables to non-volatile memory. However, having too separated checkpoints may lead to ``Sysyphean loops'', which happen when the energy harvested since the previous power cut-off is repeatedly not enough to reach the next checkpoint. This motivates the use of a watchdog timer to be triggered when the program re-executes from a checkpoint or a task boundary, e.g., to launch a checkpointing process when the timer interrupt occurs to mandatorily split previously uncompleted code group into two or multiple smaller groups \cite{Hicks.2017}.
    \item State inconsistency: complete state preservation is guaranteed when variables are exclusively stored in either volatile or non-volatile memory, not split between both. All variables in volatile memory are saved in non-volatile memory at a checkpoint, allowing consistent state recovery after power loss. Yet, most systems employ a small volatile memory for frequently used variables, placing the rest in non-volatile memory for efficiency. During rollbacks, volatile memory variables revert, while non-volatile memory variables retain new values, leading to potential data inconsistencies. This can be addressed by decomposing a longrunning program into a sequence of short and atomic tasks \cite{Maeng.2017,Sabovic.2022}, and leveraging the concept of ``idempotency''.\footnote{Specifically, an idempotent task can be executed multiple times without producing different results. To ensure idempotency, the task should not contain any idempotency violation, which is a write instruction to a non-volatile memory that was first accessed by a read instruction.}
    \item Timing inconsistency: power failures can cause timestamp discrepancies, especially in IoT devices interfaced with multiple sensors. Notice that battery-powered devices use a real-time clock to prevent this, but EH IoT devices lack consistent power, leading to inconsistent timestamps in reported sensor data. This motivates the tracking of the time elapsed between power failures, e.g., using remanence decay of SRAM and capacitor for timekeeping \cite{Rahmati.2012,Hester.2016}. Note that time tracking can also be done by receiving relevant information from the network (e.g., cellular or satellite) after recovering, but this might lead to more EC. 
\end{itemize}
The use of non-volatile RAMs combining features of RAM (fast access to any address) and flash (being non-volatile) may be appealing to avoid the above challenges (cf. Section~\ref{ML}-\ref{tinyMLsec}-\ref{futureD}).

Finally, efficient sensing and computation mechanisms for ultra-low-power EH devices can leverage i) event-driven processing, e.g., triggering computation tasks only when specific sensor data thresholds or conditions are met, reducing the need for continuous polling; ii) in-node pre-processing, e.g., essential features extraction from sensor data before transmission; iii) adaptive sampling, e.g., dynamically adjusting sampling rates based on the current operating conditions or the importance of the sensed data, conserving energy during periods of lower activity; iv)  task schedulers that prioritize sensing and computation tasks according to their urgency and importance; v) feedback loops, which may allow adjusting computation and sensing parameters based on changing environmental conditions or system performance, ensuring optimal resource utilization; and vi) HW/SW co-design to exploit the strengths of both domains for efficient computation and sensing.

\subsubsection{Security and Privacy in Energy-Sustainable IoT}
IoT ecosystems comprise low-capability devices that are responsible for collecting data, pre-processing it, and/or controlling certain processes, e.g., sensors and actuators, but also edge and network devices and cloud servers, which execute complex processing tasks to support user applications and services. IoT protocols and data management/analytics must optimize and harmonize the operation of all these heterogeneous system components, and do so while supporting security and privacy, which is inherently challenging. 

For low-capability IoT devices, the main security attacks are related to \cite{Elahi.2020,Di.2012}:
\begin{itemize}
\vspace{-2mm}  
    \item  eavesdropping, which consists of intercepting and storing messages interchanged by the authorized devices, and thus supposes a data leakage and privacy threat; 
    \item denial of service, including mainly
    jamming and flooding.
    A typical example of a jamming attack is the transmission of interfering signals that affect the performance of the communication channel. Meanwhile, a flooding attack may consist of the transmission of bogus messages, whose processing consumes energy and may ultimately deplete the IoT battery, e.g., BLE and LoRa devices may be flooding-sensitive \cite{Mikhaylov.2019}. In fact, a WuR device may experience the so-called \emph{denial of sleep}, by which it is maintained awake by a malicious system continuously sending it wake-up requests.     
    \item side-channel information leakage, where the malicious node leverages magnetic leaks, EM, EC, and/or timing information PHY features to obtain private data or secret keys;
    \item spoofing, which consists of impersonating trusted entities or devices to deceive the system, thus often leading to unauthorized access or data manipulation; 
    \item replay, which consists of retransmitting previously captured legitimate data  to gain unauthorized privileges or manipulate system behavior;     
    \item data disturbing and reprogramming, which consists of partially/totally modifying information and/or program data on a sensor, e.g., security parameters, sensing conditions, reporting destinations, by direct access to the node, or simply by feeding it with malicious data. Intermittent computing procedures such as checkpointing are typically used to support the operation of EH devices (cf. Section~\ref{integration}-\ref{techI}-\ref{sensComp}), but may, unfortunately, facilitate these security breaches, e.g., a malicious adversary could read the data on the checkpoint, checkpoint tamper after system restoration, or execute a checkpoint replay attack;    
    \item EH hampering, which consists of hampering the amount of ambient energy that the IoT device harvests, e.g., to force the device to lower its security level or to delay the transmission of highly sensitive data.
    \vspace{-2mm}
\end{itemize}
Notice also that EH-IoT devices in certain applications may go off for long periods and then back on, thus
hiding possible attacks that tamper with the device. 

Considering that performing cryptographic operations and processing/transmitting large data overhead, including signatures, keys, padding, drains valuable energy resources, extremely lightweight security and privacy-preserving mechanisms must be carefully designed \cite{Cook.2023} and operate at lower layers mostly. For instance, they could rely on adaptable security levels and waiting queues for highly confidential data such that the device transmits data when it has the energy to encrypt it \cite{Di.2012}, and TinyML-assisted approaches \cite{shafique2021towards}. In the case of \emph{denial of sleep} attacks, a simple approach is to generate/update the wake-up address of each WuR node in a pseudo-random fashion based on key material known only by authorized peers \cite{Capossele.2016}.

Unfortunately, even lightweight mechanisms may not be incorporated into such IoT devices in many cases due to their extremely low energy availability and complexity, and they must completely rely on network/edge protocol stack security to secure their data.
The IoT network/edge protocol stack must ensure compatibility and security across a wide variety, and potentially massive number, of IoT devices, platforms, and communication technologies. Moreover, keeping IoT devices up-to-date with the latest security patches and firmware updates may be challenging, especially in remote or inaccessible locations. These issues call for efficient encryption, key management, authentication, and authorization mechanisms. Also, IoT ecosystems may be divided into segments or zones, each with its own security policies and access controls, to limit the impact of a security breach and help contain potential threats. 

ML is a fundamental tool to support IoT security \cite{Cook.2023}, e.g., by helping detect anomalies and patterns indicative of security breaches or attacks. Due to its tendency to be resource-intensive, ML approaches are mainly leveraged on the network side, although edge and device sides are increasingly exploiting them as well (cf. Section~\ref{ML}). Note that ML may also help a malicious system to become smarter. For instance, a malicious system may correlate a certain IoT device activity with the surrounding environment conditions, and use such information to predict the type of sensed data or energy availability. Therefore, every system action and potential information that may be derived from it, and observed/sensed, must be carefully considered.

Finally, ML and DLT (e.g., blockchain) integration can drastically improve EE and privacy/security of transactions in 5G and beyond networks~\cite{nguyen2020privacy,nguyen2021intersection,patil2022comprehensive}. Specifically, DLT allows value transactions between parties through decentralized trust but suffers from large communication overhead and difficulty in handling massive two-way connections, which may be mitigated by ML approaches. Indeed, ML can enhance DLT security and efficiency by providing intelligent data analysis and prediction capabilities \cite{liu2020federated}. A relevant application scenario for ML and DLT integration within the context of self-sustainable and secure IoT ecosystems is that of energy trading and microgrids (cf. Section~\ref{integration}-\ref{techI}-\ref{Etrading}).

\begin{table*}[t!]
    \caption{Features and KPIs of IoT Communication Technologies$^d$}
    \vspace{-1mm}
    \label{tab:IoTConnect}
    \centering
    \begin{tabular}{L{1.5cm} L{1.35cm}  L{1.2cm} L{1.45cm} L{2.3cm} L{2.2cm} L{1.8cm} L{3.2cm}} 
        \toprule
         \textbf{Technology} & \textbf{Freq. band} & \textbf{Topology} & \ \ \textbf{Sensitivity} & \textbf{Data Rate} & \ \ \textbf{Range} & \textbf{Average PC} & \textbf{Security}  \\
         \midrule
         RFID (EPC- global Gen2) & sub-GHz & P2P & $-85$ dBm & $40\!-\!640$ kbps & $>\!10$ m & $1\!-\!10$ $\mu$W & AES-128, Grain-128A \\ \hdashline
         BLE 5 & 2.4 GHz & P2P, mesh & $-97$ dBm  & $0.125\!-\!2$ Mbps & $<\!200$ m  & $20$ mW & AES-CCM, P256 ECDH, HMAC-SHA-256 \\ \hdashline
         WiFi HaLow & sub-GHz  & star, relays & $-100$ dBm & $0.04\!-\!4$ Mbps & $<\!1$ km & $10-100$ mW & WPA3, AES-CCMP \\ \hdashline
         Z-Wave & sub-GHz & mesh & $-110$ dBm & $9.6\!-\!100$ kbps & $<\!100$ m & 1 mW & Security 2 (S2), AES-128   \\ \hdashline
         Thread & 2.4 GHz & mesh & $-105$ dBm & 250 kbps & 30 m & 0.5 mW & AES-128,IP-based security,  QR-aided authentication \\ \hdashline
         ZigBee Pro & 2.4 GHz, sub-GHz & mesh & $-100$ dBm & 250 kbps & $<\!100$ m & $0.1\!-\!0.5$ mW & CCM, IEEE 802.15.4-based\\ \hdashline
         EnOcean & sub-GHz & P2P, mesh, star & $-98$ dBm & 125 kbps & $10-30$ m & $0.05-50$ mW & rolling code, AES-128 \\ \hdashline
         NFC & HF & P2P & $-10$ dBm & 424 kbps &  10 cm & $5$ mW & AES-128, NDEF \\  \hdashline
         LoRa & sub-GHz  & star-of-stars & $-148$ dBm & $0.3\!-\!50$ kbps & $>\!10$ km & $150$ mW & AES-128 in CCM \\         \hdashline
         SigFox & sub-GHz & star & $-142$ dBm & $1\!-\!100$ bps (uplink) & $30\!-\!50$ km (rural), $3\!-\!10$ km (urban) & $100$ mW & application/manufacturer-dependent \\ \hdashline
         NB-IoT & licensed LTE bands & star & $-141$ dBm & $204.8$ kbps (uplink) & $10\!-\!35$ km & $500$ mW & 3GPP security \\ \hdashline
         LTE-M & licensed LTE bands  & star & $-108$ dBm & $4$ Mbps (downlink), $7$ Mbps (uplink) & $<\!15$ km & $>500$ mW & 3GPP security \\
         \bottomrule
    \end{tabular}
    \begin{flushleft}{\footnotesize{$^d$The data in this table come from an exhaustive search over numerous technical specifications and are intended for qualitative, rather than quantitive, comparisons. Moreover, a thorough technology comparison may require considering other indicators that are not captured in this table.}}\end{flushleft}
    \vspace{-5mm}
\end{table*}
%
\subsubsection{Tools and Testbeds}
Tools and testbeds provide controlled environments for realistic experimentation, enabling the identification/solution of potential energy inefficiencies, protocol bottlenecks, and integration issues. By simulating and analyzing the performance of IoT devices and systems, potential IoT designs/ strategies can be validated prior to real-world deployment.

Key tools for EH circuit/HW modeling and design include \href{https://www.analog.com/en/design-center/design-tools-and-calculators/ltspice-simulator.html}{LTspice}, \href{https://powersimtech.com/products/psim/capabilities-applications/}{PSIM}, \href{https://www.simplistechnologies.com/}{Simplis}, \href{https://se.mathworks.com/products/simscape-electrical.html}{MathWorks Simscape Electrical}, \href{https://www.altium.com/}{Altium Designer}, and \href{https://www.kicad.org/}{KiCad}. Meanwhile, the following tools are undoubtedly appealing for device/system-level simulation and forecasting purposes:
\begin{itemize} 
   \vspace{-2mm}
    \item \href{https://www.plexim.com/}{PLECS}, which is a simulation platform that includes support for modeling EH systems. It allows the modeling and simulation of various EH sources and circuits in both continuous-time and discrete-time domains. 
    \item \href{https://openmodelica.org/}{OpenModelica} \cite{OpenModelica.2020}, which is an open-source modeling and simulation environment supporting the development of models for various physical systems, including EH devices. It allows the modeling and analysis of the behavior of energy harvesters in complex systems. 
    \item \href{https://www.mmm.ucar.edu/models/wrf}{WRF}, which is a widely used atmospheric simulation system that provides detailed weather forecasts, especially for solar radiation, wind speed, and other weather parameters that impact energy generation from solar panels and wind turbines.
    \item \href{https://www.3ds.com/products-services/simulia/}{SIMULIA} (part of Dassault Syst\`emes), which is a suite of simulation tools that includes capabilities for studying multiphysics systems, including EH applications. It can be used to simulate the mechanical and electrical aspects of EH devices. 
    \item \href{https://pvlib-python.readthedocs.io/en/stable/}{PVLIB Python} \cite{pvlib.2018}, which is an open-source library providing functions to estimate/forecast solar irradiance, module temperature, and energy output from photovoltaic systems.
    \item EH Module in \href{https://www.nsnam.org/}{NS-3} \cite{Tapparello.2014}, which allows the simulation of the EH capabilities of IoT devices and the study of their interaction with communication protocols.
    \item \href{https://www.homerenergy.com/}{HOMER Energy}, which allows forecasting energy generation based on historical data and system specifications and provides microgrid optimization capabilities.
    \vspace{-2mm}
\end{itemize}
These tools and many others allow for investigating integration capabilities, expected performance, and valid protocols of EH-based devices and systems.

Although the above tools also provide testbed capabilities to some extent, they are often limited, thus calling for more advanced platforms for modeling and testing complex ecosystems. For instance, researchers may resort to \href{http://www.contiki-os.org/}{Contiki} or \href{https://www.riot-os.org/}{RIOT}, open-source operating systems specifically designed for IoT devices,  to develop and test IoT applications with a focus on energy optimization. Indeed, \href{http://www.contiki-os.org/}{Contiki} provides various built-in protocols, including CoAP and MQTT,\footnote{CoAP is defined in \href{https://datatracker.ietf.org/doc/html/rfc7252}{RFC 7252} as a UDP-based transport protocol, a limited HTTP, developed for devices with limited memory, storage and computing power, limited battery power, and low bandwidth. Meanwhile, MQTT is a lightweight, publish-subscribe, IoT network protocol for message queue/message queuing service.} and supports energy-efficient communication and management features. Moreover,  \href{https://networksimulationtools.com/cooja-simulator-for-iot-download/}{COOJA}, which is a network simulator specifically designed for \href{http://www.contiki-os.org/}{Contiki}, may be used to simulate large-scale IoT networks and evaluate the EC of IoT devices under different conditions. Finally, tools such as \href{https://www.openlca.org/openlca/}{OpenLCA} and \href{https://simapro.com/}{SimaPro} for life cycle assessment are becoming appealing to quantify the environmental impact of IoT products/services while considering the entire lifetime.
\vspace{-4mm}  
\subsection{Protocols, Cross-Layer Designs, and Scalability}\label{protocols}
\vspace{-1mm}
\subsubsection{IoT Connectivity Landscape}
The IoT connectivity landscape is broad, with both short- and long-RG technology solutions as illustrated in Table~\ref{tab:IoTConnect}. Also, connectivity protocols like MQTT, CoAP, and AMQP can enable seamless communication among heterogeneous IoT devices. In general,
energy-sustainable IoT connectivity solutions must i) support EH, WET, and/or energy trading, ii) achieve high EE and spectral efficiency,  iii) guarantee long BL,  iv) be scalable and support high connection density, and v) require low maintenance and intervention. Current IoT connectivity solutions are far from achieving these goals, but firm steps are currently being taken in these directions.

Short-RG IoT connectivity technologies feature protocols with a sub-km communication RG, e.g., RFID,  Bluetooth, WiFi, Z-Wave, Thread, ZigBee, EnOcean, and NFC, and target use cases such as audio streaming, well-being monitoring, and home/industry automation. RFID is a passive BC technology \cite{Alsamhi.2019,Zhao.2019,Freidl.2017,Vardakis.2021}, which often directly incorporates EH and WuR techniques, while Bluetooth and WiFi have dedicated protocols for low-power devices, namely i) BLE \cite{mikhaylov2020wake,Afaneh.2018}, which incorporates WuR, standard duty cycling, and an energy-efficient adaptive frequency hopping mechanism, and ii) IEEE 802.11ah (WiFi HaLow) \cite{Tian.2021}, which implements smart power management techniques such as distance-based power control. Z-Wave is designed specifically for smart home devices and applications, utilizing a mesh network topology such that devices can act as repeaters and may choose the most energy-efficient communication routes within the mesh network \cite{Marksteiner.2017,Cano.2018}. Thread also utilizes a mesh network topology, although supporting IPv6, which enables direct IP connectivity for devices without requiring translation or additional overhead \cite{Marksteiner.2017}. Finally, standards like EnOcean and ZigBee Green Power have defined a radio interface to specifically operate battery-less EH devices (even using RF-EH/WET in the case of EnOcean) \cite{Marksteiner.2017}, while the NFC alliance is currently focusing efforts on developing a standard for EP at short distances over the existing NFC radio interface.

Meanwhile, long-RG IoT connectivity technologies feature protocols with a communication RG from several to tens km, e.g., LoRa, ZigFox, NB-IoT, and LTE-M, and suit use cases such as smart agriculture, asset tracking, and environmental monitoring. Notably, LoRa devices are capable of dynamically re-configuring their transceivers, e.g., the spreading factor and bandwidth, so that transmissions can achieve longer distances with a given power budget \cite{Talla.2017,Hoeller.2018,Cano.2018}. SigFox and NB-IoT devices communicate over narrow bandwidth channels to reduce the received noise power and simplify the receiver's front-end design \cite{Cano.2018,3gpp161}. NB-IoT and LTE-M implement extended discontinuous reception that allows the devices to enter a sleep mode for long durations while periodically waking up to check for incoming data, thus reducing the waking-up frequency and conserving energy \cite{3gpp161,3gpp16}. A key difference is that LTE-M supports higher bandwidth (1.4 MHz) and thus cannot be deployed in guard bands, like NB-IoT. Finally, note that contrary to LoRa/Sigfox, both NB-IoT and LTE-M support IP from UE while the transceivers often implement TCP/IP, MQTT, CoAP, and even FTP and HTTP stacks. 
%
\subsubsection{Energy-aware Protocols}\label{energyProtocol}
Energy-aware protocols are crucial to meet/improve energy-sustainability KPIs (cf. Section~\ref{KPI}). The main challenges are associated with the diverse composition of IoT devices, variable network conditions, and specific application requirements. Therefore, IoT protocols must be adaptive and manage energy resources based on availability and demand patterns~\cite{lopez2021survey}. For instance, data transmissions may be triggered avoiding ultimately compromising present and future system states, which inevitably requires energy-awareness.

At the MAC and network layer, Low-Energy Adaptive Clustering Hierarchy, Time-Slotted Channel Hopping, and Routing Protocol for Low Power and Lossy Networks, designed to adapt to the data paths that are most frequently traveled, are incipient examples of energy-aware protocols. Meanwhile, content-based protocols focused on retrieving content to minimize expenditure of energy~\cite{EE_protocol1} are appealing at higher layers.

Incorporating intelligence into energy-aware protocols may promote efficient resource allocation, real-time power management, and EE increase~\cite{lopez2021survey}. Additionally, ML facilitates the recognition of common device behavior patterns, aiding the implementation of energy-saving strategies like sleep modes and adaptive power control. Ultimately, this alliance of power-driven insight can combine the efficiency of network power and the EE of the network, and promote the integration of energy-aware protocols into the IoT.

The grant-free spectrum access scenario, where the IoT devices attempt to access the transmission medium directly, without prior control signaling, thus saving energy but increasing the collision probability, constitutes an ML application opportunity. Indeed, integrating ML here may allow IoT devices to acquire optimal transmission strategies progressively, thereby minimizing collision occurrences and promoting network scalability. Numerous ML-based grant-free spectrum access mechanisms have already surfaced in existing literature, e.g., \cite{rech2021coordinated, jiang2019distributed, chandak2022learning} discussing distributed RL-based spectrum access mechanisms. Specifically, the time-slot selection task is modeled as a Markov game in \cite{rech2021coordinated} and equilibrium points within this game are learned through a reward-inaction RL algorithm. In the case of \cite{jiang2019distributed}, the spectrum access mechanism leverages the information about the inner states of the device to determine time-slot selection probabilities. Here, the RL algorithm approximates the intricate mapping between the inner states and slot selection probabilities. Meanwhile, the authors in \cite{chandak2022learning} portray the IoT devices as RL agents trained over time, while developing coordination among their spectrum access mechanisms, leading to successful message transmissions. Still, further work is needed to consider the energy availability issues and thus design energy-aware MAC protocols.

Finally, note that these protocols are not only limited by the IoT devices' characteristics but also the network side's and the variety of emerging technologies such as distributed MIMO, IRS, and LIS. Such technologies may not only have advantages but also intrinsic limitations regarding resource allocation and interoperability, which must be considered.
%
\subsubsection{Cross-Layer Optimization}
An IoT ecosystem mainly includes the following layers \cite{Kumar.2017}: i) HW/sensing/actuation, ii) local processing and storage, iii) communication, iv) cloud computing (and storage), and v) application. Each of them has its own sub-layers and typical EE-promoting techniques. For instance, sleep/wake-up, self-organization, and approximate computing in TinyML \cite{shafique2021towards} are HW/sensing layer techniques, while EH and intermittent computing-related techniques, including NAS, network pruning, quantization, and ROMANet in TinyML \cite{shafique2021towards}, correspond to the local processing and storage layer. TinyML MDS is decided by NAS, while TinyML peak SRAM  during inference can be reduced by optimizing the memory scheduling using ROMANet. The communication layer comprises the (sub-)layers of the communication protocol stack, including scheduling, access, and routing protocols. Meanwhile, centralized optimization, virtual machines, and caching correspond to the cloud computing layer, and data aggregation, predictive analytics, and user-centric customization techniques pertain to the application layer. By considering the interdependencies and interactions between layers and optimizing accordingly, the system performance may largely improve, increasing lifetime, reducing costs, enhancing QoS in dynamic environments, and improving sustainability.

\begin{figure}
        \centering
        \includegraphics[width=1\columnwidth]{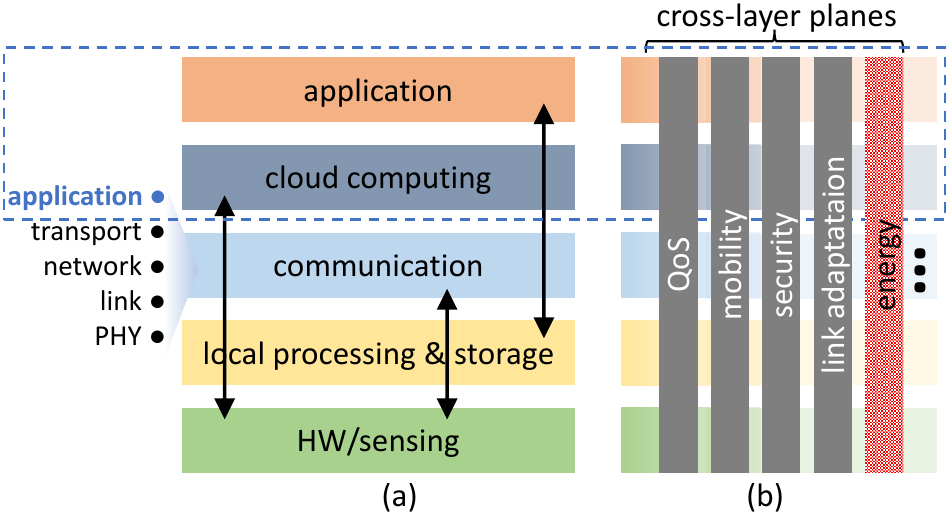}
        \caption{Cross-layer designs: (a) direct layer interactions and (b) layer interactions via coordination planes.}
        \label{fig:layers}
        \vspace{-6mm}
    \end{figure}

The traditional cross-layer perspective is illustrated in Figure~\ref{fig:layers}a and lies in creating new interfaces establishing direct paths between non-adjacent layers (or sub-layers) for parameter sharing. Many of the energy-aware protocols from Section~\ref{integration}-\ref{protocols}-\ref{energyProtocol} exploit this approach. However, this can lead to tightly coupled layers as changes in one layer may require corresponding adjustments in multiple other layers, thus compromising modularity and making the system complex and difficult to manage. Also, excessive information sharing between distant layers might raise security and privacy concerns. This may be mitigated by introducing coordination planes, which are cross-section views of the protocol stack on which interlayer coordination algorithms can be applied. This was proposed in   \cite{Carneiro.2004} within the communication protocol stack environment/layer while identifying four key coordination planes, namely security, QoS, mobility, wireless link adaptation, and energy. This framework must evolve to incorporate energy and sustainability aspects over the layers and sub-layers as illustrated in Figure~\ref{fig:layers}b. 

All in all, holistic approaches are needed for orchestrating the wide spectrum of technical approaches discussed throughout the paper across the IoT ecosystem layers.

%
\subsubsection{Topology, Data, and Load Management}
Topology, data, and load management play crucial roles in achieving scalable energy-sustainable IoT networks. Indeed, by optimizing the physical layout and data processing tasks of the network, EC, EH/EP capabilities, and connectivity performance can be improved. For instance, clustering nodes into groups or hierarchies can alleviate scalability issues and minimize EC \cite{Zhu.2021}, while an appropriate and dynamic deployment of RF PBs can improve the EH statistics of RF-EH IoT networks and/or lower the deployment/maintenance costs of the PBs network \cite{Rosabal.2021,Rosabal.2023}. Also, the expansion of MEC (cf. Section~\ref{ML}-\ref{edgeML}-\ref{MEC}) and distributed data storage capabilities can help reduce latency, communication, and energy costs, supporting scalability and EE. Energy-aware self-organizing networks and routing protocols (cf. Section~\ref{integration}-\ref{protocols}-\ref{energyProtocol}) are appealing as well. In terms of data management,  techniques such as i) data fusion, integrating data collected from multiple sensors while filtering out redundant or irrelevant information; ii) resource-constrained data storage models, optimizing storage and computing resources; iii) distributed and replicated storage, reducing reliance on centralized data centers; and iv) energy-aware compression techniques reducing the size of data being transmitted; may be leveraged depending on the scenario characteristics and performance requirements \cite{Bassirou.2020,Krishnamurthi.2020}. Note that data management is intrinsically related to load management as well, which comprises i) load balancing techniques, distributing tasks and data processing load evenly across devices; ii) edge and fog computing offload workloads to devices and fog nodes, reducing the need for data transmission to remote clouds \cite{Wan.2018}; iii) dynamic resource allocation, adjusting network topology and parameters based on real-time conditions. Regarding the latter, ML techniques (cf. Section~\ref{ML}) can be exploited to analyze data patterns and predict workload demands for efficient resource allocation.

Finally, DLTs provide a tamper-proof and transparent ledger that is crucial for data fusion and storage \cite{nguyen2021intersection,Sun.2020}. Also, the corresponding consensus mechanisms optimize resource allocation and facilitate load balancing, thus promoting EE. This synergy between IoT strategies and DLT integration promises enhanced data security, efficient ET and trading (cf. Section~\ref{integration}-\ref{techI}-\ref{Etrading}), and optimized resource utilization, advancing energy-sustainability goals.

\begin{table*}[h!]
    \caption{Real-world IoT implementations comprising EP/ET}
    \vspace{-1mm}
    \label{tab:realImplement}
    \centering
    \begin{tabular}{L{2.1cm} L{1.8cm} L{2.7cm} L{3.8cm} L{5.3cm}}
        \toprule
         \textbf{Solution} & \textbf{EP/ET} & \textbf{Communication} & \textbf{Intelligence} & \textbf{Use Cases} \\
         \midrule
         Track Extreme & light & BLE, NB-IoT, LTE-M. multi-RAT & N/A & asset tracking \\ \hdashline
         enerSENSE & indoor light & LoRa, NFC. Multi-RAT & edge ML for data processing, visualization, and control &  industry 4.0, smart buildings \\ \hdashline
         Series S2 & light & 2G, LTE-M, 2.4 GHz. Multi-RAT & N/A &  industrial monitoring, asset tracking \\ \hdashline
         Jack & light & BLE & cloud ML for damage evaluation and location & fleet management \\ \hdashline
         EnOcean & vibration, heat, light  & EnOcean\textregistered, Bluetooth, Zigbee. WuR support & N/A & smart spaces, EE, smart homes \\ \hdashline
         ONiO.zero & vibration, heat, RF-EH, light  & BLE. WuR support & N/A & ultra-low PC applications \\ \hdashline
         IoT Pixels & RF-EH & Bluetooth & cloud ML for mapping measu- rements into the physical world & smart healthcare, supply chain \\ \hdashline 
         MIMO tag \cite{Eid.2023} & RF-EH & MIMO and backscattering & N/A &  digital twins, RFID tags \\ \hdashline
         Infinity & heat, light & BLE & cloud ML & machine health monitoring \\ \hdashline 
         AirCord\texttrademark & laser-ET & WiFi & cloud ML for power allocation & industrial IoT, healthcare, gaming, retail \\ \hdashline
         Cota\textregistered\ Real Wireless Power\texttrademark\ & RF-ET & N/A & cloud ML for collection and analysis of device data & building automation, smart homes, lighting control, and industrial applications \\ \hdashline
         Emrod & RF-ET & N/A & N/A & energy distribution, space-based energy infra- structure, connecting renewables to the grid \\
         \bottomrule
    \end{tabular}
\end{table*}
\vspace{-4mm}
\subsection{Real-world Case Studies and Applications}
\vspace{-2mm}
Herein, we delve into real-life implementations that pertain to the surveyed technologies on the road to achieving energy-sustainable IoT. In particular, we focus on those relying on the discussed EP and ET technologies to supply energy. Table~\ref{tab:realImplement} summarizes our discussion.

Light-based EH is a popular technology for commercial EH-powered solutions due to its high availability/harvestability compared to other solutions. Inspired by this, SODAQ\footnote{Please refer to \url{https://sodaq.com/} for more details.} developed Track Extreme, a device for asset tracking. Every time the Tracker Extreme detects the asset moves, it obtains its location based on WiFi and GPS signals which then can report over BLE, NB-IoT, or LTE-M radio solutions. Similarly, Xeelas\footnote{Please refer to \url{https://www.xeelas.nl/?lang=en} for more details.} developed Series S2, a light-based EH device with applications in industrial monitoring and asset tracking that can communicate over 2G, LTE-M, and 2.4 GHz. In the transportation sector, AGC Automotive\footnote{Please refer to \url{https://www.smart-jack.com/} for more details about their Jack project.} has developed the Jack device for fleet management operations. The Jack utilizes vibration sensors to detect impacts on the vehicles' windshields while the measurements are uploaded to the ML-powered cloud application via a BLE gateway for damage evaluation and reporting. Accounting for the duration of indoor human activities, enerthing\textregistered\ has developed a thin-film PV specifically optimized for indoor EH. The solution is utilized for powering the enerSENSORS devices for Industry 4.0 and smart building applications. The enerSENSORS utilize a LoRa radio to upload the measurements to a cloud-based service that runs ML algorithms to aid decision-making processes based on the received measurements. The enerSENSORS also support over-the-air configurations via an NFC radio \footnote{Please refer to \url{https://www.enerthing.com/en/products/} for more details about enerthing's portfolio.}.

Undoubtedly, RF-EH stands over its technological counterparts in achieving small form factor implementations due to its relative simplicity. This technology is exploited by Wiliot\footnote{Please refer to \url{https://www.wiliot.com/} for more details.} for the IoT Pixels, which is a credit card-sized smart label intended for smart healthcare and supply chain management applications. The device communicates over Bluetooth with a cloud service to report temperature measurements and location information. The cloud service transforms the raw sensory information into the corresponding physical domains using ML algorithms. Meanwhile, researchers at the Georgia Institute of Technology developed a multi-antenna solution in \cite{Eid.2023} for a card-sized device to boost the output power of RF-EH. Their prototype implements a low-power MIMO solution using Rotman lenses. Rotman
lenses achieve beam scanning by adjusting the time delays created in the microstrips network that feeds the antenna
elements; hence, eliminating the need for phase shifters.

EnOcean\footnote{Please refer to \url{https://www.enocean.com/} for more details.} designs devices capable of harvesting energy from vibrations, light, and heat, and communicating over Bluetooth, Zigbee, and EnOcean\textregistered. EnOcean's solutions also provide support for adding an external WuR circuitry as an energy-saving mechanism. The company targets smart spaces (occupation maximization), EE, and smart homes. Similarly, Sensemore\footnote{Please refer to \url{https://sensemore.io/} for more details.} developed Infinity, a self-sustained device that can harvest energy from heat and light. The Infinity device communicates over BLE to an ML-powered cloud service, which can evaluate the condition of industrial machinery based on sensory data. Aiming at reducing cost and form factor, ONiO\textregistered\ developed ONiO.zero, a tiny MCU that integrates power management, memory, BLE transceiver, and support for WuR circuitries. Currently, ONiO.zero targets batteryless applications, such as a marketed remote control, which can be sustained by EH from vibrations, multiband RF-EH, light, and thermal energy\footnote{Please refer to \url{https://www.onio.com/} for more details.}.

Efforts to develop commercial ET applications are also substantial. For example, Wi-Charge\footnote{Please refer to \url{https://www.wi-charge.com/} for more details.} developed the laser-based AirCord\texttrademark\ technology, which aims at meeting the energy demands of industrial IoT, healthcare, gaming, and retail applications. Wi-Charge transmitters are hybrid nodes capable of transmitting energy and communicating with the devices via WiFi. The transmitters communicate through a cloud-based ML controller that allocates power to the different devices based on their power requirements, battery level, power availability, and customer-set priority. Similarly, Ossia\textregistered\ is working on an RF-WET/EH MIMO system. Ossia's Cota\textregistered\ Real Wireless Power\texttrademark\ technology also relies on an ML-powered cloud service for user charging management\footnote{Please refer to \url{https://www.ossia.com/} for more details.}. Finally, a company named Emrod is aiming to solve some of the challenges faced in the current power distribution grid network via an RF-WET solution. Emrod envisions a system for transferring energy over long distances by utilizing a network of relays in the path between the transmitter and the receiver to guarantee LOS radio link. Each node is equipped with a massive antenna array to enhance coverage. Currently, the company is also targeting space-based energy applications that redirect the solar energy harvested in space to power earth applications via RF-WET and aims at using this technology as a means for connecting hard-to-reach renewable energy generators to the main power grid.
\vspace{-3mm}
	\section{\uppercase{Conclusions \& Outlook}} \label{conclusion}

 \begin{table*}[t!]
            \caption{Summary of the Main Challenges and Associated Research Directions of the Overviewed Technologies}
            \centering          
            \begin{tabular}{L{1.5cm}L{5.4cm}L{9.3cm}}
                \toprule
                \textbf{Tech.} & \textbf{Challenges} & \textbf{Research Directions} \\
                \midrule
              Light/ & Energy-intensive manufacturing & Exploitation of new materials such as perovskite and organic PVs\\
              Laser  & Efficient operation & Multi-junction PVs, CPV\\
              & LOS dependency & UAVs, supporting reflectors, hybrid RF and light-EH/laser-WET \\
        \hhline{~--}
              Heat & Limited temperature variations & Transducer arrays, large surface area, materials engineering \\
              & High thermal stress/strain & Exploitation of new materials for the contacts \\
        \hhline{~--}
              MFC & Limited lifespan & Genetically engineered bacteria for improved activity and lifetime \\
              & Poor cost-effectiveness & Cheap, yet durable, materials and stacked or multilayer MFCs \\
        \hhline{~--}
              Vibration & Reduced resonant frequency & Frequency up-conversion and low-stiffness materials \\
              & Frequency matching & Passive frequency tuning and bandwidth broadening schemes \\
        \hhline{~--}
              Flow & Footprint and threat to wildlife & Compact designs, contained/vertical blades, bladeless generators\\
              & Operation in variable wind conditions & Wind prediction, real-time optimization of cut-in/cut-off speed \\
        \hhline{~--}
              RF-EH/ & Very low ambient EH & Multi-antenna, metasurface-aided, and self-reconfigurable receivers\\
              RF-WET & Low end-to-end efficiency & CSI-free beamforming, robotic WET, holistic optimization, distributed MIMO \\
        \hhline{~--}
               Inductive & Sensitive to coils' misalignment & Efficient coils design and deployment \\
               & Large form-factor & Meta-materials-aided design for EM field focusing \\
        \hhline{~--}
               Capacitive & Requires high/variable reactive compensation & Active variable reactance rectifier \\
               & Limiting fringing fields & Phased multi-plate charging \\ 
        \hhline{~--}
               Acoustic & Limited charging distance/efficiency & Phased acoustic arrays, optimal deployment of acoustic transmitters \\
               \hhline{~--}
               BC & Imperceptible integration to everyday objects & Exploitation of meta-materials \\
               &  High-frequency operation & Low-complexity protocols for beam search/exploitation \\
               & Wide-band designs & Advanced, yet affordable, modulation schemes \\
               & Frequency-agnostic designs & HW and algorithms for proper frequency tuning \\
               & Security &  Quantum backscattering mechanisms \\
               & Enhanced RF-EH sensitivity & Advanced leakage power reduction and technology and voltage scaling\\
               \hhline{~--}
               IRS/LIS & Low-cost/energy control & Trade-offs studies and low-cost/energy-effective control mechanisms \\
               & CSI acquisition, especially for active IRS & Low-complexity/overhead training protocols \\
               & Advanced metasurface implementations &  SW-controlled techniques exploiting diverse metamaterial functions \\
               & Data-driven optimization & Low-complexity/energy ML mechanisms, including TinyML \\
               \hhline{~--}
               Radio  & Costly signaling between antenna elements  &  Efficient distributed processing architectures \\
              stripes & Advanced HW and protocols & Novel circuits/prototypes and optimized resource allocation\\
              \hhline{~--}
              WuR & Radio resource management and scheduling & Characterization of EE, latency, reliability, and robustness trade-offs\\
              & HW complexity and cost & Inband and RFIC-embedded WuR implementation\\
              & Beamformed WuS at mm-waves & Beamforming design for WuS, how many beams in a single WuS burst?\\
              \hhline{~--}
              O-RAN & Lack of results for envisioned goals & Practical and real-world trials on the virtualized RAN\\
              \hhline{~--}
              MEC & Complexity due to coordination issues & Distributed coordination frameworks\\
              & Heterogeneity and standardization & Edge-native frameworks to provide a unified and standardized approach to managing devices and data\\
              & Security and privacy concerns & Secure communication protocols and data encryption techniques\\
              & Accuracy \& latency in real-time applications & Data pre-processing, caching, and filtering, QoS management\\
              & Designing a green MEC & Dynamic right-sizing, GLB, and renewable energy-powered MEC systems\\
              \hhline{~--}
              FL & Expensive communication \& synchronism & Reducing communication rounds and vanishing gradients in each round\\
              & Security/Privacy issues & Protection against malicious attacks and privacy-enhancing\\
              & Robustness issues & Heterogeneous HW support and robust aggregation algorithms\\
              & MDS & Efficient training and inference for massive heterogeneous networks\\
              \hhline{~--}
              TinyML & Limited processing power of devices & Novel HW to reduce the computational burden of ML algorithm\\
              & Noise sensitivity of in-memory computing & Error-correcting codes \& advanced, yet affordable, signal-processing \\
                \bottomrule
            \end{tabular}
         \label{challengeTable}  \vspace{-5mm}       
            \end{table*}

 \begin{table*}[p!]
            \caption{Summary of Key (Sub-)System-Level Technical Challenges and Associated Research Directions}
            \centering          
            \begin{tabular}{L{5.5cm}L{7.95cm}L{3.84cm}}
                \toprule
                \textbf{Challenge} & \textbf{Research Directions} & \textbf{Tools}  \\
                \midrule
              \emph{Quantifying the energy-sustainability grade of IoT ecosystems:} IoT ecosystems are composed of heterogeneous subsystems and devices with complex dynamics/interactions. Also, energy sustainability comprises economic, societal, and environmental aspects, which must be captured and is especially challenging since both short and long-term perspectives must be considered.  & A unifying quantitative metric for assessing the energy-sustainability grade of an IoT ecosystem may be unfeasible. Instead, key metrics separately for economic, societal, and environmental aspects could be used and they should be readily understandable by the general public, e.g., labels. Their expected evolution over time should be provided. The definition of such key quantities and measurement and forecasting procedures requires intensive interdisciplinary research. 
              It should be up to the specific stakeholders, regulators, administrations, etc to establish comparisons and targets. & Energy-related performance metrics, including those in Section~\ref{KPI}. Big-data collection and analysis, e.g., via ML. Life cycle assessment SW, e.g., SimaPro and OpenLCA. Economic and societal modeling tools.    \\   \hline
              \emph{Intelligent energy management:} EC must be minimized without compromising QoS, calling for intelligent energy management mechanisms, which must cope with EH/EC variability;  dynamic network topologies (since devices may frequently connect and disconnect); and real-time decision-making requirements. & Developing context-aware low-energy EH/EC predictive approaches (at network, edge, and device), tailored energy-efficient HW designs, and multi-EH transducers with efficient HW/SW-level integration.  Developing energy-aware routing protocols and efficient self-organizing and self-healing network architectures. Characterizing cloud- vs edge- vs local-based decision-making trade-offs and proposing novel related protocols. & EH forecasting/simulation tools (cf. Section~\ref{integration}). Statistical and optimization tools, including ML.   HW simulation/design tools, IoT-tailored operating systems \& network simulators (cf. Section~\ref{integration}). Power profilers.       \\ \hline
              \emph{Network extreme-heterogeneity and scalability support:}
              How to efficiently support the operation of the increasingly massive and heterogeneous (in terms of characteristics, supported protocols, and performance requirements) IoT devices? 
              & Developing protocols that enable seamless interoperability among heterogeneous IoT devices (e.g., by strengthening MQTT, CoAP, and AMQP protocols). Developing mechanisms to distribute tasks/data processing among the devices (e.g., based on their energy levels, computational capabilities, and proximity to each other). Developing O-RAN functionalities that facilitate serving such extremely massive and heterogeneous IoT networks. & Protocol analyzers and stacks. IoT middleware platforms. IoT -tailored operating systems and network simulators (cf. Section \ref{integration}).  Game theory.  Statistical and optimization tools, including ML (especially FL).     
              \\ \hline
             \emph{Distributed energy trading:} Energy trading in microgrids is increasingly gaining traction. The current infrastructure-based model limits the flexibility and scalability of the technology, motivating distributed implementations. The latter are challenged by a greater coordination and interoperability complexity, difficulty in guaranteeing grid stability, reduced resilience, and data management and security issues.  & Developing control algorithms that can handle many distributed energy resources and heterogeneous ET mechanisms (including WET), manage energy trading autonomously at the microgrid level, and scale efficiently. Designing multi-agent systems that can negotiate energy transactions, optimize energy flows, and adapt to dynamic supply/demand changes. DLT is promising for the latter, but further research is needed to reduce the large communication overhead and support massive two-way connections and EE. Developing and standardizing practical WET protocols.
             & Game theory. DLT tools. Statistical and optimization tools, including ML (e.g., RL, FL, and evolutionary algorithms). Graph theory and network analysis tools. EH forecasting/simulations tools (cf. Section~\ref{integration}). Quantum computing.
             \\ \hline
             \emph{Low-power signal processing and task execution:} Signal processing and task execution procedures directly affect the EC of IoT devices but are inevitably necessary for supporting sensing, communication, management, and interoperability capabilities and specific high-level applications. How to minimize their energy footprint without QoS degradation?     & 
             Integrating EH forecasting capabilities on computation offloading decision mechanisms and MEC systems. Developing novel techniques for optimizing TinyML platforms, e.g., based on bio-inspired optimization, spiking and analog NN, and in-memory computing. Developing baseline TinyML models and efficient (HW-aided) intermittent computing mechanisms. Developing intelligent energy management mechanisms (cf. 2nd row). Developing efficient context sensing techniques from EH patterns. Developing low-overhead protocols (cf. 3rd row). & EH forecasting/simulation tools (cf. Section~\ref{integration}). Game and control theory.  Markov decision processes. MCU, cloud computing, and MEC (e.g., OpenFog and EdgeX) platforms. IoT middleware. ML (especially clustering, pattern recognition, and evolutionary algorithms).     \\ \hline
             \emph{Cross-layer design and holistic integration:} Jointly optimizing different layers, technologies, and subsystems is crucial for realizing energy-sustainable IoT ecosystems. However, cross-layer designs and holistic integration require a deep understanding of the layers' interactions and trade-offs to preserve security and privacy, prevent unforeseen interactions, enable real-time adaptation, and mediate over conflicting objectives. & Characterizing system layers' interactions and potential overheads from implementing cross-layer optimizations. Developing adaptive and self-organizing mechanisms that allow IoT devices to adjust their behavior across layers based on changing conditions. Designing communication protocols that explicitly consider interactions and trade-offs between different layers. Exploring methods for achieving real-time adaptation across layers, including rapid decision-making mechanisms and efficient feedback loops to enable timely adjustments. Developing metrics and evaluation frameworks that capture the energy-sustainability benefits of cross-layer designs. & IoT-tailored operating systems \& network simulators (cf. Section~\ref{integration}). Queueing and control theory. Protocol and network analysis tools. Statistical and optimization tools, including ML (especially RL and NN). MEC platforms (e.g., OpenFog and EdgeX). DLT tools. \\
                \bottomrule
            \end{tabular}
         \label{challengeTable2}         
            \end{table*}

In this work, we provided valuable insights into energy-sustainable IoT and claimed it could only be supported by the harmonious coexistence of EP, ET, and EE processes. These processes refer to charging exploiting green energy sources, the intentional movement of energy from one device/system to another, and the ability of a device/system/process to perform its intended function with minimal energy. We overviewed the main technologies corresponding to these processes together with their use cases, recent advances, challenges, and research directions. Specifically, EH technologies based on light, heat, MFCs, vibration, flow, and RF were discussed within the EP processes. In the case of ET, the focus was on RF, inductive/capacitive coupling, laser, and acoustic-based technologies. Meanwhile,  in terms of EE support, we focused on technologies enabling low-power communication, namely BC, metasurface-aided communication, radio stripes, and WuR, and also ML approaches aiming to reduce the EC burden of application tasks at the device/edge/network side. 
In general, the appropriateness of a specific technology/technique depends on the characteristics and performance requirements of the target application, and increasingly on their sustainability support level. This may be assessed by considering proper performance metrics. Indeed, we discussed relevant performance metrics to assess energy-sustainability potential and listed some relevant target values for specific technologies in the next generation of wireless systems. Moreover, we discussed protocol, integration, and implementation issues at the technology and system level.

\vspace{-3mm}
Table~\ref{challengeTable} compiles a summary of the main challenges and corresponding research directions of all the discussed technologies, while Table~\ref{challengeTable2} does the same but from a system-level perspective instead of technology-wise. Observe that although the focus of this work was on IoT energy-sustainability aspects during the operation phase, a truly self-sustainable ecosystem must consider sustainability aspects along the entire product lifecycle, i.e., planning, manufacturing, deployment, operation (including the target use case), maintenance, and disposal. The natural progression of our work is to expand in this direction. 

Finally, policy and regulatory frameworks play a pivotal role in shaping the adoption and deployment of truly sustainable IoT solutions. Indeed, ensuring effective government policies, incentivizing sustainable manufacturing practices, and fostering international cooperation are essential. This is often challenging since policy and regulatory factors i) vary significantly across different regions, ii) are subject to frequent changes, and iii) can be context-specific and dependent on the specific application of IoT technologies. Additionally,  the establishment of robust data privacy laws and standardized regulations must strike a balance between responsible IoT deployment and safeguarding user information. Going forward, it is imperative to address these policy and regulatory challenges and explore their intersection with technical advancements to promote the development of self-sustainable IoT ecosystems along the entire product lifecycle.
\vspace{-3mm}
\bibliographystyle{IEEEtran}
\bibliography{IEEEabrv,references}

\begin{IEEEbiography}[{\includegraphics[width=1in,height=1.25in,clip,keepaspectratio]{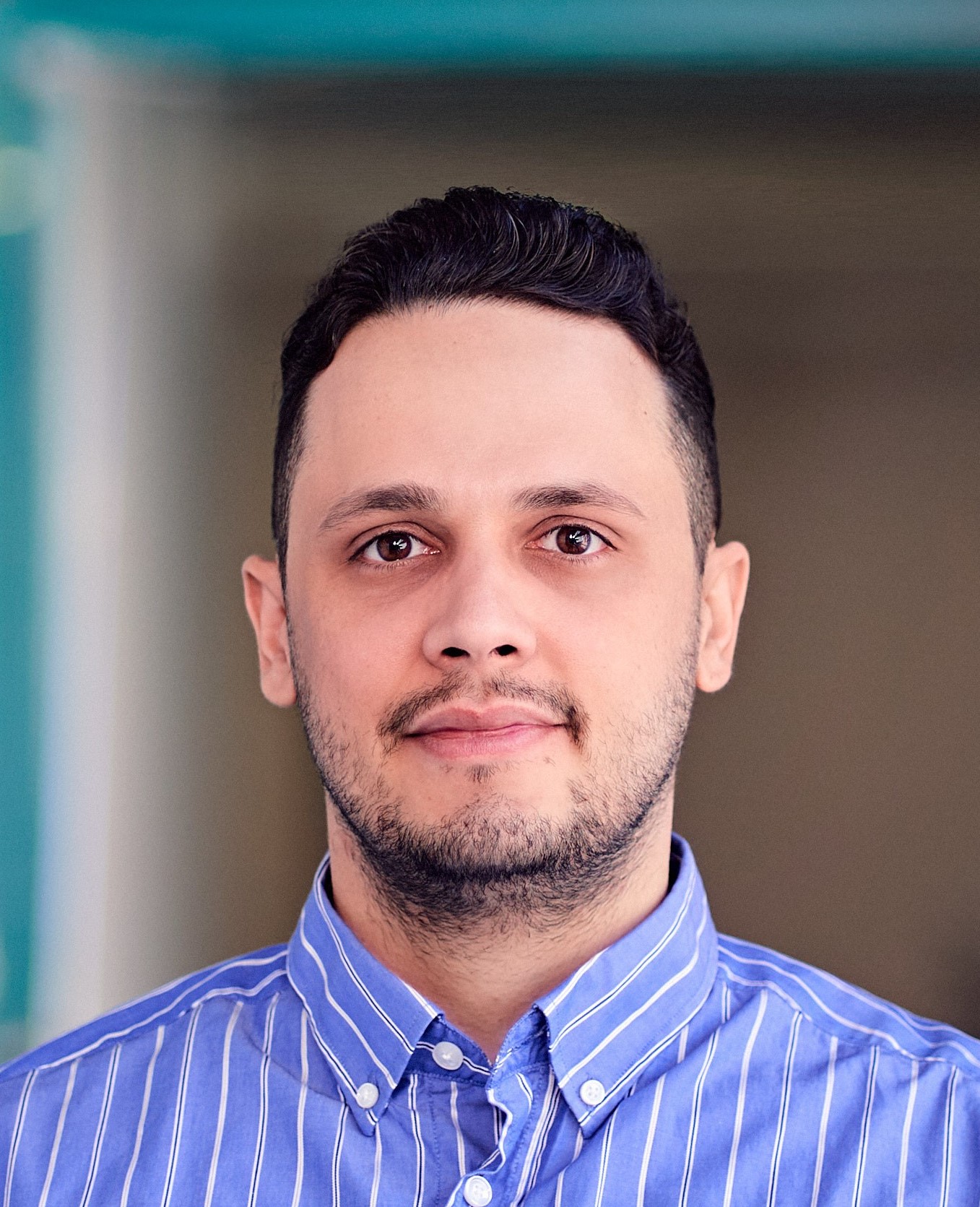}}]
{Onel L. A. L\'opez }  (Member, IEEE) (S'17-M'20) received the B.Sc. (1st class honors, 2013), M.Sc. (2017), and D.Sc. (with distinction, 2020) degree in Electrical Engineering from the Central University of Las Villas (Cuba),  the Federal University of Paran\'a (Brazil), and the University of Oulu (Finland), respectively.  

He is a collaborator to the 2016 Research Award given by the Cuban Academy of Sciences, a co-recipient of the 2019 IEEE EuCNC Best Student Paper Award, and the recipient of the 2020 Best Doctoral Thesis Award granted by Finland TEK and TFiF in 2021. He is co-author of the book entitled ``Wireless RF Energy Transfer in the Massive IoT Era: Towards Sustainable Zero-energy Networks'', Wiley, Dec 2021. He currently holds an Assistant Professorship (tenure track) in sustainable wireless communications engineering at the Centre for Wireless Communications (CWC), Oulu, Finland. His research interests include wireless communications, signal processing, sustainable IoT, and wireless RF energy transfer.\par
\end{IEEEbiography}

\begin{IEEEbiography}
[{\includegraphics[width=1in,height=1.25in,clip,keepaspectratio]{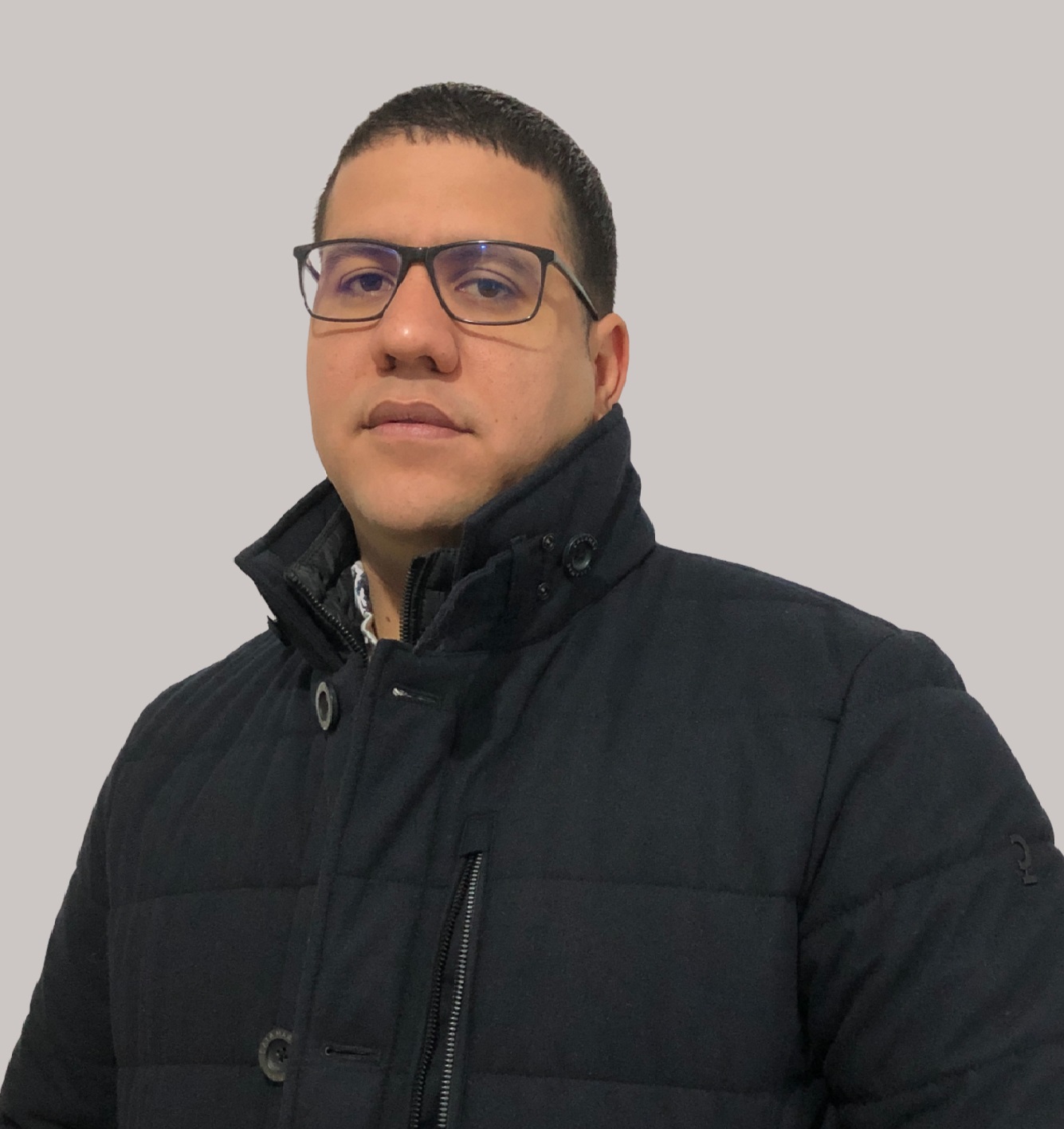}}]
{Osmel M. Rosabal } (Student Member, IEEE) 
received the B.Sc. (Hons., 2013) degree in Electrical Engineering and M.Sc. (Hons., 2021) degrees from the Central University of Las Villas (Cuba) and the University of Oulu (Finland), respectively, both in Electrical Engineering. From 2013-2018 he served as a specialist in telematics at the Cuban telecommunications company (ETECSA). He is currently pursuing his Ph.D. in Wireless Communications Engineering with a focus on sustainable wireless energy transfer systems, at the Centre for Wireless Communications (CWC), University of Oulu, Finland.\par
\end{IEEEbiography}

\begin{IEEEbiography}[{\includegraphics[width=1in,height=1.25in,clip,keepaspectratio]{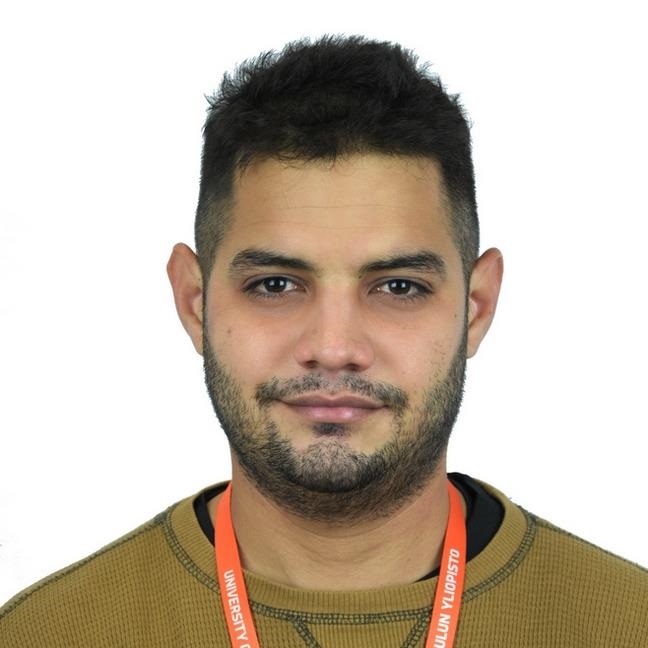}}]
{David E. Ruiz-Guirola } (Student Member, IEEE) received the B.Sc. (1st class honors, 2018) and M.Sc. (with distinction, 2019) degree in Telecommunications and Electronic Engineering from the Central University of
Las Villas (UCLV), Santa Clara, Cuba. From 2018-2021 he served as an Associate Professor at UCLV. He joined the
Centre for Wireless Communications, University of Oulu, Finland, in 2021. He is currently pursuing a Ph.D. degree at the University of Oulu. His research interests include sustainable IoT, energy harvesting, wireless RF energy transfer, machine-type communications, machine learning, and traffic prediction.\par
\end{IEEEbiography}

\begin{IEEEbiography}
[{\includegraphics[width=1in,height=1.25in,clip,keepaspectratio]{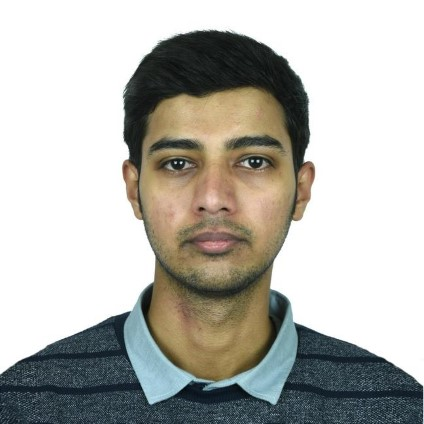}}]
{Prasoon Raghuwanshi } (Student Member, IEEE) received the B.Tech and M.Tech degrees in electronics $\&$ communication engineering from the National Institute of Technology Hamirpur, Hamirpur, India, in 2020. He is currently pursuing a D.Sc. degree in communications engineering from the University of Oulu, Oulu, Finland. He joined the Centre for Wireless Communications, University of Oulu, in 2022. His research interests include random access protocols for IoT networks, machine-type communications, deep reinforcement learning, and TinyML. \par
\end{IEEEbiography}

\begin{IEEEbiography}[{\includegraphics[width=1in,height=1.25in,clip,keepaspectratio]{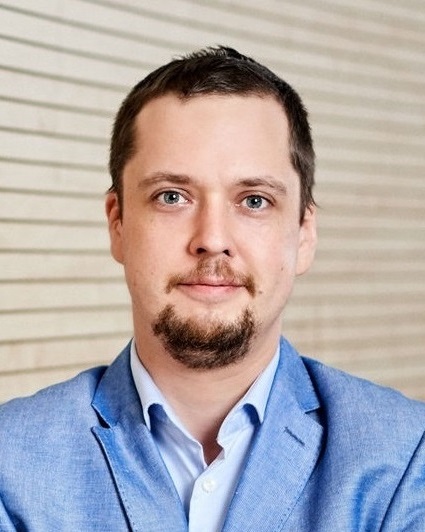}}]
{Konstantin Mikhaylov } (Senior Member, IEEE) is the Assistant Professor for Convergent IoT Communications with the Centre for Wireless Communications at the University of Oulu, Oulu, Finland. His research focuses on radio access and beyond-access technologies and protocols for massive and dependable IoT and the matters related to the design and practical use of IoT systems. He has authored or co-authored more than one hundred research papers on wireless connectivity for IoT, IoT system and device design, and applications. \par
\end{IEEEbiography}

\begin{IEEEbiography}[{\includegraphics[width=1in,height=1.25in,clip,keepaspectratio]{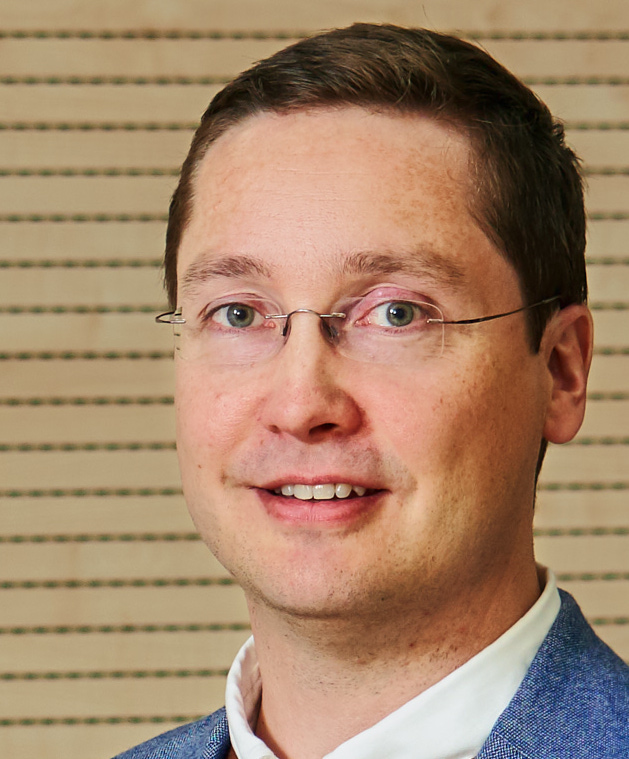}}]
{Lauri Lov\'en } (Senior Member, IEEE) D.Sc.(Tech), is a postdoctoral researcher in the 6G Flagship research program at the Center for Ubiquitous Computing (UBICOMP), University of Oulu, in Finland. He received his D.Sc. at the University of Oulu in 2021, and has worked at the Distributed Systems Group, TU Wien, in 2022. His current research concentrates on edge intelligence, and in particular on the orchestration of resources as well as distributed learning and decision-making in the computing continuum. He has co-authored 2 patents and ca. 40 research articles in international peer-reviewed journals, conferences, and workshops.\par
\end{IEEEbiography}

\begin{IEEEbiography}[{\includegraphics[width=1in,height=1.25in,clip,keepaspectratio]{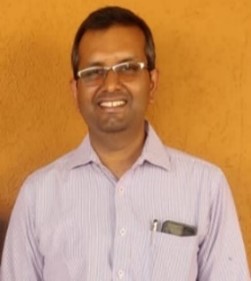}}]
{Sridhar Iyer } (Senior Member, IEEE) received the M.S. degree from New Mexico State University, U.S.A in 2008, and the Ph.D. degree from Delhi University, India in 2017. He received the Young Scientist award from the DST/SERB, Govt. of India in 2013, and Young Researcher Award from Institute of Scholars in 2021. He is the Recipient of the `Protsahan Award' from IEEE ComSoc, Bangalore Chapter as a recognition to his contributions towards paper published/tutorial offered in recognized conferences/journals, during Jan 2020-Sep 2021, and during Oct 2021-Oct 2022). His current research focus includes semantic communications and spectrum enhancement techniques for Intelligent Wireless Systems. Currently, he serves as Professor in the Dept. of CSE(AI), KLE Technological University, Dr M.S. Sheshgiri Belagavi Campus, Karnataka, India. \par 
\end{IEEEbiography}

\end{document}